\documentclass{jfm}

\usepackage
[
bookmarks=true,
pdfborder={0 0 0},
linktocpage,
breaklinks=true,
colorlinks,
citecolor=blue,
filecolor=black,
linkcolor=red,
urlcolor=red
pdftex
]{hyperref}

\usepackage{multirow}
\usepackage{array}
\usepackage{subfiles}
\usepackage{graphicx}
\usepackage{epstopdf,epsfig,subfloat,subfig,floatrow}
\usepackage{newtxtext}
\usepackage{newtxmath}
\usepackage{natbib}
\usepackage{amsfonts}
\usepackage{amsmath}
\usepackage{amssymb,color,mathtools}
\usepackage{bm}

\shorttitle{Gas kinetic model with velocity-dependent collision frequency }
\shortauthor{Ruifeng Yuan and Lei Wu}

\title{Capturing the influence of intermolecular potential in rarefied gas flows by a kinetic model with velocity-dependent collision frequency}

\author{     Ruifeng Yuan
	\and
         Lei Wu
        \corresp{\email{wul@sustech.edu.cn}}
        }

\affiliation{
 Department of Mechanics and Aerospace Engineering, Southern University of Science and Technology, Shenzhen 518055, China
  }

\usepackage{lineno}

\begin{document}
\maketitle

\begin{abstract}
	A kinetic model called the $\nu$-model is proposed to replace the complicated Boltzmann collision operator in the simulation of rarefied flows of monatomic gas. The model follows the relaxation-time approximation, but the collision frequency (i.e, inverse relaxation time) is a function of the molecular velocity to reflect part of the collision details of the Boltzmann equation, and the target velocity distribution function (VDF) to which the VDF relaxes is close to that used in the Shakhov model. Based on the numerical simulation of strong non-equilibrium shock waves, a half-theoretical and half-empirical collision frequency is designed for different intermolecular potentials: the $\nu$-model shows significantly improved accuracy, and the underlying mechanism is analysed. The $\nu$-model also performs well in canonical rarefied micro-flows, especially in the thermal transpiration, where the conventional kinetic  models  with velocity-independent collision frequency lack the capability to distinguish the influence of intermolecular potentials. 
\end{abstract}

\section{Introduction}\label{sec:introduction}

The Boltzmann equation is the fundamental equation in the study of rarefied gas dynamics that has found applications in space vehicle re-entry~\citep{Ivanov1998Review}, microelectromechanical system processing~\citep{Karniadakis2005}, vacuum technology~\citep{sharipov1998data,Sone2002Book}, and shale gas extraction~\citep{Wu2016JFM,Wu2017JFM2}. In Boltzmann’s description, all molecules move in straight lines with fixed velocities until they encounter elastic collisions with other molecules. The free transport is described by the streaming operator, while the binary collision is modelled by the Boltzmann collision operator, which is a nonlinear function of the  velocity distribution function (VDF) and incorporates the effect of intermolecular potential. In the past century, the complicated structure of the Boltzmann collision operator has stimulated the development of kinetic models that strive to imitate as closely as possible the behaviour of the Boltzmann equation. In gas kinetic modelling, the streaming operator remains unchanged, while the Boltzmann collision operator is replaced by simpler expressions, not only making the problems tractable, but also reducing the computational cost. For example, in the deterministic solver, the computational complexity of the Boltzmann collision operator solved by the fast spectral method is about $O(M^2N^3\log{N})$, where $N$ is the number of discretized velocity grid in each velocity direction, and $M^2\sim{N}$ is the number of discretized solid angle~\citep{Lei2013}. However, the computational cost for the kinetic models is only $O(N^3)$.

Several basic considerations are taken into account when simplifying the Boltzmann collision operator~\citep{henning}. First, the conservation laws of mass, momentum and energy must be satisfied. Second, the VDF must be reduced to the Maxwellian equilibrium distribution when the gas system reaches equilibrium. Third, transport coefficients such as the shear viscosity and thermal conductivity derived from the kinetic model equation should coincide with those from the Boltzmann equation. Fourth, the H-theorem, which states that the production of entropy is always positive and vanishes only if the system is in equilibrium, should be satisfied. Note that while the first two are basic physical requirements, and the third one is crucial as it yields consistent solutions with the Boltzmann equation in the continuum flow regime (governed by the Navier-Stokes-Fourier equations), 
the fourth requirement can be loosen. This is because in most rarefied gas flows the fulfilment of H-theorem does not necessary guarantee the accuracy of kinetic models: if a kinetic model is exactly the same as the Boltzmann equation, then the entropy production rate should be the same as well; however, this is in general impossible as so far no kinetic model satisfies this condition. In fact, as we will see later, the \cite{Shakhov1968,Shakhov_S} kinetic model, where the H-theorem has not been proven in nonlinear cases, usually performs better than the ellipsoidal-statistical model (ESBGK) that satisfies the H-theorem~\citep{Holway1966}.


Since the Boltzmann collision operator can be decomposed into the gain term $Q^+$ and loss term $\nu f$ as $Q=Q^+ -\nu f$, the modelled collision operator is often formulated in the relaxation-time approximation: 
\begin{equation}\label{overall_kinetic_model}
Q=\nu\left[f_{r}(t,\bm{v},\bm{x})-f(t,\bm{v},\bm{x})\right],
\end{equation}
where $t$ is the time, $\bm{x}=(x_1,x_2,x_3)$ is the spatial coordinate, $\bm{v}=(v_1,v_2,v_3)$ is the molecular velocity,  $\nu$ is the collision frequency (inverse relaxation time), and $f_{r}$ is the target VDF. Therefore, the two terms to be modelled are $f_{r}$ and $\nu$, which are connected with the gain and loss terms of the Boltzmann collision operator, respectively. Many relaxation-type kinetic models assume $\nu$ to be a constant throughout the molecular velocity space and concentrate on the modelling of $f_{r}$. Three popular kinetic models of this kind 
are the BGK model~\citep{Bhatnagar1954}, ESBGK model~\citep{Holway1966}, and the Shakhov model~\citep{Shakhov1968}. The BGK model cannot recover the shear viscosity and thermal conductivity simultaneously, hence it will not be discussed in this paper. The ESBGK model satisfies the H-theorem, while the Shakhov model satisfies the H-theorem only in linearised flows; nevertheless, the latter often predicts better results~\citep{Chensz2013,liu2014investigation} over a wide range of Knudsen number ($Kn$, the ratio of molecular mean free path to  characteristic flow length). 

It is noted that although these models assume velocity-independent collision frequency, the collision frequency of the Boltzmann collision operator depends on the molecular velocity and this dependence influences the rarefied gas dynamics~\citep{Cercignani2000Book,Zheng2005}. For example, in the linearised Poiseuille flow and thermal transpiration, the Boltzmann equation yields different solutions for different intermolecular potentials even when the viscosity is same~\citep{Sharipov2009,Takata2011,lei_Jfm,wuPoF2015}. However, the ESBGK model and the Shakhov model do not have this capability: after linearisation their collision operators are only determined by the value of shear viscosity at some reference temperature.

To increase the accuracy of kinetic models, it would be highly desirable to add more information to the collision frequency and target VDF. Based on the eigenvalues and eigenfunctions of the linearised Boltzmann collision operator for Maxwellian molecules (see~\eqref{power_law_potential} below), \cite{GrossJackson1959} proposed a systematic way to construct kinetic models with arbitrary order of accuracy. However, this is only limited to the linearised flow of Maxwellian gas. To be more general, the relaxation model~\eqref{overall_kinetic_model} with velocity-dependent collision frequency becomes a natural consideration. To this end, kinetic models based on eigenfunctions of linearised Boltzmann operator combined with variable collision frequency have been proposed by \cite{Cercignani1966} and \cite{Loyalka1967,Loyalka1968}; however the flow cases considered in these researches are limited to the simple velocity and temperature slip problems where the variation of collision frequency has very limited influence on the slip coefficients. Relevant work has also been done by \cite{larina2007models}, but the linearised variable-collision-frequency model performs even worse than the constant-collision-frequency one. For the nonlinear case, \cite{krook1959continuum} and \cite{cercignani1975theory} have mentioned a BGK-type model with velocity-dependent collision frequency and Maxwellian-type $f_r$. This model is further developed by  \cite{Struchtrup_velocity} and \cite{Mieussens2004},  where the collision frequency $\nu$ is some power-law functions of the molecular velocity and the model is called  the $\nu$-BGK model.  The $\nu$-BGK model, however, fails to  satisfactory predict the normal shock wave and the Couette flow. \cite{Zheng2005} then  developed the $\nu$-ESBGK model, where a more physically-meaningful collision frequency derived from the loss term of the Boltzmann collision operator is applied. It performs better than the $\nu$-BGK in the normal shock wave, but shows worse  accuracy  than the standard ESBGK model in the simulation of Couette flow. 

Besides the above relaxation-time approximations, the Fokker-Planck model~\citep{Jenny2010JCP,Gorji2011,Gorji2013} is another popular kinetic model. This model is applied to rarefied gas dynamics because, when compared to the direct simulation Monte Carlo method~\citep{Bird1994}, it allows much larger time step in the near-continuum flow regimes where $Kn\ll1$, and hence  reduces the computational cost significantly. In terms of the model accuracy, despite its more complicated formulation, the Fokker-Planck model does not to have absolute advantage over relaxation-type models in the transition flow regime where $Kn\sim1$. For instance, in the simulation of normal shock waves, it is found that the Fokker-Planck model works well for the argon gas where the viscosity index (see~\eqref{temperature_dependence} below) is $\omega=0.81$, but its predication capability deteriorates for hard-sphere and Maxwell molecules~\citep{Liu2019PRE,Fei2020AIAA}, where $\omega=0.5$ and 1, respectively. Moreover, like the BGK, ESBGK, and Shakhov models, this model does not distinguish the influence of different intermolecular potentials in the simulation of Poiseuille flow and thermal transpiration~\citep{Sharipov2009}, as well as the Rayleigh-Brillouin scattering~\citep{LeiJFM2015}.

In view of the above facts, we aim to further develop the relaxation model~\eqref{overall_kinetic_model}, with  velocity-dependent collision frequency  to recover more  details of the Boltzmann collision operator, while keep the computation complexity in an affordable level. The $\nu$-model we propose adopts the velocity-dependent collision frequency based on the equilibrium collision frequency of the Boltzmann collision operator with empirical modification. The influence of intermolecular potential~\citep{Sharipov2009} is appropriately accounted for, including the Lennard-Jones potential which is accurate in a wide range of temperature. To recover the correct Prandtl number, considering the fact that the Shakhov model often performs better than the ESBGK model~\citep{Chensz2013,liu2014investigation}, a Shakhov-type target VDF is employed. With this two critical improvements, we find that the model accuracy is greatly improved; moreover, the multiscale numerical method that is efficient from the continuum to free-molecular flow regimes can be adopted, and the computational cost only increases slightly when compared to conventional kinetic models.

The rest of the paper is organized as follows. The Boltzmann equation, as well as the transport coefficients and equilibrium collision frequency, are introduced in section~\ref{sec:Boltzmann}. In section~\ref{sec:kineticmodel}, the BGK, ESBGK, and Shakhov models are introduced and our $\nu$-model is proposed. In section~\ref{sec:numerical_method}, a multiscale numerical method is developed to solve the proposed model equation deterministically. In sections~\ref{sec:numerical_results} and~\ref{num_micro_flows}, the accuracy of our model is assessed by numerous canonical test cases and the underlying mechanisms on how the $\nu$-model improves the results are discussed. The summary and outlooks are given in section~\ref{sec:conclusion}.

\section{The Boltzmann equation}\label{sec:Boltzmann}

A fundamental theory at the mesoscopic level that bridges the microscopic and mesoscopic behaviours is highly demanded to describe the rarefied gas dynamics. As we are not interested in the individual dynamics of gas molecules but their collective behaviours, the VDF $f(t,\bm{x},\bm{v})$ is introduced to describe the state of gaseous system. It is defined in such a way that the quantity $f(t,\bm{x},\bm{v})d\bm{x}d\bm{v}$ is the molecular number in the phase-space volume $d\bm{x}d\bm{v}$, therefore,  macroscopic quantities such as the molecular number density $n(t,\bm{x})$, flow velocity $\bm{u}(t,\bm{x})$, temperature $T(t,\bm{x})$, pressure tensor $p_{ij}(t,\bm{x})$, and heat flux $\bm{q}(t,\bm{x})$ can be calculated as:
\begin{equation}\label{macroscopic_origin}
\begin{aligned}[b]
[n,\bm{u}, T, p_{ij}, \bm{q}]=\int{}\left[1,\frac{\bm{v}}{n},\frac{m}{3k_Bn(t,x)}c^2,mc_ic_j,\frac{m}{2}c^2\bm{c}\right]f(t,\bm{x},\bm{v})d\bm{v}, 
\end{aligned}
\end{equation}
where $\bm{c}=\bm{v}-\bm{u}$ is the peculiar velocity, $k_B$ is the Boltzmann constant, and $m$ is the molecular mass. Note that the ideal gas law holds for dilute gas, where the gas pressure is $p=nk_BT$. Also, we introduce the pressure deviation tensor $\sigma_{ij}$ as $\sigma_{ij}=p_{ij}-p\delta_{ij}$, where $\delta$ is the Kronecker function. 

In the absence of external force, the Boltzmann equation reads
\begin{equation}  \label{Boltzmann}
\begin{aligned}[b]
\frac{\partial {f}}{\partial{t}}+{\bm{v}}\cdot\frac{\partial
	{f}}{\partial{\bm{x}}}=
\iint
B(|\bm{v}_r|,\theta)
[f(t,\bm{x},\bm{v}'_\ast)f(t,\bm{x},\bm{v}')-f(t,\bm{x},\bm{v}_\ast)f(t,\bm{x},\bm{v})]d\Omega d{\bm{v}}_\ast,
\end{aligned}
\end{equation}
where the term in the right-hand side is the Boltzmann collision operator. The subscript $\ast$ represents the second molecule in the binary collision, the superscript $'$ stands for quantities after the collision, $\bm{v}_r=\bm{v}-\bm{v}_\ast$ is the relative pre-collision velocity, and $\theta$ is the deflection angle. The post-collision molecular velocities are given by $\bm{v}'=\bm{v}+\frac{|\bm{v}_r|\Omega-\bm{v}_r}{2}$ and $
\bm{v}'_\ast=\bm{v}_\ast-\frac{|\bm{v}_r|\Omega-\bm{v}_r}{2}$.
where $\Omega$ is the solid angle. The deflection angle $\theta$ between the pre- and post-collision relative velocities satisfies $\cos\theta= \Omega\cdot\bm{v}_r/|\bm{v}_r|$, $0\le\theta\le\pi$.

The collision kernel $B(|\bm{v}_r|,\theta)$ in the Boltzmann collision operator is a product of the differential cross-section $\sigma_D$ and the relative collision speed:
\begin{equation}\label{cross_section0}
B(|\bm{v}_r|,\theta)=\sigma_D|\bm{v}_r|\equiv\frac{b|db|}{\sin\theta|d\theta|}|\bm{v}_r|,
\end{equation}
which is always non-negative. Given the intermolecular potential $\phi$ and the aiming distance $b$ between two colliding molecules, the deflection angle can be calculated either from the classical mechanics or quantum mechanics. When the gas temperature is not too low, both methods yield the same transport coefficients~\citep{Sharipov_quantum_DCS2017}. Therefore, we take the classical mechanics:
\begin{equation}\label{deflection}
\theta(b,{v}_r)=\pi-2\int_0^{W_1}\left[1-W^2-\frac{4\phi(r)}{m{v}_r^2}\right]^{-1/2}dW,
\end{equation}
where $W=b/r$ with $r$ being the intermolecular distance, and $W_1$ is positive root of  the term in brackets. In gas kinetic theory, the inverse power-law potentials are normally considered:
\begin{equation}\label{power_law_potential}
\phi(r)=\frac{K}{\eta-1}r^{1-\eta},
\end{equation}
although the Lennard-Jones potential is more realistic (it is widely used in the molecular dynamics simulation):
\begin{equation}\label{Lennard_Jones_chapter}
\phi(r)=4\epsilon\left[\left(\frac{d_{LJ}}{r}\right)^{12}-\left(\frac{d_{LJ}}{r}\right)^6\right],
\end{equation}
where $\epsilon$ is the potential depth,  and $d_{LJ}$ is the distance between two molecules where the potential is zero. The power-law potentials are called hard- and soft-potentials when $\eta>5$ and $\eta<5$, respectively. Maxwell molecules have the potential with $\eta=5$. Another special case is the hard-sphere gas, where the repulsive potential is infinity (and zero) when $r$ is less (larger) than the molecular diameter $\sigma$.


For the power-law potential, it is seen from ~\eqref{deflection} that the deflection angle is only a function of $s=\left[\frac{m(\eta-1)}{4K}\right]^{\frac{1}{\eta-1}}bv^{\frac{2}{\eta-1}}_r$.
That is, $\theta=\theta(s)$. Thus, the differential cross-section is
\begin{equation}\label{DCS_chapter1}
B(|\bm{v}_r|,\theta)=\left(\frac{m(\eta-1)}{4K}\right)^{\frac{2}{1-\eta}}v^{\frac{\eta-5}{\eta-1}}_r\times\underbrace{\frac{sds}{\sin\theta{d\theta}}}_{\Theta(\theta)}.
\end{equation}
For Maxwell molecules, the collision kernel is independent of the relative collision speed, while for hard-sphere gas the collision kernel is independent of the deflection angle: 
$
B(|\bm{v}_r|,\theta)=\frac{\sigma^2}{4}|\bm{v}_r|.
$

\subsection{Transport coefficients and modelled collision kernel}

The collision kernel $B(|\bm{v}_r|,\theta)$ determines the transport coefficients such as the shear viscosity and thermal conductivity. In the continuum flow regime, the Navier-Stokes-Fourier equations can be derived from the Chapman-Enskog expansion of the Boltzmann equation, where the shear viscosity is given by~\citep{CE}
\begin{equation}\label{shear_CE_viscosity0}
\mu=\frac{5\sqrt{\pi{m}k_BT}}{8D},
\quad
D=\left(\frac{m}{4k_BT}\right)^4\int_0^\infty
v_r^7\sigma_{\mu}\exp\left(-\frac{mv^2_r}{4k_BT}\right)dv_r,
\end{equation}
with $
\sigma_{\mu}=2\pi\int_0^\pi\sigma_D{\sin^3\theta}d\theta$.
The corresponding thermal conductivity is given by
\begin{equation}\label{shear_CE_thermal0}
{\kappa}=\frac{15}{4}\frac{k_B}{m}\mu,
\end{equation}
which results in a Prandtl number of $\Pr=\frac{5k_B}{2m}\frac{\mu}{\kappa}=\frac{2}{3}$.

Therefore, for the inverse-power potential, we have $\mu\propto{T^\omega}$, where
\begin{equation}\label{temperature_dependence}
\quad
\omega=\frac{\eta+3}{2(\eta-1)}
\end{equation}
is the viscosity index; for the Lennard-Jones potential, the viscosity is not a power-law function of the temperature, since $D$ is approximated by~\citep{Lei2013} 
\begin{equation}\label{LJ_D}
\frac{D}{d_{LJ}^2}=b_1\left(\frac{k_BT}{\epsilon}\right)^{-0.4} +b_2\left(\frac{k_BT}{\epsilon}\right)^{-0.45}+b_3\left(\frac{k_BT}{\epsilon}\right)^{-0.5},
\end{equation}
where $b_1=407.4, b_2=-811.9$, and $b_3=414.4$; each term can be viewed as the inverse power-law potential with the viscosity indices $\omega_1=0.9$, $\omega_2=0.95$, and $\omega_3=1$, respectively. This expression is accurate when $1<k_BT/\epsilon<25$. It should be noted that this viscosity is more accurate than the Sutherland's formula.

In the direction simulation Monte Carlo method~\citep{Bird1963} and the fast spectral approximation of the Boltzmann collision operator~\citep{Lei2013}, the modelled collision kernels such as the variable hard-sphere and variable soft-sphere models are used: the transport coefficients are recovered, but the detailed form of $\Theta(\theta)$ in~\eqref{DCS_chapter1} is modified to make the computation simple. For example, in the inverse power-law potential and Lennard-Jones potential, the modelled collision kernel are respectively
\begin{equation}
\begin{aligned}[b]
 B&=\frac{5\sqrt{\pi{m}k_BT_0}(4k_BT_0/m)^{(1-\alpha)/2}}{64\pi\mu(T_0)\Gamma[(3+\alpha+\gamma)/2]\Gamma(2-\gamma/2)}\sin^{\alpha+\gamma-1}\left(\frac{\theta}{2}\right)
\cos^{-\gamma}\left(\frac{\theta}{2}\right)
|\bm{v}_r|^\alpha,\\
B&=\frac{d_{LJ}^2}{8\pi}\sum_{j=1}^3 \frac{({m}/{4\epsilon})^{(\alpha_j-1)/2}}{\Gamma\left(\frac{3+\alpha_j}{2}\right)}b_j
\sin^{\alpha_j-1}\left(\frac{\theta}{2}\right)
|\bm{v}_r|^{\alpha_j},
\end{aligned}
\end{equation}
where $\Gamma$ is the gamma function, $\mu(T_0)$ is the shear viscosity at the reference temperature $T_0$,
\begin{equation}
\alpha=\frac{\eta-5}{\eta-1}=2(1-\omega),
\end{equation}
and $\alpha_1=0.2, \alpha_2=0.1$, and $\alpha_3=0$. Note that $\gamma$ is a free parameter, the different value of which leads to different value of equilibrium collision frequency but always the same value of shear viscosity.



\subsection{Equilibrium velocity distribution and collision frequency}\label{equi_coll_chapter2}

It is well-known that in equilibrium the Boltzmann collision operator vanishes,  and the VDF takes the form of Maxwellian distribution
\begin{equation}\label{equilibrium_Maxwellian}
F_{eq}=n\left(\frac{m}{2\pi   k_BT}\right)^{3/2}\exp\left(-\frac{mc^2}{2k_BT}\right).
\end{equation}

If the total cross-section is finite (either through the cut-off of aiming distance or from the quantum calculation of differential cross-section), the Boltzmann collision operator can be separated into a gain term $Q^+ $ and a loss term $\nu{f}$ as
$
Q(f,f_*)=Q^+ -\nu(|\bm{v}|){f}$,
where the collision frequency is
\begin{equation}
\nu(|\bm{v}|)=\iint
B(|\bm{v}_r|,\theta)
f(\bm{v}_{\ast})
d\Omega
d\bm{v}_\ast.
\end{equation}

For inverse power-law potentials, the equilibrium collision frequency corresponding to the collision kernel~\eqref{DCS_chapter1} and equilibrium VDF~\eqref{equilibrium_Maxwellian} is~\citep{henning}:
\begin{equation}\label{equi_fre_chapter2}
\begin{aligned}[b]
\nu_{eq}(|\bm{v}|)=2\pi\frac{\eta-1}{3\eta-7}\nu_\eta^0
\underbrace{\int_0^\infty \frac{\xi_\ast}{\xi}\exp(-\xi^2_\ast)
\left[ (\xi+\xi_\ast)^{\frac{3\eta-7}{\eta-1}} -|\xi-\xi_\ast|^{\frac{3\eta-7}{\eta-1}}\right]
d\xi_\ast}_{\nu^0_{eq}},
\end{aligned}
\end{equation}
where 
\begin{equation}
\begin{aligned}
\nu_\eta^0&=\left(\frac{m(\eta-1)}{4K}\right)^{\frac{2}{1-\eta}}
\frac{2n}{\sqrt{\pi}}\sqrt{\frac{2k_BT}{m}}^{\frac{\eta-5}{\eta-1}}
\int sds,\\
\nu_{HS}^0&=n\sqrt{\frac{2k_BT}{m\pi}}\sigma^2,
\end{aligned}
\end{equation}
and $\xi=\frac{c}{v_m}$ with
\begin{equation}
v_m=\sqrt{\frac{2k_BT}{m}}
\end{equation}
being the most probable speed at temperature $T$.

Specifically, for Maxwellian molecules with $\eta=5$, the collision frequency is independent of the molecular velocity, and independent of the temperature:
$
\nu_{eq}=\pi^{3/2}\nu_5^0$,
while for hard-sphere molecules,
\begin{equation}\label{equi_fre_HS}
\nu_{eq}(|\bm{v}|)=n\sqrt{\frac{2\pi{k_B}T}{m}}\sigma^2
\underbrace{ \left[ \exp\left(-\xi^2\right)+\frac{\sqrt{\pi}}{2}\left(\frac{1}{\xi}+2\xi\right)\text{erf}\left(\xi\right) \right]}_{2\nu^0_{eq}/3},
\end{equation}
where $\text{erf}(x)$ is the Gauss error function.

\section{Kinetic models}\label{sec:kineticmodel}

In this section we first introduce the popular kinetic models with velocity-independent collision frequency, then we propose a new kinetic model where the collision frequency is a function of the molecular velocity. All these models share the same form of~\eqref{overall_kinetic_model}.

\subsection{Velocity-independent collision frequency}\label{GGK_model}

From~\eqref{overall_kinetic_model} we can see that in the velocity-independent collision-frequency model the only term to be modelled is the target VDF $f_{r}$, which is connected with the gain term of the Boltzmann collision operator and directly determines the velocity distribution of the post-collision molecules. The BGK model~\citep{Bhatnagar1954} adopts the local Maxwellian $F_{eq}$ to approximate $f_{r}$ and is the simplest kinetic model being widely used. One can easily verify that it satisfies the conservation laws. Also, in the equilibrium where the collision operator vanishes, we have $f=F_{eq}$, which fulfils the second requirement of kinetic modelling. The H-theorem can also be proven. However, from the Chapman-Enskog expansion, it can be found that the shear viscosity and thermal conductivity are 
\begin{equation}\label{collision_frequency_shear}
\mu=\frac{p}{\nu}, \quad
\kappa=\frac{p}{\nu}\frac{5k_B}{2m},
\end{equation}
which results in a Prandtl number of unity. That is to say, the BGK model cannot recover the viscosity and thermal conductivity simultaneously in the continuum limit.  Therefore, many kinetic models have been proposed to correct the Prandtl number, among which the  ESBGK model~\citep{Holway1966} and the Shakhov model~\citep{Shakhov1968,Shakhov_S} are two of the most popular kinetic models. 

In the  ESBGK model of~\cite{Holway1966}, the target VDF is obtained by maximizing the entropy function $H=-\int{f\ln}fd\bm{v}$ under the given information of mass, momentum, energy, and the stress tensor. This can be finished by the Lagrange multipliers method and the target VDF finally has a form of an anisotropic Gaussian:
\begin{equation}\label{ellipsoidal_model}
\begin{split}
f_r^{ES}=\frac{n}{\sqrt{\operatorname{det}
		[2\pi\lambda_{ij}]}}\exp\left(-\frac{1}{2}\lambda_{ij}^{-1}c_i{c_j}\right),
\end{split}
\end{equation}
where
\begin{equation}\label{ellipsoidal_model_11}
\lambda_{ij}=\frac{k_BT(1-b)}{m}\delta_{ij}+\frac{bp_{ij}}{nm}=\frac{p\delta_{ij}+b\sigma_{ij}}{nm},
\end{equation}
with a constant $b$. If $b=0$, the tensor $\lambda_{ij}$ becomes diagonal, and the BGK model is recovered. According to the Chapman-Enskog expansion, the transport coefficients are
\begin{equation}
\mu=\frac{p}{\nu(1-b)},
\quad
\kappa=\frac{p}{\nu}\frac{5k_B}{2m}.
\end{equation}
Therefore, $b$ should take the value of $-\frac{1}{2}$ to produce a Prandtl number of $\frac{2}{3}$ for monatomic gas.

The ESBGK model satisfies the mass, momentum and energy conservations, as well as the H-theorem~\citep{Andries2000}. On the other hand, although at first sight it may appear that the VDF is guided toward the target one which is not the equilibrium distribution. However, in spatial-homogeneous problems we have
\begin{equation}
\frac{\partial \sigma_{ij}}{\partial t}=-\frac{p}{\mu}\sigma_{ij},
\end{equation} 
which means that the deviational stress will be decayed to zero. 
Thus, the route to equilibrium of the ESBGK model is as follows: as $f$ approaches $f_r^{ES}$, $f_r^{ES}$ itself approaches $F_{eq}$ as per equations~\eqref{ellipsoidal_model} and~\eqref{ellipsoidal_model_11}; eventually $f=F_{eq}$ when the equilibrium state is reached.  Therefore,  the ESBGK model satisfies all the four requirements of kinetic modelling (see section~\ref{sec:introduction}), and it has attracted great attentions. 

In contrast to the ESBGK model where the stress tensor is introduced in the target VDF, in the Shakhov model the heat flux is introduced on top of the BGK model~\citep{Shakhov1968,Shakhov_S} through the Hermit polynomial: 
\begin{equation}\label{smodel}
f_r^S=F_{eq}\left[1+(1-\operatorname{Pr})\frac{2m\bm{q}\cdot
	\bm{c}}{5n(k_BT)^2}\left(\frac{mc^2}{2k_BT}-\frac{5}{2}\right)\right],
\end{equation}
where the two transport coefficients are
\begin{equation}
\mu=\frac{p}{\nu}, \quad
\kappa=\frac{1}{\text{Pr}}\frac{5k_B}{2m}\frac{p}{\nu}.
\end{equation}
Thus, the correct value of Prandtl number is recovered.  The route to equilibrium of the Shakhov model is as follows: as $f$ approaches $f_r^{S}$, $f_r^{S}$ itself approaches $F_{eq}$ since in spatial-homogeneous problems the heat flux decays to zero according to the equation
\begin{equation}\label{heat_relax_Shakhov}
\frac{\partial \bm{q}}{\partial t}=-\frac{2}{3}\frac{\mu}{p}\bm{q}.
\end{equation}
Eventually $f=F_{eq}$ when the equilibrium state is reached.  

Comparing with the ESBGK model, theoretically the Shakhov model has two shortcomings. First, the H-theorem can be proved only for linearised flows, while one can neither prove nor disprove the H-theorem in nonlinear flows. Second, the VDF may become negative, which is not physical. However, despite the two deficiencies, the Shakhov model has been widely used, and often  performs better than the ESBGK model. 


\subsection{Velocity-dependent collision frequency}

Kinetic models with velocity-dependent collision frequency have been investigated in very early history. 
For the linearised Boltzmann equation, \cite{Cercignani1966} and \cite{Loyalka1967,Loyalka1968} have presented variable-collision-frequency models based on  eigenfunctions of the linearised operator. These models are applied to simple velocity and temperature slip problems, and a limited influence on the slip coefficient  due to the variation of collision frequency has been found.  \cite{larina2007models} have also developed a linearised model with velocity-dependent collision frequency, but in the simulation of normal shock wave,  their model performs even worse than its constant-collision-frequency counterpart. For the nonlinear case, \cite{krook1959continuum} and \cite{cercignani1975theory} have mentioned a variable-collision-frequency model where the target VDF is approximated as a Maxwellian with modified density, velocity and temperature determined by the collision conservation condition. Further investigations about this model have been done by \cite{Struchtrup_velocity} and \cite{Mieussens2004}, where the collision frequency  is some power-law functions of the molecular velocity; the model is called the $\nu$-BGK, but the numerical results for normal shock wave are not satisfactory.


It is interesting to note that although the original motivation of developing kinetic models with velocity-dependent collision frequency is to correct the  Prandtl number of the standard BGK model, it is found that setting $\nu$ to be the equilibrium collision frequency of the Boltzmann equation in the $\nu$-BGK model leads to an approximate unit Prandtl number~\citep{Mieussens2004}. This suggests that the wrong Prandtl number of the standard BGK model is mainly due to the error in target VDF $f_{r}$ (the gain term), but not the error of collision frequency (the loss term). Therefore, it may be not reasonable to adjust the Prandtl number through modifying the collision frequency $\nu$. In contrast, one should modify the target VDF to guarantee a right Prandtl number while applying a physically meaningful collision frequency. This has been done by~\cite{Zheng2005} in their $\nu$-ESBGK model, where the equilibrium collision frequency is applied and an ESBGK-type target VDF is adopted to adjust the Prandtl number. The $\nu$-ESBGK model performs better  than the $\nu$-BGK and ESBGK models in the shock wave simulation, but performs worse than standard ESBGK in Couette flow \citep{Zheng2005}.


In view of the fact that the Shakhov model often performs  better than the ESBGK model~\citep{Chensz2013,liu2014investigation},  we design the target VDF as
\begin{equation}\label{beyond_SM2}
\begin{aligned}[b]
f_r=\left[\hat{\varrho}+\hat{\Gamma}c^2+\gamma_ic_i
+\beta\frac{2m\bm{q}\cdot
	\bm{c}}{5n(k_BT)^2}\left(\frac{mc^2}{2k_BT}-\frac{5}{2}\right) \right] F_{eq},
\end{aligned}
\end{equation}
where $\hat{\varrho}$, $\hat{\Gamma}$ and $\gamma_i$ are velocity-independent, which can be solved directly from the conservation condition. Note that in the $\nu$-BGK model~\citep{Mieussens2004} $\hat{\Gamma}$ and $\gamma_i$ appear in the exponential function so Newton’s iteration method should be applied; here we put them in the brackets to avoid the use of Newton’s iteration method in the numerical simulation. The heat flux term as that in the Shakhov model is used, and the velocity-independent parameter $\beta$ is used to adjust the Prandtl number. Thus, the collision frequency can be arbitrary function of the molecular velocity. When $\nu$ is velocity-independent, this model will be reduced to the \cite{Shakhov1968,Shakhov_S} model.

Note that in the current work, the velocity-dependent collision frequency $\nu(|\bm v|)$ is isotropic. Applying the Chapman-Enskog expansion, the VDF to the first-order approximation reads
\begin{equation}\label{f1_MS_beyond}
\begin{aligned}[b]
f=&F_{eq}(\hat{\varrho}+\hat{\Gamma}c^2+\gamma_ic_i)
+F_{eq}
\beta\frac{2m\bm{q}\cdot
	\bm{c}}{5n(k_BT)^2}\left(\frac{mc^2}{2k_BT}-\frac{5}{2}\right) \\
-&\frac{F_{eq}}{\nu(|\bm v|)}
\left\{ \frac{m}{k_BT}\frac{\partial u_{\langle i}}{\partial x_{j\rangle}} c_{\langle i}c_{j\rangle} +\frac{1}{T}\frac{\partial T}{\partial x_i}c_i\left(\frac{mc^2}{2k_BT}-\frac{5}{2}\right) \right\}.
\end{aligned}
\end{equation}
where $\hat{\varrho}=1, \hat{\Gamma}=0$, and 
\begin{equation}\label{gammi_i}
\gamma_i=\frac{8}{3\sqrt{\pi}}\frac{1}{T}\frac{\partial{T}}{\partial{x_i}} 
\int_0^\infty\frac{\xi^4}{{\nu}(\xi)}\left(\xi^2-\frac{5}{2}\right)\exp(-\xi^2)d{\xi}.
\end{equation}
 Therefore, the shear viscosity and thermal conductivity are
\begin{equation}\label{shear_beyond}
\begin{aligned}[b]
\mu=&\frac{16p}{15\sqrt{\pi}}\int_0^\infty \frac{\xi^6}{\nu(\xi)}\exp(-\xi^2)d{\xi},\\
\kappa=&\frac{1}{1-\beta}\frac{16p}{15\sqrt{\pi}}\frac{5k_B}{2m}\int_0^\infty \frac{\xi^4\left(\xi^2-\frac{5}{2}\right)^2}{\nu(\xi)}\exp(-\xi^2)d{\xi}.
\end{aligned}
\end{equation}
Thus, to recover the viscosity, an arbitrary positive collision frequency function $v'(\xi )$ can be used in the $\nu$-model with the normalization
\begin{equation}\label{eq:nu_normalization}
\begin{aligned}
\nu (\xi ) = A\frac{p}{\mu }\nu '(\xi ), \quad
A = \frac{{16}}{{15\sqrt \pi  }}\int_0^\infty {\frac{1}{{\nu '(\xi )}}{\xi ^6}{e^{ - {\xi ^2}}}d\xi },
\end{aligned}
\end{equation}
and to recover the thermal conductivity, the parameter $\beta$ can be calculated based on the Pr number
\begin{equation}\label{eq:nu_beta}
\beta  = 1 - \Pr\frac{{ {{\int_0^\infty {\frac{{{\xi ^4}}}{{\nu ' (\xi )}}\left( {{\xi ^2} - \frac{5}{2}} \right)} }^2}\exp ( - {\xi ^2})d\xi }}{{\int_0^\infty {\frac{{{\xi ^6}}}{{\nu ' (\xi )}}\exp ( - {\xi ^2})d\xi } }}.
\end{equation}
As for the velocity-dependent collision frequency $\nu(\xi )$, as analysed above there are many forms to be chosen. In the current work, a half-theoretical and half-empirical formula has been established for $\nu(\xi )$, which will be discussed in section~\ref{sec:shock_wave}.  

It is clear that the $\nu$-model satisfies the conservation laws. Also, the VDF can be properly relaxed to the Maxwellian distribution~\eqref{equilibrium_Maxwellian}, because when the equilibrium is reached the heat flux vanishes in a way similar to~\eqref{heat_relax_Shakhov} and according to \eqref{gammi_i} there will be $\hat{\varrho}=1, \hat{\Gamma}=0, \gamma_i=0$ and finally the target VDF \eqref{beyond_SM2} turns to a Maxwellian. On the other hand, as is similar to the situation of the Shakhov model, we can neither prove nor disprove the H-theorem for the $\nu$-model. Nevertheless, according to \eqref{shear_beyond}, the $\nu$-model recovers the correct viscosity and thermal conductivity, and thus satisfies the H-theorem in the small $Kn$ number naturally.

\section{Numerical method}\label{sec:numerical_method}

For practical calculations, it is convenient to introduce the following dimensionless variables:
\begin{equation}\label{normalization}
\begin{aligned}[b]
\widetilde{f}&=\frac{v_m^3}{n_0}f, \ \ \ \widetilde{\bm{x}}=\frac{\bm{x}}{L}, \ \ \ (\widetilde{\bm{v}},\widetilde{\bm{u}},\widetilde{\bm{c}})=\frac{(\bm{v},\bm{u},\bm{c})}{v_m}, \ \ \
\widetilde{t}=\frac{v_m}{L}t,  \\
\widetilde{n}&=\frac{n}{n_0}, \ \ \  \widetilde{T}=\frac{T}{T_0}, \ \ \ \
\widetilde{p}_{ij}=\frac{{p}_{ij}}{n_0k_BT_0}, \ \ \
\widetilde{\bm{q}}=\frac{\bm{q}}{n_0k_BT_0v_m},
\end{aligned}
\end{equation}
where $n_0$ is the average number density of gas molecules, $L$ is the characteristic flow length, $v_m=\sqrt{2k_BT_0/m}$ is the most probable speed at the reference temperature $T_0$. For simplicity, the tildes on normalized quantities will be omitted hereafter. 

Under these normalization, the Boltzmann equation for inverse power-law potentials takes the following form
\begin{equation}  \label{Boltzmann_dimensionless}
\begin{aligned}
\frac{\partial {f}}{\partial{t}}+{\bm{v}}\cdot\frac{\partial
	{f}}{\partial{\bm{x}}}=\frac{1}{Kn'}\iint
\sin^{\alpha+\gamma-1}\left(\frac{\theta}{2}\right)\cos^{-\gamma}\left(\frac{\theta}{2}\right){v}_r^\alpha
[{f}({\bm{v}}'_{\ast}){f}({\bm{v}}')-{f}({\bm{v}}_{\ast}){f}({\bm{v}})] d\Omega d{\bm{v}}_\ast.
\end{aligned}
\end{equation}
where
\begin{equation}\label{Knudsen}
{Kn'}=\frac{64\sqrt{2}^\alpha}{5}\Gamma\left(\frac{\alpha+\gamma+3}{2}\right)
\Gamma\left(2-\frac{\gamma}{2}\right)Kn,
\end{equation}
with
\begin{equation}\label{Knudsen_number}
Kn=\frac{\mu(T_0)}{n_0{L}}\sqrt{\frac{\pi}{2mk_BT_0}},
\end{equation} 
being the unconfined Knudsen number, with $n_0$ the reference molecular number density, and $T_0$ the reference temperature. For the Lennard-Jones potential, the term
$\sin^{\alpha+\gamma-1}(\theta/2)\cos^{-\gamma}(\theta/2){v}_r^\alpha/Kn'$
in ~\eqref{Boltzmann_dimensionless} should be replaced by
\begin{equation}\label{LJ_kernel}
\frac{5\sum_{j=1}^3{}b_j  (k_BT_0/2\epsilon)^{(\alpha_j-1)/2} \sin^{\alpha_j-1}({\theta}/{2})
	{v}_r^{\alpha_j}/{\Gamma(\frac{\alpha_j+3}{2})}}
{64\sqrt{2}Kn\sum_{j=1}^3b_j(k_BT_0/\epsilon)^{(\alpha_j-1)/2}}.
\end{equation}

Considering the above normalization, the normalized macroscopic quantities are related to the normalized VDF as $
[{n},\bm{u},T,p_{ij},q_i]=\int \left[1,\frac{1}{n},\frac{4}{3{n}} |{\bm{c}}|^2, 2c_ic_j, |{\bm{c}}|^2c_i\right] {f}d{\bm{v}}$,
and the ideal gas law is $p=nT$. The collision operator for the $\nu$-model with the collision frequency~\eqref{nu_freq_power} or~\eqref{nu_freq_LJ} becomes
\begin{equation}\label{normalized_model}
\begin{aligned}[b]
Q=&\frac{\sqrt{\pi}}{2Kn}\times\frac{\mu(T_0)}{n_0k_BT_0}\times\nu\left(\frac{c}{T}\right)\\
&\times\left\{  \left[\hat{\varrho}+\hat{\Gamma}c^2+\gamma_ic_i
+\beta\frac{4\bm{q}\cdot
	\bm{c}}{5nT^2}\left(c^2-\frac{5}{2}\right) \right] \frac{n}{(\pi{}T)^{3/2}}\exp\left(-\frac{c^2}{T}\right)-f \right\}.
\end{aligned}
\end{equation}

%

\subsection{Multi-scale implicit scheme for steady state solution}

A multiscale numerical method is proposed to solve the $\nu$-model deterministically, the merit of which is that the streaming and collision is handled simultaneously so  (i) the numerical cell size can be much larger than the molecular mean free path while keeping the numerical dissipation small~\citep{WANG201833}, and (ii) the time step is not limited by the CFL condition. Comparing to the corresponding method with constant collision frequency~\citep{yang2019numerical,yuan2019conservative}, main improvements of the current algorithm are, 
(i) the velocity-dependent collision frequency is updated for every discrete velocity point at every cell centre and cell interface,
(ii) the three variables $\hat \varrho ,\hat \Gamma,\gamma$ in the target VDF \eqref{beyond_SM2} are interpolated to calculate  the target VDF at the cell interface,
(iii) after the discrete VDF has been updated, $\hat \varrho ,\hat \Gamma,\gamma$ are updated through a simple algorithm satisfying the conservation laws in the discrete level. 
The whole computation process is detailed below.

Discretizing the physical space by the finite volume method, applying the implicit backward Euler formula for the time, and discretizing the velocity space into discrete velocity points, the implicit discrete equation for the $\nu$-model can be written as
\begin{equation}\label{eq:discmic}
\frac{{{V_i}}}{{\Delta t}_i^{n+1}}\left( {f_{i,k}^{n + 1} - f_{i,k}^n} \right) + \sum\limits_{j \in N\left( i \right)} \left({{A_{ij}}{\bm{v}_k} \cdot {\bm{n}_{ij}}f_{ij,k}^{n + 1}}\right)  = {V_i}\nu _{i,k}^{n + 1}\left( {f_{{r},i,k}^{n + 1} - f_{i,k}^{n + 1}} \right),
\end{equation}
where $i,n,k$ correspond to the discretization in physical space, time and velocity space, respectively. $j$ denotes the neighbouring cell of cell $i$ and $N\left( i \right)$ is the set of all of the neighbours of $i$. $ij$ denotes the variable at the interface between cell $i$ and $j$. $A_{ij}$ is the interface area, ${\bm{n}_{ij}}$ is the outward normal unit vector of interface $ij$ relative to cell $i$, and $V_i$ is the volume of cell $i$. ${\Delta t}_i^{n+1}$ is the local time step and can be handled by various of traditional implicit time step control techniques. 

Equation~\eqref{eq:discmic} can be rearranged into the incremental form as
\begin{equation}\label{eq:discmic_inc}
\begin{aligned}
\left( {\frac{{{V_i}}}{{\Delta t}_i^{n+1}} + {V_i}\nu_{i,k}^{n+1}} \right)\Delta f_{i,k}^{n + 1} &+ \sum\limits_{j \in N(i)} {{A_{ij}}{{\bm v}_k} \cdot {{\bm n}_{ij}}\Delta f_{ij,k}^{n + 1}} \\
 = & {V_i}\nu_{i,k}^{n+1}\left({f_{{{r}},i,k}^{ n + 1} - f_{i,k}^n}\right) - \sum\limits_{j \in N(i)} {{A_{ij}}{{\bm v}_k} \cdot {{\bm n}_{ij}}f_{ij,k}^n},
\end{aligned}
\end{equation}
where terms on the left-hand side of the equal sign are the increments and will converge to zero when the steady state is reached. In the following paragraphs, the terms on the right-hand side of~\eqref{eq:discmic_inc} are determined first, and then the increment of the distribution function $\Delta f_{i,k}^{n + 1}$ can be worked out to update the variables for one time step. 

It is well known that the conventional discrete velocity method will suffer from excessive numerical viscosity and yield over-dissipating result in the case of small $Kn$ number. To avoid this problem and ensure good accuracy both in the collisionless limit as well as the hydrodynamic limit, the calculation of the interface distribution function $f_{ij,k}^{n}$ should be carefully handled.
Here, the construction of the interface distribution function proposed by~\cite{yuan2019conservative} is adopted to ensure the multi-scale property of the scheme:
\begin{equation}\label{eq:interfacef}
{f_{ij,k}} = \frac{1}{{1 + {\nu _{ij,k}}{h_{ij}}}}f\left( {{\bm{ x}_{ij}} - {\bm{ v}_k}{h_{ij}},{\bm{ v}_k}} \right) + \frac{{{\nu _{ij,k}}{h_{ij}}}}{{1 + {\nu _{ij,k}}{h_{ij}}}}{f_{{{r}},ij,k}},
\end{equation}
where 
\begin{equation}
f({\bm{x}_{ij}} - {\bm{v}_k}{h_{ij}},{\bm{v}_k}) = \left\{ {\begin{array}{*{20}{l}}
{f_{i,k} + ({{\bm{x}}_{ij}} - {{\bm{x}}_i} - {{\bm{v}}_k}{h_{ij}})\cdot\nabla f_{i,k},\; {{\bm{v}}_k} \cdot {{\bm{n}}_{ij}} \ge 0,}\\
{f_{j,k} + ({{\bm{x}}_{ij}} - {{\bm{x}}_j} - {{\bm{v}}_k}{h_{ij}})\cdot\nabla f_{j,k},\; {{\bm{v}}_k} \cdot {{\bm{n}}_{ij}} < 0.}
\end{array}} \right.
\end{equation}
The calculation of the terms in the above equations is detailed as follows. $\nabla f_{i,k}^{n+1}$ and $\nabla f_{j,k}^{n+1}$ are gradients of the VDF and can be obtained by reconstruction based on the initial VDF data. $f_{{{r}},ij,k}$ is the target VDF at the cell interface, and according to~\eqref{beyond_SM2} the target VDF should be determined by the macroscopic variables including the conserved variables $\bm{{W}} = {(\rho ,\rho \bm u,\rho E)^T}$, the heat flux $\bm q$, and the parameters $\bm{w}=(\hat \varrho ,\hat \Gamma ,\gamma )^T$. For $\bm q_{ij}$ and $\bm{w}_{ij}$ at the interface $ij$, they are simply calculated via interpolation 
\begin{equation}
{\bm q_{ij}} = \frac{{{V_j}}}{{\left| {{V_i} + {V_j}} \right|}}{\bm q_i} + \frac{{{V_i}}}{{\left| {{V_i} + {V_j}} \right|}}{\bm q_j},
\quad
{\bm w_{ij}} = \frac{{{V_j}}}{{\left| {{V_i} + {V_j}} \right|}}{\bm w_i} + \frac{{{V_i}}}{{\left| {{V_i} + {V_j}} \right|}}{\bm w_j}.
\end{equation}
For the conserved variable $\bm{{W}}_{ij}$, it is calculated based on the idea of upwind splitting
\begin{equation}\label{eq:interfaceW}
{\bm { W}_{ij}} = \int_{\bm v \cdot {{\bm n}_{ij}} \ge 0} {\bm \psi F_{{eq},ij}^{\rm{l}}d\bm v}  + \int_{\bm v \cdot {{\bm n}_{ij}} < 0} {\bm \psi F_{{eq},ij}^{{\rm{r}}}d\bm v}  ,
\end{equation}
where $\bm \psi=(2,2\bm v,\bm v^2)^T$ is the vector of moments, $ F_{{eq},ij}^{\rm{l}}$ and $F_{{eq},ij}^{{\rm{r}}}$ are Maxwellian distributions determined by the conserved variables on the left/right sides of the interface, and these conserved variables are obtained by the data reconstruction. ${h_{ij}} = \min ({h_i},{h_j})$ in~\eqref{eq:interfacef} is the physical local time step to evolve the interface distribution $f_{ij,k}$ to match the scale of the local cell size, and is calculated by the local CFL condition as
\begin{equation}
{h_i} = \frac{{{V_i}}}{{\mathop {\max }\limits_k \left( {\sum\limits_{j \in N(i)} {\left( {{{\bm v}_k} \cdot {{\bm n}_{ij}}{A_{ij}}{\rm{H}}[{{\bm v}_k} \cdot {{\bm n}_{ij}}]} \right)} } \right)}}{\rm{CFL}},
\end{equation}
where ${\rm{H}}[x]$ is the Heaviside function.
The collision frequency $\nu_{ij,k}$ in~\eqref{eq:interfacef} is calculated considering the artificial viscosity to stabilize the scheme in the region of discontinuity:
\begin{equation}
{\nu _{ij,k}} = \frac{{{\nu _{ij,k,{\rm{physical}}}}}}{{1 + {\mathcal{K}_{ij,{\rm{artificial}}}}}},
\end{equation}
where ${{\nu _{ij,k,{\rm{physical}}}}}$ is the collision frequency calculated from~\eqref{eq:nu_normalization} based on the interface conserved variables ${\bm { W}_{ij}}$. ${{\mathcal{K}_{ij,{\rm{artificial}}}}}$ is calculated as
\begin{equation}
{\mathcal{K}_{ij,{\rm{artificial}}}} = \frac{{\left| {p_{ij}^{\rm{l}} - p_{ij}^{{\rm{r}}}} \right|}}{{\left| {p_{ij}^{\rm{l}} + p_{ij}^{{\rm{r}}}} \right|}}{h_{ij}}{\nu _{ij,k,{\rm{physical}}}},
\end{equation}
in which ${p_{ij}^{\rm{l}}}$ and ${p_{ij}^{{\rm{r}}}}$ are the reconstructed pressure values on the two sides of the interface. More details including the idea of constructing such  interface distribution function are discussed in~\cite{yuan2019conservative}.

For the target VDF $f_{{{r}},i,k}^{n+1}$ at the $(n+1)$-th step on the right-hand side of~\eqref{eq:discmic_inc}, it is handled by the macroscopic variable prediction technique \citep{zhuyajun2016} to guarantee fast convergence of the scheme in both rarefied and continuum flow regimes. As stated above the target VDF $f_{{r}}$ should be determined by $\bm{{W}}$, $\bm q$ and $\bm{w}$. Here, a predicted value $\tilde f_{{{r}},i,k}^{n+1}$ is used to approximate $f_{{{r}},i,k}^{n+1}$ on the right-hand side of~\eqref{eq:discmic_inc}, which is calculated by $\bm q_i^n$, $\bm{w}_i^n$ and a predicted conserved variable $\tilde{\bm{{W}}}_i^{n+1}$. To calculate the predicted $\tilde{\bm{{W}}}_i^{n+1}$, taking the moment of the $\nu$-model for $\bm \psi$ and the corresponding discrete macroscopic governing equation can be expressed as
\begin{equation}\label{eq:discmac}
\frac{{{V_i}}}{{\Delta t}_i^{n+1}}\left( {{\bm{{W}}}_i^{n + 1} - \bm{{W}}_i^n} \right) + \sum\limits_{j \in N(i)} {{A_{ij}}\bm{\mathcal{F}}_{ij}^{n + 1}}  = \bm 0.
\end{equation}
Then replacing $\bm{{W}}_i^{n + 1}$ with the predicted $\tilde{\bm{{W}}}_i^{n+1}$, and rearranging~\eqref{eq:discmac} into the incremental form
\begin{equation}\label{eq:discmac_inc}
\frac{{{V_i}}}{{\Delta t}_i^{n+1}}\Delta \tilde {\bm{{W}}}_i^{n + 1} + \sum\limits_{j \in N(i)} {{A_{ij}}\Delta \tilde{\bm{\mathcal{F}}}_{ij}^{n + 1}}  =  - \sum\limits_{j \in N(i)} {{A_{ij}}\bm{\mathcal{F}}_{ij}^n},
\end{equation}
where the symbol $\sim$ denotes the predicted variables for the $(n+1)$-th step. The flux $\bm{\mathcal{F}}_{ij}^n$ on the right-hand side of~\eqref{eq:discmac_inc} is obtained by the numerical integration of the interface distribution function $f_{ij,k}^n$ in the discrete velocity space, i.e.
\begin{equation}\label{eq:discinc_flux}
\bm{\mathcal{F}}_{ij}^n = \sum\limits_{k} {{{\bm \psi }_k}{{\bm v}_k} \cdot {{\bm n}_{ij}}f_{ij,k}^n\Delta \bm v_k },
\end{equation}
where the interface distribution function $f_{ij,k}^n$ is just calculated by (\ref{eq:interfacef}). The variation of the flux $\Delta \tilde{\bm{\mathcal{F}}}_{ij}^{n + 1}$ on the left-hand side of (\ref{eq:discmac_inc}) is handled like in the traditional macroscopic implicit scheme based on Navier-Stokes equation, i.e.
\begin{equation}\label{eq:deltamacflux}
\Delta \tilde{\bm{\mathcal{F}}}_{ij}^{n + 1} = \tilde {\bm{\mathsf{F}}}_{ij}^{n + 1} - {\bm{\mathsf{F}}}_{ij}^n,
\end{equation}
where $\bm{\mathsf{F}}_{ij}$ has the form of the well-known Roe's flux function
\begin{equation}\label{eq:roeflux}
\bm{\mathsf{F}}_{ij} = \frac{1}{2}\left( {{\bm{\mathbb{F}}_{ij}}({{\bm{ {W}}}_i}) + {\bm{\mathbb{F}}_{ij}}({\bm{ {W}}_j}) + {\mathfrak{r}_{ij}}{{\bm{ {W}}}_i} - {\mathfrak{r}_{ij}}{{\bm{ {W}}}_j}} \right).
\end{equation}
Here ${\bm{\mathbb{F}}_{ij}}({\bm{ {W}}})$ is the Euler flux
\begin{equation}
{\bm{\mathbb{F}}_{ij}}({\bm{ {W}}}) = \left( \begin{array}{c}
\rho \bm u \cdot {{\bm n}_{ij}}\\
\rho \bm u \bm u \cdot {{\bm n}_{ij}} + p{\bm n_{ij}}\\
(\rho E + p)\bm u \cdot {{\bm n}_{ij}}
\end{array} \right),
\end{equation}
and $\mathfrak{r}_{ij}$ is
\begin{equation}
{\mathfrak r_{ij}} = \left| {{{\bm u}_{ij}} \cdot {{\bm n}_{ij}}} \right| + {a_{ij}} + 2\frac{{{\mu _{ij}}}}{{{\rho _{ij}}\left| \bm x_i -\bm x_j\right|}},
\end{equation}
in which $a_{ij}$ is the acoustic speed. Substituting (\ref{eq:deltamacflux}) and (\ref{eq:roeflux}) into (\ref{eq:discmac_inc}), and noting that $\sum\limits_{j \in N(i)} {{A_{ij}}{\bm{\mathbb{F}}_{ij}}({{\bm{ {W}}}_i})}  = \bm 0$ holds, the equation for the increment $\Delta \tilde {\bm{{W}}}_i^{n + 1}$ can be then expressed as
\begin{equation}\label{eq:refreshmac_conserve}
\begin{aligned}
\left( {\frac{{{V_i}}}{{\Delta t}_i^{n+1}} + \frac{1}{2}\sum\limits_{j \in N(i)} {{\mathfrak{r}_{ij}^n}{A_{ij}}} } \right)\Delta \tilde {\bm{{W}}}_i^{n + 1} =  &  - \sum\limits_{j \in N(i)} {{A_{ij}}\bm{\mathcal{F}}_{ij}^n}  + \frac{1}{2}\sum\limits_{j \in N(i)} {{\mathfrak{r}_{ij}^n}{A_{ij}}\Delta \tilde {\bm{{W}}}_j^{n + 1}} \\&
 - \frac{1}{2}\sum\limits_{j \in N(i)} {{A_{ij}}\left( {\bm{\mathbb{F}}_{ij}(\tilde {\bm{{W}}}_j^{n + 1}) - \bm{\mathbb{F}}_{ij}(\bm{{W}}_j^{n})} \right)}
\end{aligned}.
\end{equation}
Equation~\eqref{eq:refreshmac_conserve} is solved by the Symmetric Gauss-Seidel (SGS) method, or also known as the Point Relaxation Symmetric Gauss-Seidel (PRSGS) method \citep{Rogers1995Comparison,Yuan2002Comparison}. The SGS method includes several times of forward/backward sweep from the first/last cell to the last/first cell, during which the conserved variable $\tilde {\bm{{W}}}_i^{n + 1}$ (or the increment $\Delta \tilde {\bm{{W}}}_i^{n + 1}$) of the cell $i$ is always updated by the latest data of its neighbouring cells by (\ref{eq:refreshmac_conserve}), and after several times of iteration an estimation for $\tilde {\bm{{W}}}_i^{n + 1}$ can be obtained with certain accuracy. After $\tilde {\bm{{W}}}_i^{n + 1}$ is determined, the predicted target VDF $\tilde{f}_{{{r}},i,k}^{n+1}$ can be calculated, and a prediction for the collision frequency $\tilde{\nu}_{i,k}^{n+1}$ can be calculated for (\ref{eq:discmic_inc}) as well.

Since the terms on the right-hand side of (\ref{eq:discmic_inc}) have all been determined, approximating the variation of the interface distribution function ${\Delta f_{ij,k}^{n + 1}}$ on the left-hand side by the first-order upwind scheme and then the equation for the increment $\Delta f_{i,k}^{n+1}$ can be written as
\begin{equation}\label{eq:refreshmic}
\begin{aligned}
\left( {\frac{{{V_i}}}{{\Delta t}} + {V_i}\tilde{\nu}_{i,k}^{n+1} + \sum\limits_{j \in N_k^ + (i)} {{A_{ij}}{{\bm v}_k} \cdot {{\bm n}_{ij}}} } \right)\Delta f_{i,k}^{n + 1} = &{V_i}\tilde{\nu}_{i,k}^{n+1}\left({\tilde{f}_{{{r}},i,k}^{n + 1} - f_{i,k}^n}\right) - \sum\limits_{j \in N(i)} {{A_{ij}}{{\bm v}_k} \cdot {{\bm n}_{ij}}f_{ij,k}^n} \\
 &- \sum\limits_{j \in N_k^ - (i)} {{A_{ij}}{{\bm v}_k} \cdot {{\bm n}_{ij}}\Delta f_{j,k}^{n + 1}},
\end{aligned}
\end{equation}
where $ N_k^ + (i)$ is the set of cell $i$'s neighbours satisfying ${\bm v_k} \cdot {\bm n_{ij}} \ge 0$ while for $ N_k^ - (i)$ it satisfies ${\bm v_k} \cdot {\bm n_{ij}} < 0$. Likewise, (\ref{eq:refreshmic}) is solved by the SGS method. After several times of SGS iteration, the increment $\Delta f_{i,k}^{n+1}$ can be obtained and the distribution function $f_{i,k}^{n+1}$ for the next time step can be updated. Once $f_{i,k}^{n+1}$ has been determined, the conserved variable $\bm{{W}}_i^{n+1}$ and the heat flux $\bm q_i^{n+1}$ can also be updated through numerical integration in the velocity space \eqref{macroscopic_origin}, and the remaining procedure to do is the update of the parameter $\bm{w}_i^{n+1}$. This can be finished by solving the collision conservation constraint equation at the discrete level, i.e.
\begin{equation}\label{eq:w_disc0}
\sum\limits_k {{\bm{\varphi} _k}\nu _{i,k}^{n + 1}\left( { f_{{{r}},i,k}^{n + 1} - f_{i,k}^{n + 1}} \right)\Delta {{\bm v}_k}}=\bm 0,
\end{equation}
where $\bm{\varphi}$ is defined as $\bm{\varphi}=(1,\vec c,{{\vec c}^2})^T$. Substituting the expression of the target VDF ~\eqref{beyond_SM2} into (\ref{eq:w_disc0}) will yield
\begin{equation}\label{eq:w_disc1}
\begin{aligned}
\left(\sum\limits_k {{{\bm \varphi }_k}\bm \varphi _k^T\nu _{i,k}^{n + 1}F_{{eq,}i,k}^{n + 1}\Delta {{\bm v}_k}}\right)\bm w_i^{n + 1}  = \sum\limits_k {{{\bm \varphi }_k}\nu _{i,k}^{n + 1}\left( {f_{i,k}^{n + 1} - \beta \frac{{4\bm q_i^{n + 1} \cdot {{\bm c}_k}}}{{5n_i^{n + 1}{{(T_i^{n + 1})}^2}}}\left( {\frac{{\bm c_k^2}}{{T_i^{n + 1}}} - \frac{5}{2}} \right)F_{{eq,}i,k}^{n + 1}} \right)\Delta {{\bm v}_k}},
\end{aligned}
\end{equation}
which is actually a linear set of five equations and can be easily solved out. The above conservation treatment can guarantee the conservation laws  in the discrete level \citep{Luc2000JCP,mieussens2000discrete}, which can significantly reduce the requirement for the discrete velocity point number. Furthermore, according to the conservative compensation technique proposed by~\cite{yuan2019conservative}, there is also an alternative approach to calculate $\bm{w}_i^{n+1}$. That is, first calculate the moments of the target VDF $f_{{{r}},i}^{n + 1}$ as
\begin{equation}\label{eq:w_compst0}
\begin{aligned}
\int {\bm {\varphi} \nu _i^{n + 1}f_{{{r}},i}^{n + 1}d\bm v}  = \sum {{{\bm {\varphi} }_k}\nu _{i,k}^{n + 1}f_{i,k}^{n + 1}\Delta {{\bm v}_k}}  + \int {\bm {\varphi} \nu _i^nf_{{{r}},i}^nd\bm v}  - \sum {{{\bm {\varphi} }_k}\nu _{i,k}^nf_{{{r}},i,k}^n\Delta {{\bm v}_k}}, 
\end{aligned}
\end{equation}
where the last two terms on the right-hand side are just the integral error for the moments of the target VDF due to the discretization of the velocity space. Then according to the analytical integral of the target VDF~\eqref{beyond_SM2}, $\bm w_i^{n + 1}$ can be solved out easily as explicit expressions
\begin{equation}\label{eq:w_compst1}
\begin{aligned}
\hat \varrho _i^{n + 1} &= \frac{{\sqrt \pi  }}{{4n_i^{n + 1}}}\left( {\frac{{{\varepsilon _2}}}{{{\varepsilon _0}{\varepsilon _2} - \varepsilon _1^2}}\int {\nu _i^{n + 1}f_{{{r}},i}^{n + 1}d\bm v}  - \frac{{{\varepsilon _1}}}{{{\varepsilon _0}{\varepsilon _2} - \varepsilon _1^2}}\frac{1}{{T_i^{n + 1}}}\int {{c^2}\nu _i^{n + 1}f_{{{r}},i}^{n + 1}d\bm v} } \right),\\
\hat \Gamma _i^{n + 1} &= \frac{{\sqrt \pi  }}{{4n_i^{n + 1}}}\left( {\frac{{{\varepsilon _0}}}{{{\varepsilon _0}{\varepsilon _2} - \varepsilon _1^2}}\frac{1}{{{{(T_i^{n + 1})}^2}}}\int {{c^2}\nu _i^{n + 1}f_{{{r}},i}^{n + 1}d\bm v}  - \frac{{{\varepsilon _1}}}{{{\varepsilon _0}{\varepsilon _2} - \varepsilon _1^2}}\frac{1}{{T_i^{n + 1}}}\int {\nu _i^{n + 1}f_{{{r}},i}^{n + 1}d\bm v} } \right),\\
\bm{ \gamma} _i^{n + 1} &= \frac{1}{{n_i^{n + 1}T_i^{n + 1}}}\left( {\frac{{3\sqrt \pi  }}{{4{\varepsilon _1}}}\int {\bm c\nu _i^{n + 1}f_{{{r}},i}^{n + 1}d\bm v}  - \beta \frac{2}{5}\frac{{(2{\varepsilon _2} - 5{\varepsilon _1})}}{{{\varepsilon _1}}}\frac{1}{{T_i^{n + 1}}}\bm {q}_i^{n + 1}} \right),
\end{aligned}
\end{equation}
where $[{\varepsilon _0},{\varepsilon _1}, {\varepsilon _2}]= \int_0^\infty [{\xi^2},{\xi^4},{\xi^6}]{\nu (\xi){e^{ - {\xi^2}}}d\xi}$.

%

When the whole algorithm converges, (\ref{eq:w_compst0}) will turn into
\begin{equation}
\begin{aligned}
\sum {{{\bm {\varphi} }_k}\nu _{i,k}f_{i,k}\Delta {{\bm v}_k}}   - \sum {{{\bm {\varphi} }_k}\nu _{i,k} f_{{{r}},i,k} \Delta {{\bm v}_k}} = \bm 0, 
\end{aligned}
\end{equation}
which is in fact the same as (\ref{eq:w_disc0}). Thus this compensation approach, ~\eqref{eq:w_compst0} combined with~\eqref{eq:w_compst1}, is just as accurate as (\ref{eq:w_disc0}) with less computational cost.

In summary, the computation procedure from the time step $n$ to $n+1$ is listed as follows:
\begin{description}
    \item[Step 1.]~Reconstruct the data and calculate $f_{ij,k}^n$ at the interface by (\ref{eq:interfacef}).
    \item[Step 2.]~Calculate the flux $\bm{\mathcal{F}}_{ij}^n$ on the right-hand side of (\ref{eq:refreshmac_conserve}) based on the numerical integration of $f_{ij,k}^n$ in the discrete velocity space.
    \item[Step 3.]~Solve (\ref{eq:refreshmac_conserve}) by SGS iterations to get the predicted $\tilde {\bm{{W}}}_i^{n + 1}$.
    \item[Step 4.]~Calculate $\tilde f_{{{r}},i,k}^{n + 1}$ and $\tilde{\nu}_{i,k}^{n+1}$ in (\ref{eq:refreshmic}) based on the predicted $\tilde {\bm{{W}}}_i^{n + 1}$.
    \item[Step 5.]~Solve (\ref{eq:refreshmic}) by SGS iterations to obtain ${f_{i,k}^{n + 1}}$ at the $(n+1)$-th time step.
    \item[Step 6.]~Integrate ${f_{i,k}^{n + 1}}$ numerically in the discrete velocity space to obtain ${\bm{{W}}}_i^{n + 1}$ and ${\bm{q}}_i^{n + 1}$ at the $(n+1)$-th time step.
    \item[Step 7.]~Calculate $\bm{w}_i^{n+1}$ by (\ref{eq:w_disc0}) or (\ref{eq:w_compst0}).
\end{description}

\begin{figure}
	\centering
		\vskip 0.3cm
	\subfloat[Maxwell gas: $\omega=1$]{	\includegraphics[width=5cm]{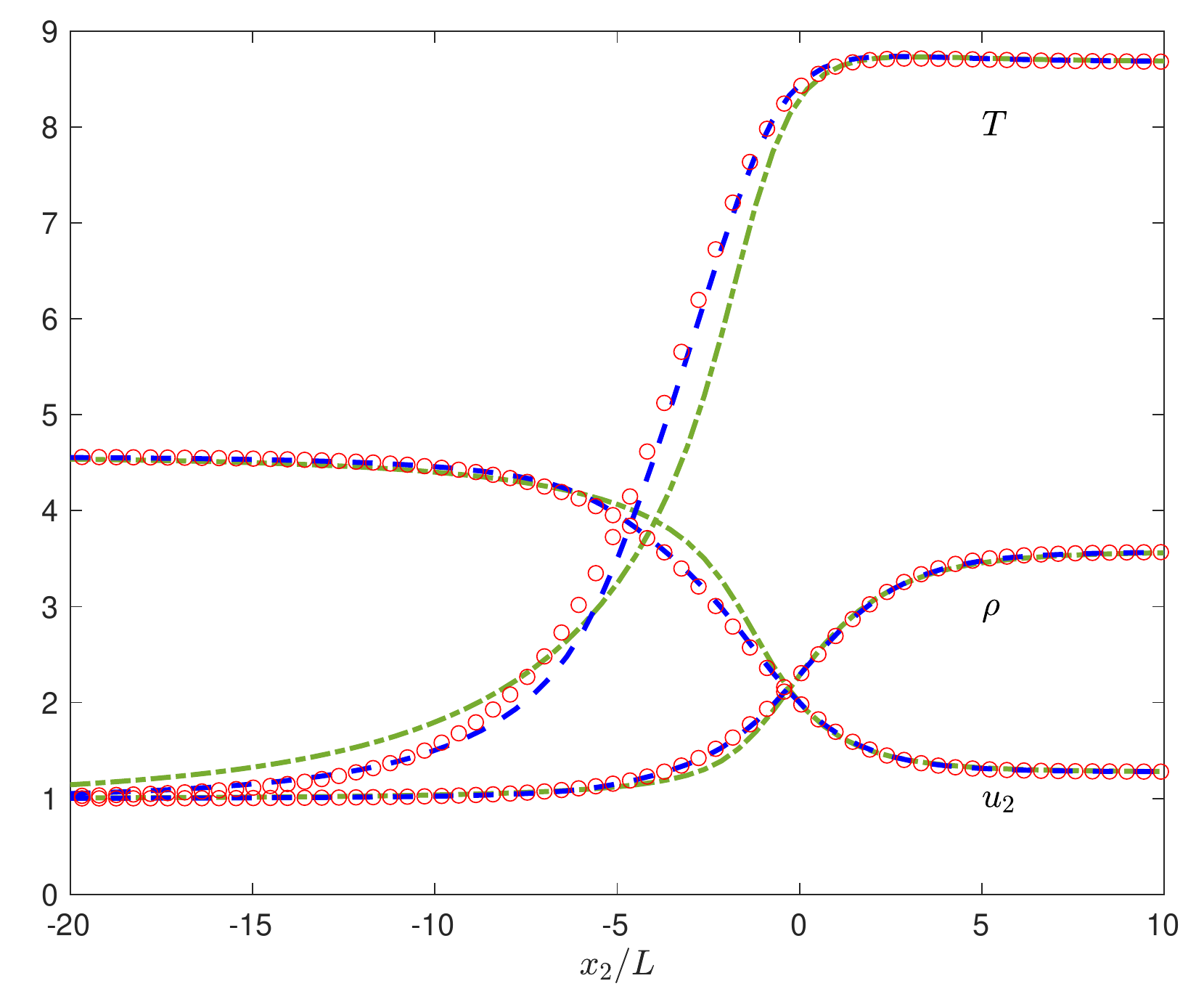}
		\hskip 0.5cm
		\includegraphics[width=5cm]{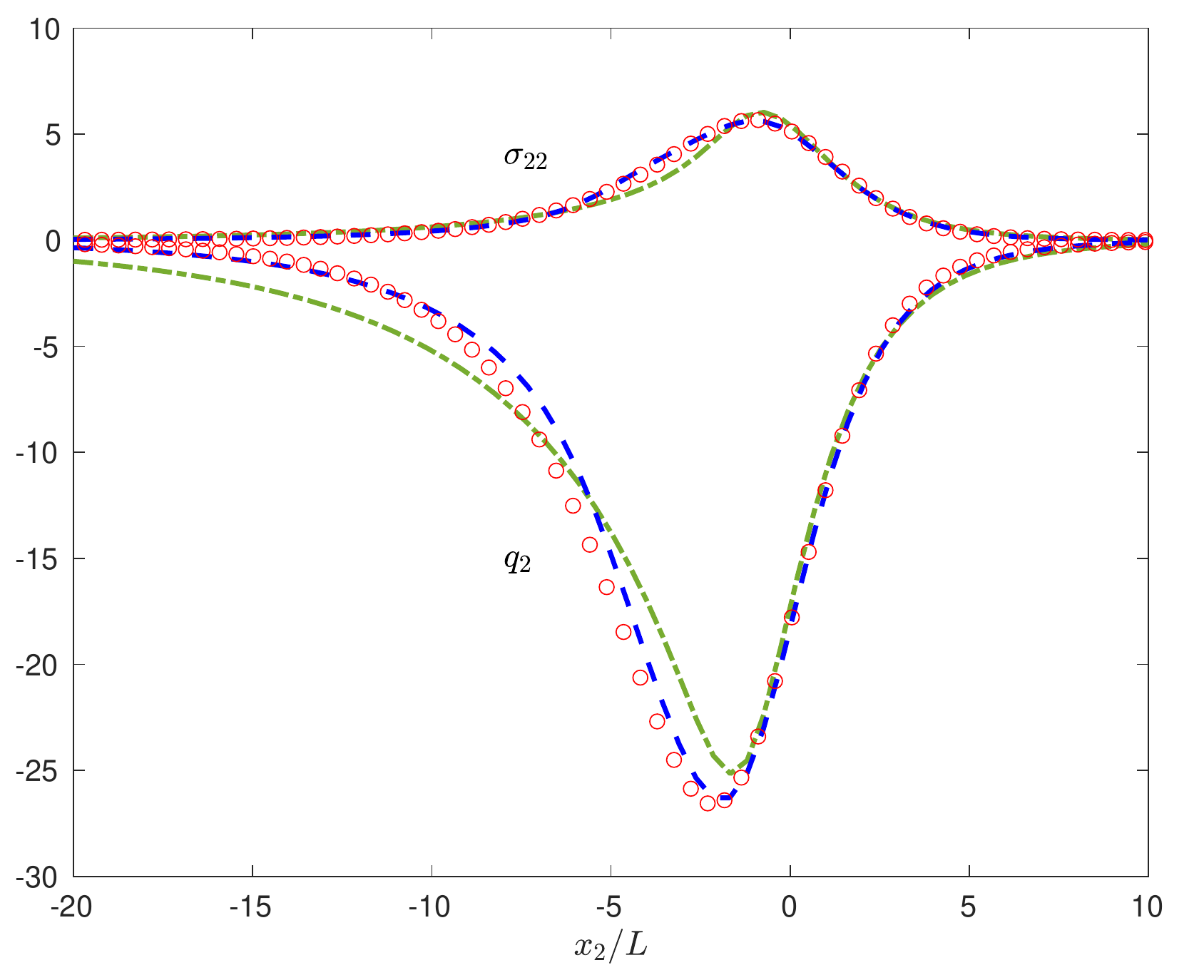}
	}\\
	\vskip 0.3cm
	\subfloat[Inverse power-law potential: $\omega=0.75$]{	\includegraphics[width=5cm]{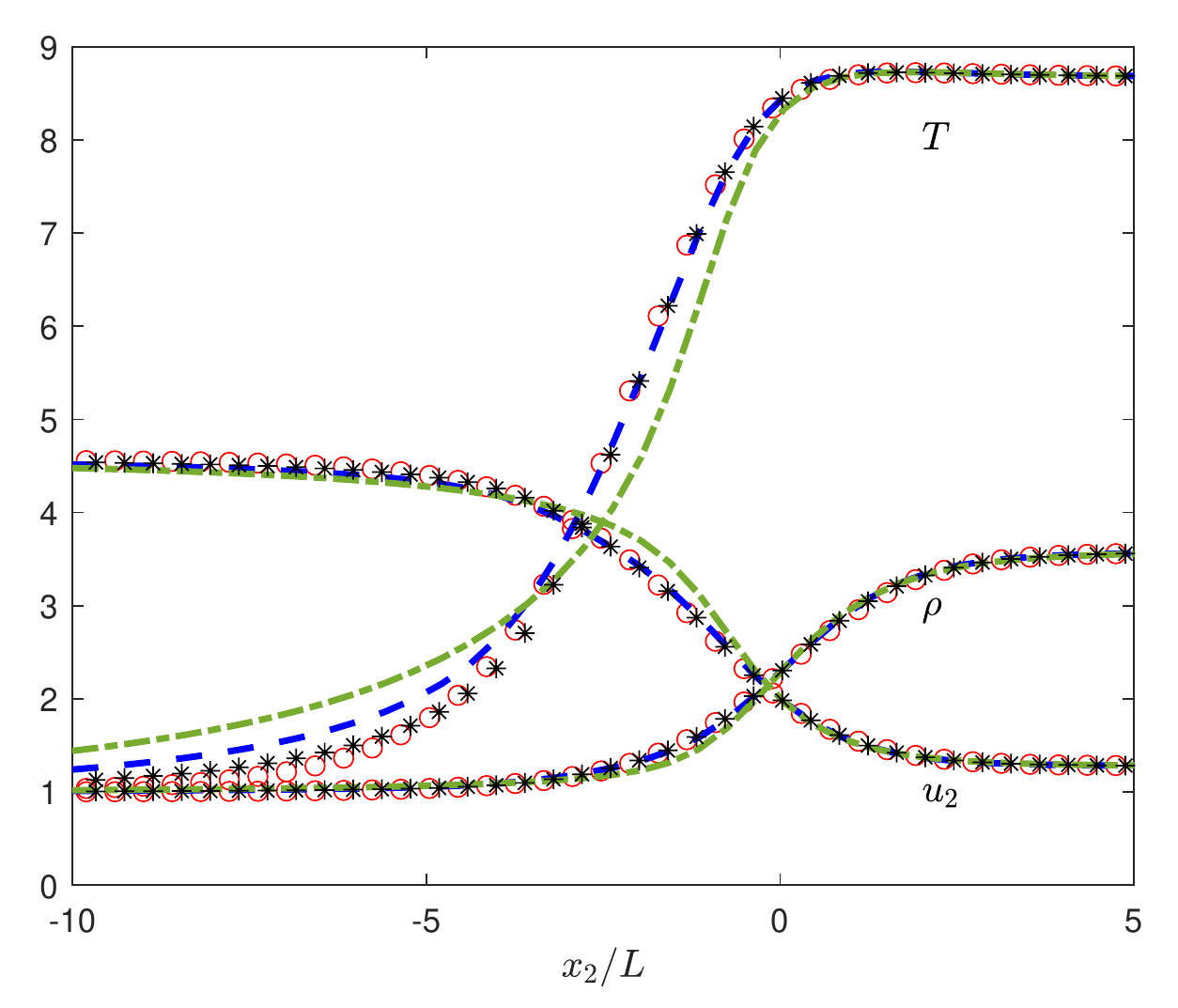}
		\hskip 0.5cm
		\includegraphics[width=5cm]{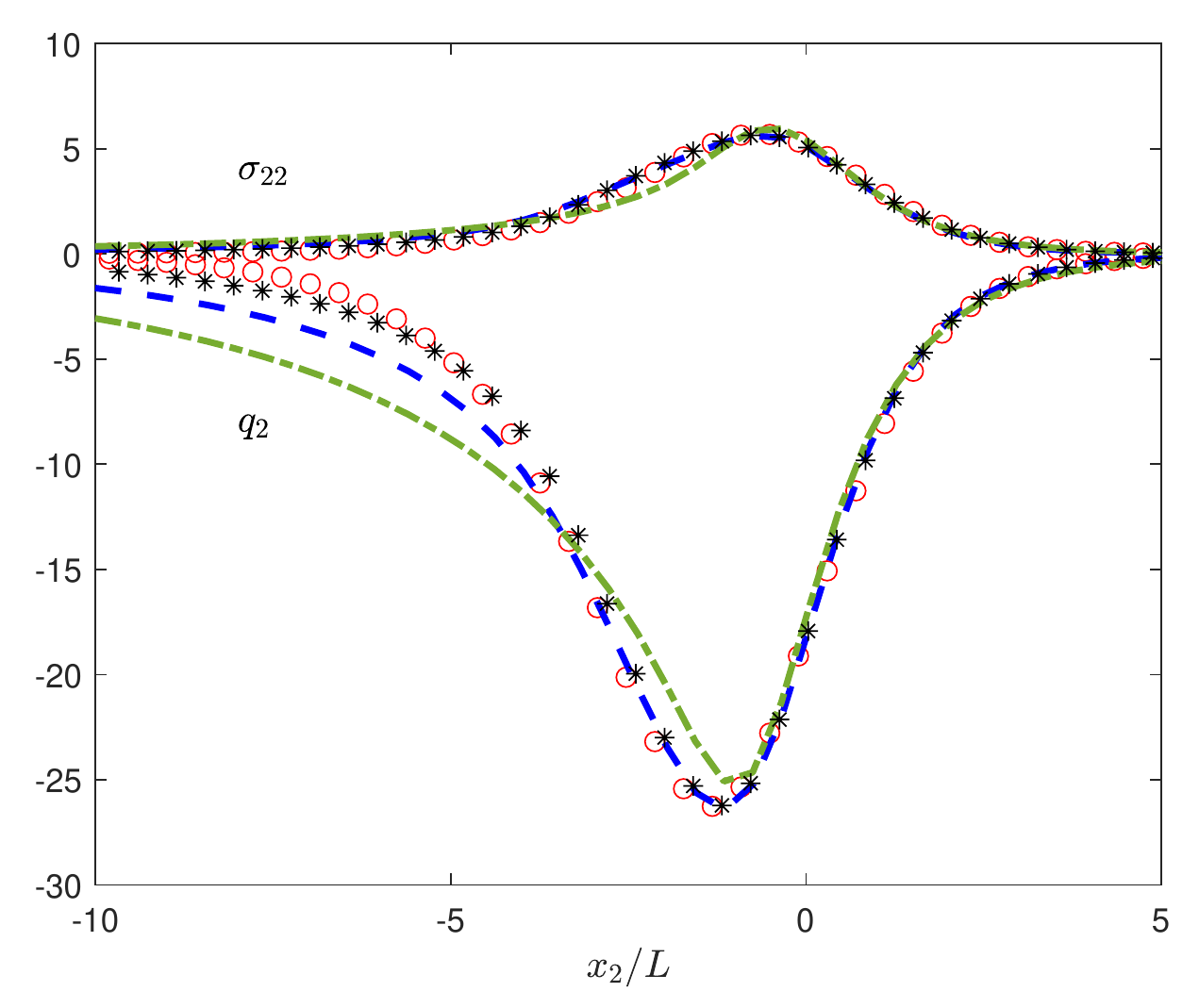}
	}\\
	\vskip 0.3cm
	\subfloat[Hard-sphere gas: $\omega=0.5$]{	\includegraphics[width=5cm]{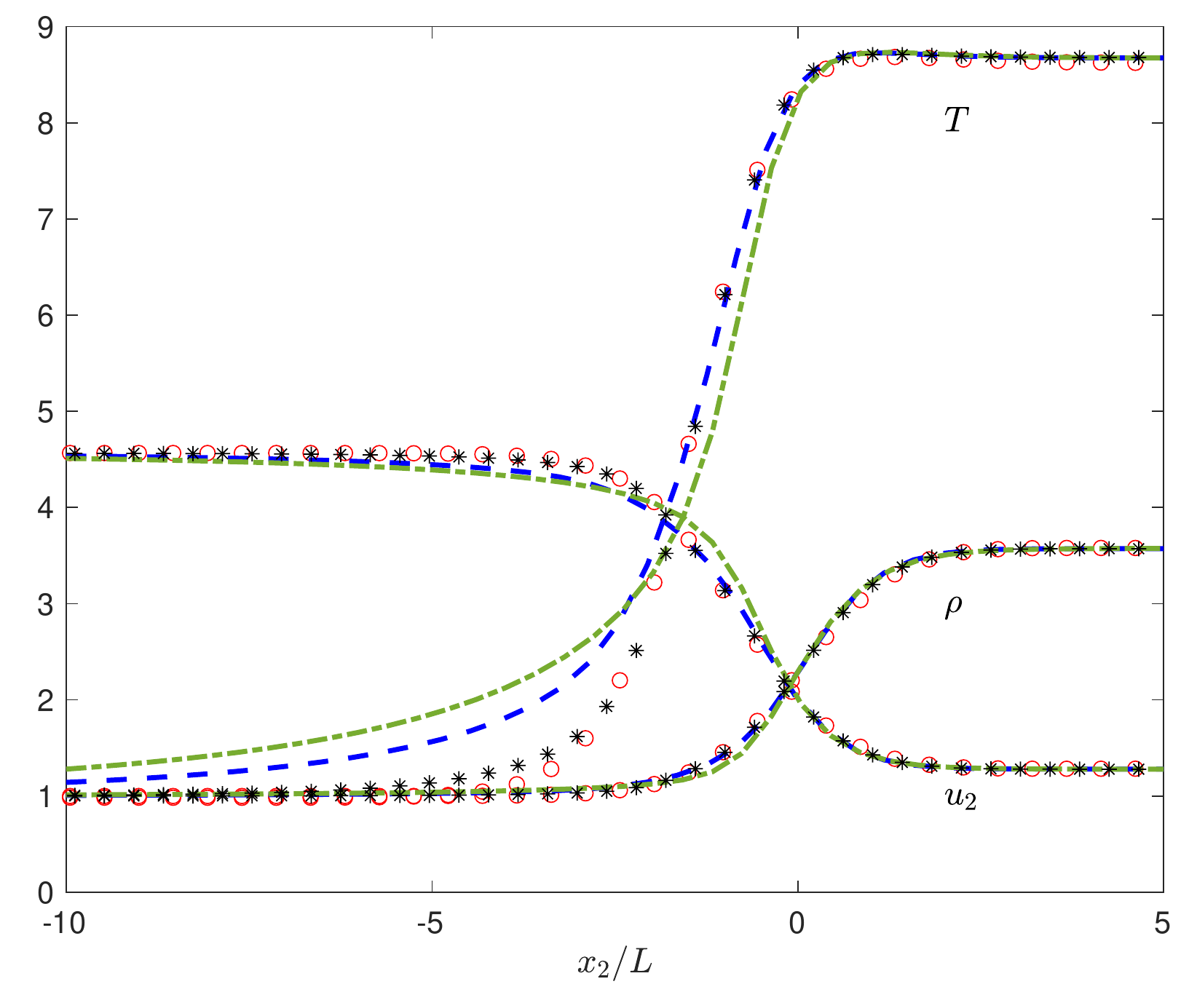}
		\hskip 0.5cm
		\includegraphics[width=5cm]{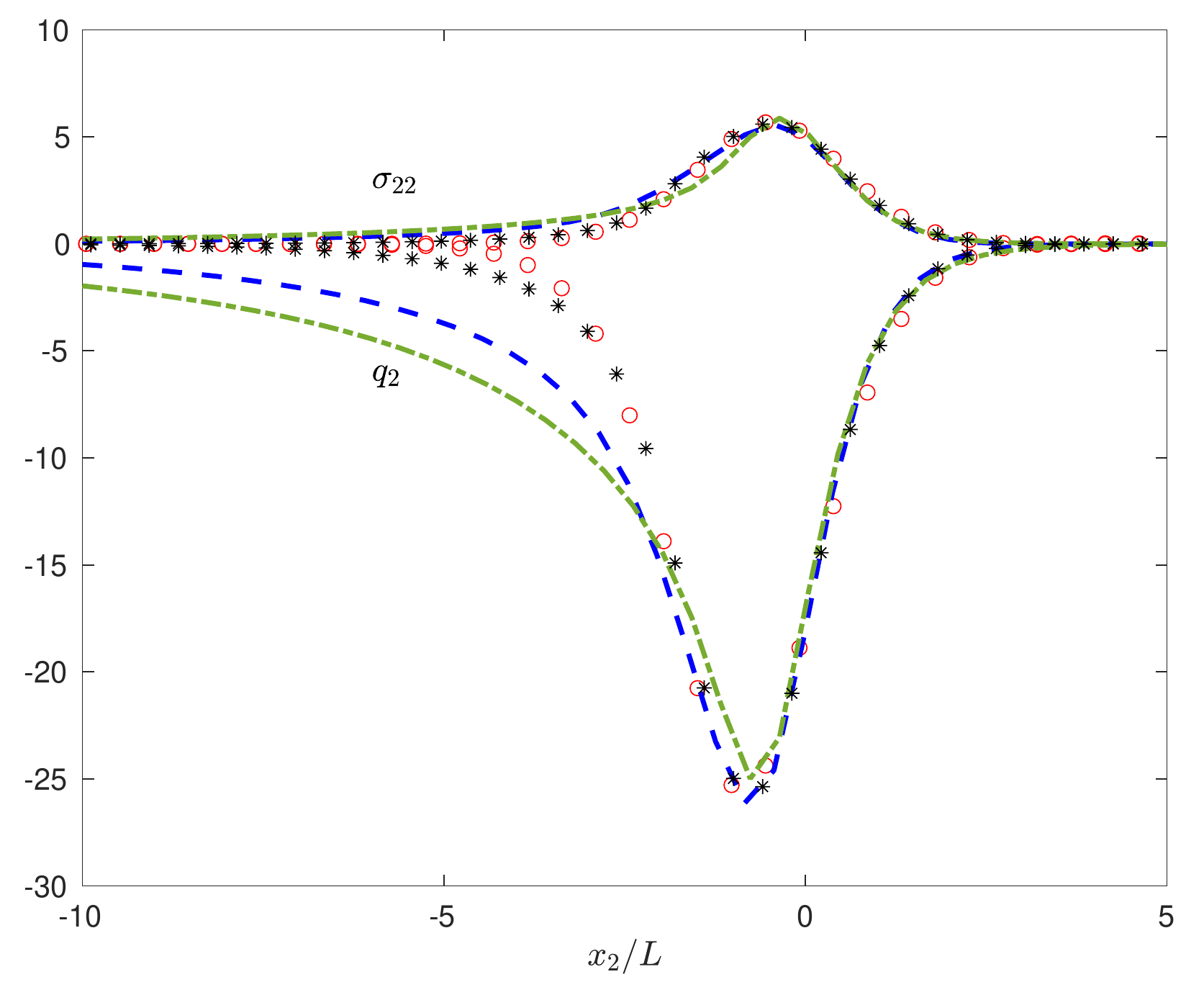}
	}
	\caption{
		The Mach 5 shock wave structures for molecules interacting through inverse power-law potentials. Shakhov model: blue dashed line; ESBGK model: green dash-dotted lines;  Boltzmann equation: red circles; $\nu$-model: Asterisks.	Note that the characteristic length is chosen to be the mean free path  $L=\frac{16}{5\pi}\sqrt{\frac{\pi}{2mk_BT_0}} \frac{\mu(T_0)}{n_0}$ in the upstream part of the normal shock wave, so we take $Kn=5\pi/16$ in the numerical simulation. The shock density centre is at $x_2=0$. 
	}
	\label{shock_Ma5_Maxwell}
\end{figure} 

\section{Numerical results in hypersonic flows}\label{sec:numerical_results}

In this section, we determine the collision frequency of the $\nu$-model by comparing its solution of the normal shock wave with that of the Boltzmann equation. Then the $\nu$-model is compared with the DSMC in the simulation of two-dimensional hypersonic flows passing through a disc.

\subsection{Normal shock waves}\label{sec:shock_wave}

\subsubsection{Inverse power-law potential}

Figure~\ref{shock_Ma5_Maxwell} compares the shock wave structures obtained from the Boltzmann equation, the Shakhov model, and the ESBGK model, when the upstream Mach number is $5$. Different inverse power-law potentials, reflected through the viscosity index $\omega$ in~\eqref{temperature_dependence}, are considered. The Boltzmann equation is solved by the fast spectral method~\citep{Lei2013}. For the Maxwellian gas with $\omega=1$, it is found that the Shakhov model gives a very good prediction of the shock structure, while the ESBGK model overpredicts the temperature and heat flux in the upstream part. When the viscosity index decreases to 0.75 and eventually to 0.5 of the hard-sphere gas, the Shakhov model still predicts the density and velocity profiles well but significantly overpredicts the temperature and heat flux in the upstream part: the smaller the value of $\omega$, the larger the deviation. For the ESBGK model, the deviations of temperature and heat flux from those of the Boltzmann equation are large for all values of $\omega$, and similarly the over-prediction of the upstream temperature and heat flux can be clearly observed. The better performance of the Shakhov model over the ESBGK model suggests the importance of including the heat flux in the gain term of the modelled collision operator~\eqref{beyond_SM2}.

\begin{table}[t]
	\centering
	\begin{tabular}{cccccc|cc}
		\hline
		$\omega$& $c_4$   & $c_3$     & $c_2$    & $c_1$    & $c_0$    & A        & $\beta$\\
		$0.50$  &0.0145   &-0.2019    &1.0561 	 &0.0753    &2.9774    &0.0871    &0.3486\\
		$0.55$  &0.0132   &-0.1793    &0.8826 	 &0.0863    &2.8180    &0.0944    &0.3470\\
		$0.60$  &0.0117   &-0.1557    &0.7261    &0.0929    &2.6691    &0.1022    &0.3453\\
		$0.65$  &0.0101   &-0.1320    &0.5859    &0.0950    &2.5300    &0.1107    &0.3436\\
		$0.70$  &0.0085   &-0.1089    &0.4615    &0.0926    &2.3999    &0.1197    &0.3419\\
		$0.75$  &0.0069   &-0.0867    &0.3522    &0.0859    &2.2782    &0.1293    &0.3403\\
		$0.80$  &0.0054   &-0.0659    &0.2571    &0.0752    &2.1642    &0.1395    &0.3387\\
		$0.85$  &0.0039   &-0.0467    &0.1753    &0.0608    &2.0572    &0.1505    &0.3372\\
		$0.90$  &0.0025   &-0.0292    &0.1059    &0.0432    &1.9566    &0.1622    &0.3358\\
		$0.95$  &0.0012   &-0.0137    &0.0477    &0.0227    &1.8619    &0.1747    &0.3345    \\										
		\hline
	\end{tabular}
\caption{
	Numerical fitting of the equilibrium collision frequency by the quartic function $\nu_{eq}(\xi)=\sum_{j=0}^4c_j\xi^j$, when  $\xi\le5$, as well as the constants $A$ in~\eqref{eq:nu_normalization} and $\beta$ in~\eqref{eq:nu_beta}. When $\xi>5$, the collision frequency $\nu_{eq}$ can be approximated when the first-order Taylor expansion is applied to~\eqref{equi_fre_chapter2}, resulting in $\nu_{eq}= \left(2-\omega\right)\sqrt{\pi}\xi^{2(1-\omega)}$. In the calculation of $\beta$  we take the Prandtl number to be $\text{Pr}=2/3$, while the collision frequency is given by~\eqref{nu_freq_power}.
 } \label{table_coll} 
\end{table}

\begin{figure}
	\centering
	\includegraphics[width=0.36\textwidth]{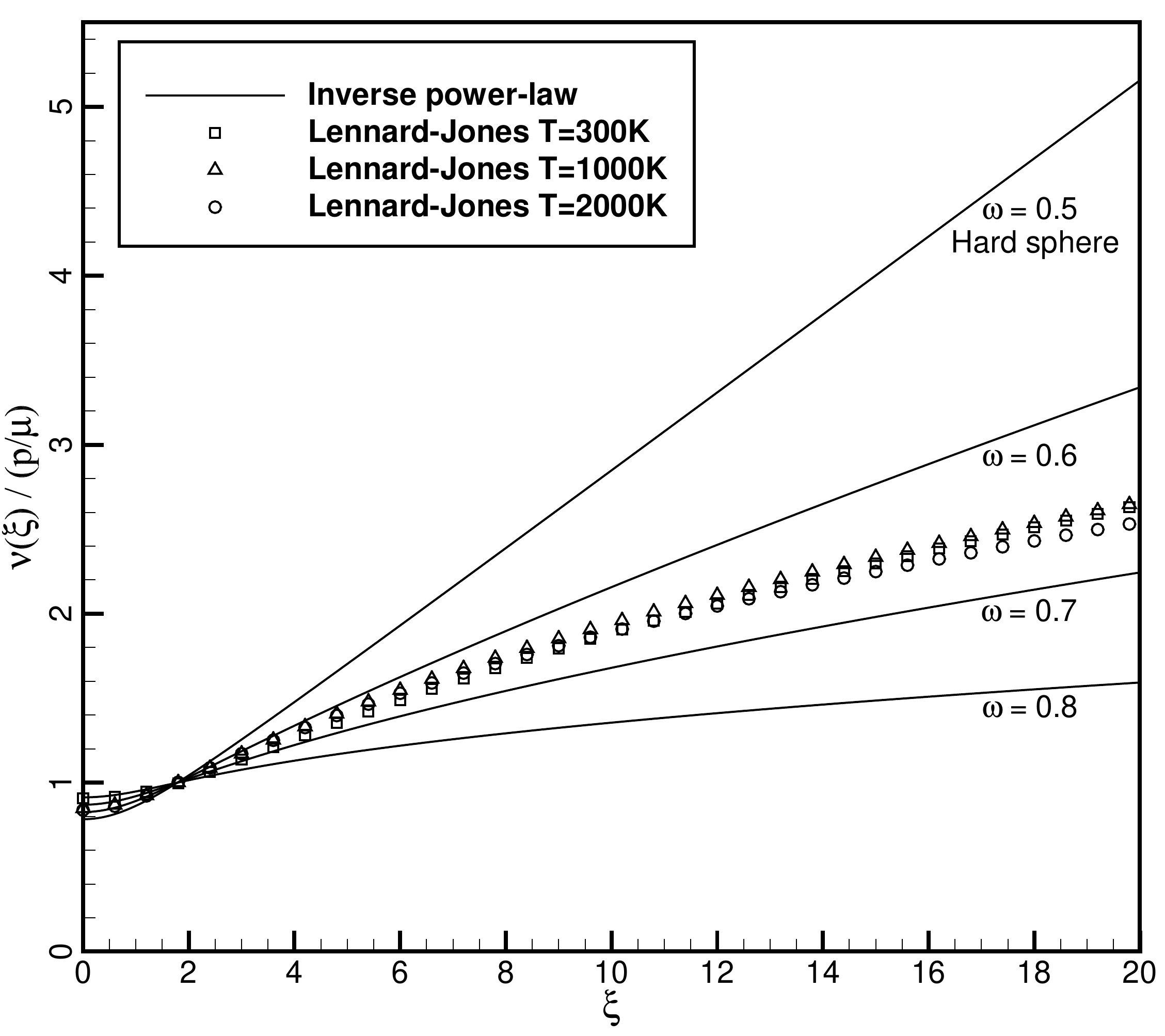}
	\caption{
		Molecular-velocity-dependent collision frequency based on the semi-empirical formulas ~\eqref{nu_freq_power} and ~\eqref{nu_freq_LJ} for different intermolecular potentials. The Lennard-Jones potential for argon with the potential depth $\epsilon=119.2k_B$ is considered.
	}
	\label{fig:frequency}
\end{figure}


To determine the velocity-dependent collision frequency $\nu(\xi)$ in the $\nu$-model, we first use the equilibrium collision frequency $\nu_{eq}(\xi)$ defined in~\eqref{equi_fre_chapter2} and~\eqref{equi_fre_HS} with the normalization~\eqref{eq:nu_normalization}, and find that the upstream temperature is underestimated (not shown). Therefore, a flatter collision frequency curve is required; after a few trial-and-errors we find that good agreement in the shock structures can be achieved (see figure~\ref{shock_Ma5_Maxwell}) when the following semi-empirical formula is used:
\begin{equation}\label{nu_freq_power}
\nu_\omega(\xi)=A\frac{p}{\mu}[\nu_{eq}(\xi)+2\nu_{eq}(0)],
\end{equation} 
where $A$ is determined from~\eqref{eq:nu_normalization}. 

The collision frequency~\eqref{nu_freq_power} for typical inverse power-law potential is shown in Figure~\ref{fig:frequency}. In numerical simulations, $\nu_{eq}(\xi)$ can be calculated by fitting functions and the parameters for typical values of viscosity index are summarized in Table~\ref{table_coll}. The term $2\nu_{eq}(0)$ is an empirical parameter, which makes the collision frequency curve flatter and accounts for the deviation of collision frequency in non-equilibrium state from that in the Maxwellian distribution. The semi-empirical formula~\eqref{nu_freq_power} is implemented in all of the test cases performed in this paper. It will be demonstrated that this semi-empirical formula works well not only in normal shock waves, but also in other test cases and has a certain universality. It is also worth noting that for Maxwellian molecules the collision frequency $\nu(\xi )$ is velocity-independent, so the $\nu$-model reduces to the Shakhov model.

\begin{figure}
	\centering
	{\includegraphics[width=0.36\textwidth]{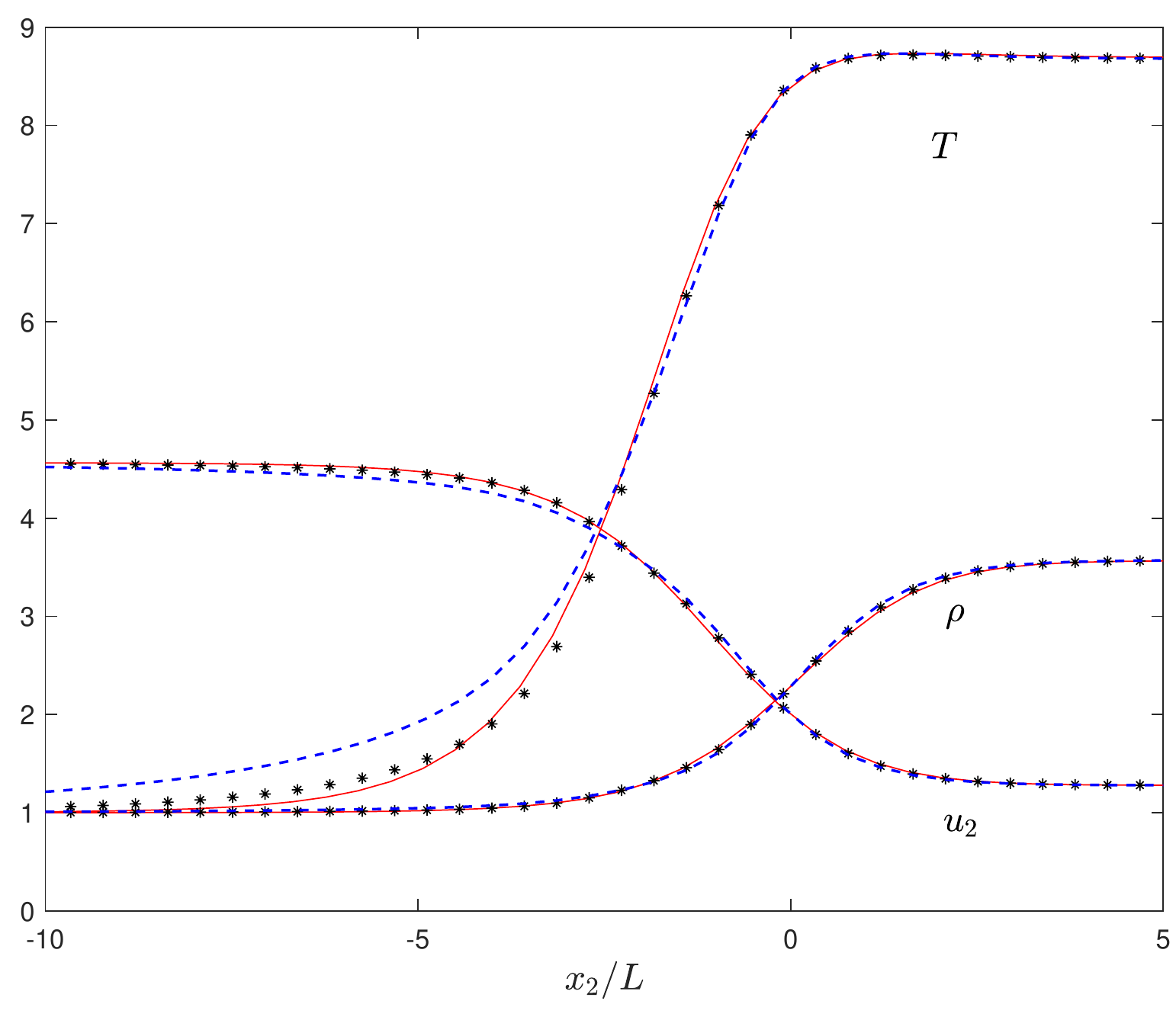}
	}\hskip 0.5cm
	{\includegraphics[width=0.36\textwidth]{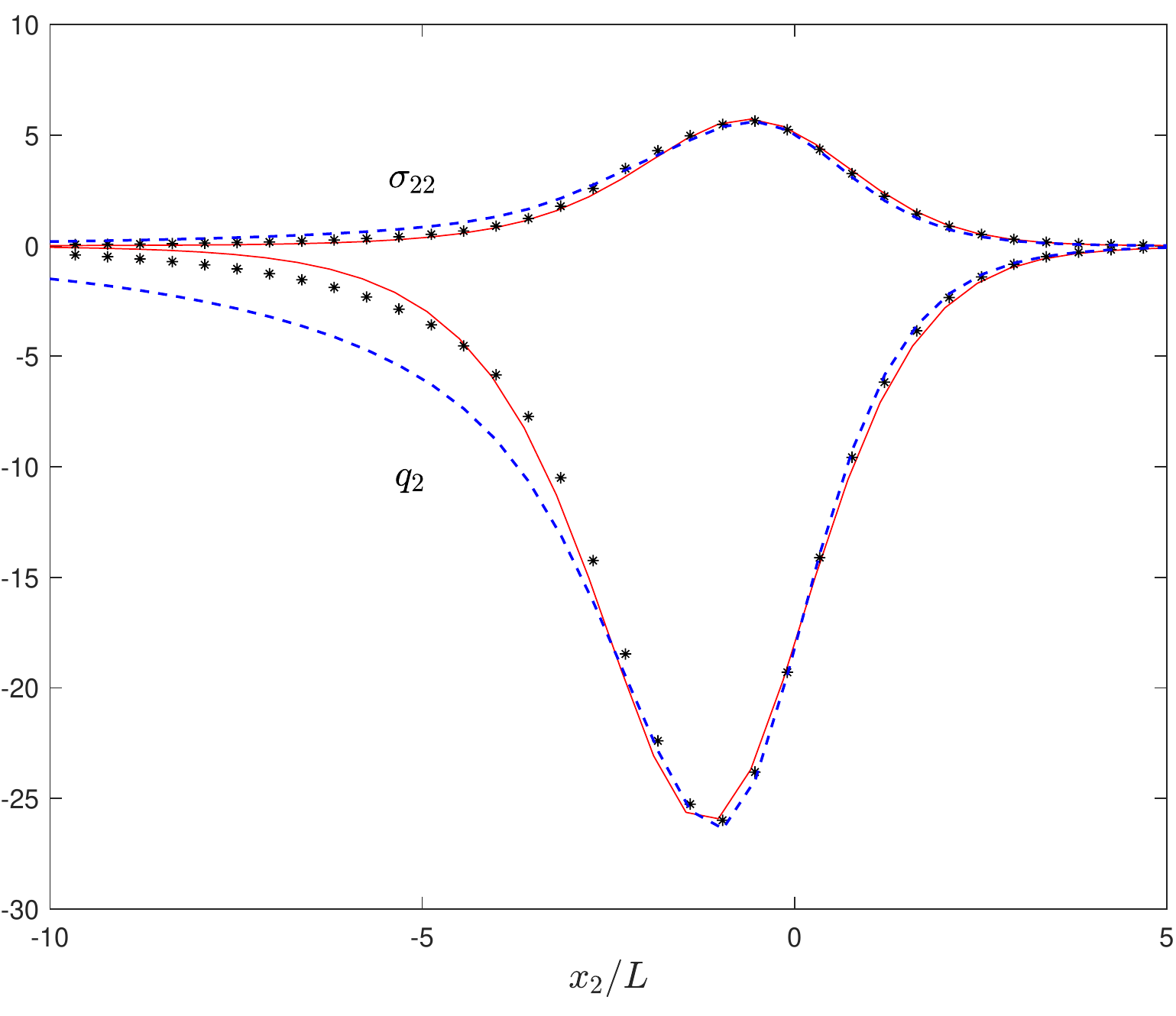}
	}
	\caption{ The Mach 5 normal shock wave in argon, using the Lennard-Jones potential. Solid lines: Boltzmann solutions with the collision kernel~\eqref{LJ_kernel}. Asterisks: the $\nu$-model with the collision frequency~\eqref{nu_freq_LJ}. Blue dashed lines: Shakhov model. 
	}
	\label{compare_shock_wave_LJ}
\end{figure}

\subsubsection{Lennard-Jones potential}

The $\nu$-model for the Lennard-Jones potential can be proposed straightforwardly, where the velocity-dependent collision frequency is designed to be a linear combination of those based on the inverse power-law potentials, in accordance with~\eqref{LJ_D}: 
\begin{equation}\label{nu_freq_LJ}
\nu_{LJ}(v)=A_{LJ}\sum_{j=1}^3b_j\left(\frac{k_BT}{\epsilon}\right)^{0.5-\omega_j}
\times\nu_{\omega_j}(\xi),
\end{equation}
 and $A_{LJ}$ can be determined from~\eqref{eq:nu_normalization}. Figure~\ref{fig:frequency} shows the typical collision frequency curves calculated by~\eqref{nu_freq_LJ} for the Lennard-Jones potential. Unlike the inverse power-law potential,  the shape of the collision frequency curve is different at different temperature for the Lennard-Jones potential.

 \begin{figure}
 	\centering
 	\includegraphics[width=0.7\textwidth]{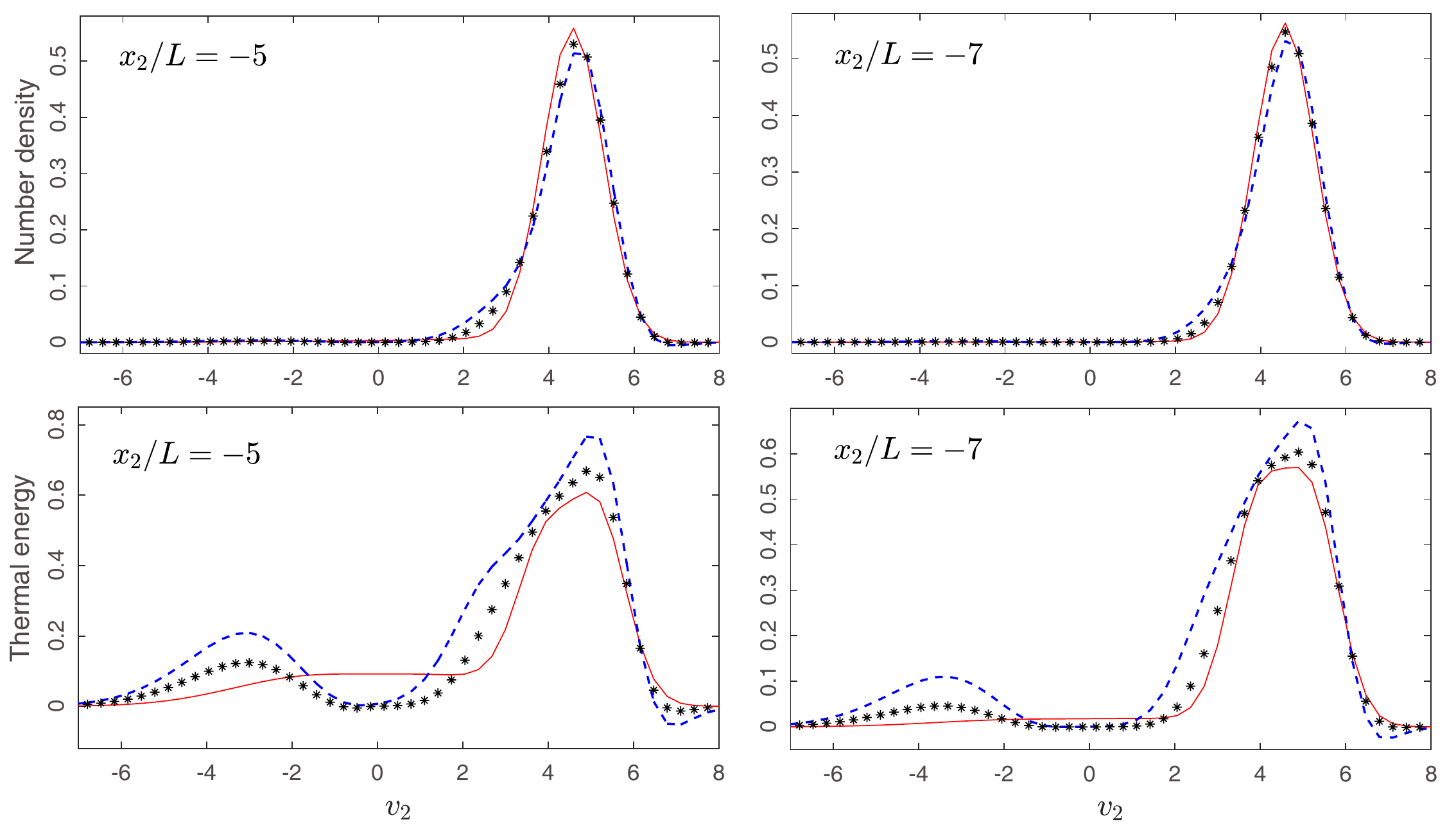}
 	\caption{ 
 		Marginal velocity distribution functions in the argon normal shock wave of $\text{Ma}=5$, using the Lennard-Jones potential. (Top row) Number density distribution $\int {f}dv_1dv_3$ and (bottom row) thermal energy distribution $\int {c^2f}dv_1dv_3$. Solid lines: Boltzmann solutions. Asterisks: $\nu$-model with the collision frequency~\eqref{nu_freq_LJ}. Blue dashed lines: Shakhov model.
 	}
 	\label{compare_shock_wave_LJ_VDF2}
 \end{figure}

\begin{table}
\centering
\caption{\label{tab:compare_shock_wave_LJ} Proportions of the molecular number density $\int_{{v_2} < 0} {fd\bm v}/\int{fd\bm v} $ and thermal energy $\int_{{v_2} < 0} {c^2fd\bm v} /\int {c^2fd\bm v}$ occupied by molecules with $v_2<0$, in the argon normal shock wave with $\text{Ma}=5$.}
\begin{tabular}{m{0.08\textwidth}<{\centering} m{0.18\textwidth}<{\centering}  m{0.09\textwidth}<{\raggedleft} m{0.09\textwidth}<{\raggedleft} m{0.09\textwidth}<{\raggedleft}}
\hline
\multicolumn{1}{c}{} &\multicolumn{1}{c}{Proportions of} & \multicolumn{1}{c}{Boltzmann} & \multicolumn{1}{c}{$\nu$-model} & \multicolumn{1}{c}{Shakhov}\\
\hline
$x_2/L$ & Number density & 	$0.03\%$ & 		$0.07\%$ &		 	$0.25\%$  \\ \cline{2-5}
=-9& Thermal energy & 		$1.01\%$ & 	 	$3.45\%$ &		 	$9.93\%$  \\ \hline 
$x_2/L$ & Number density & 	$0.13\%$ & 	 	$0.18\%$ &		 	$0.45\%$  \\ \cline{2-5}
=-7& Thermal energy & 	 	$4.20\%$ & 	 	$7.62\%$ &		 	$14.85\%$  \\ \hline 
$x_2/L$ & Number density & 	$0.68\%$ & 	 	$0.55\%$ &		 	$0.97\%$  \\ \cline{2-5}
=-5& Thermal energy & 	 	$15.66\%$ & 	 	$16.26\%$ & 	 	$21.94\%$  \\ \hline 
$x_2/L$ & Number density & 	$3.76\%$ & 	 	$3.11\%$ &		 	$3.61\%$  \\ \cline{2-5}
=-3& Thermal energy &	 	$34.28\%$ & 	 	$32.09\%$ &	 	$32.84\%$  \\
\hline
\end{tabular}
\end{table}

For the normal shock wave with Mach number 5 and upstream temperature $T_0=300$~K, the downstream temperature is 2604~K. For argon with the potential depth $\epsilon=119.2k_B$ in~\eqref{Lennard_Jones_chapter}, the viscosity given by~\eqref{shear_CE_viscosity0} and~\eqref{LJ_D} works well when the temperature is between 100~K and 3000~K.  Figure~\ref{compare_shock_wave_LJ} shows the macroscopic variable distributions along the flow direction calculated by different kinetic models. It is seen that the $\nu$-model yields consistent results with those from Boltzmann equation, while the Shakhov model significantly overpredicts the temperature and heat flux in the upstream area. Note that~\cite{Lei2013} have shown that the density, velocity and  temperature from the Boltzmann equation with the collision kernel~\eqref{LJ_kernel} agree with those from the molecular dynamics simulations of~\cite{Valentini2009}.

To further assess the accuracy of different kinetic models, figure~\ref{compare_shock_wave_LJ_VDF2} compares the marginal velocity distributions, especially the 
thermal energy distribution 
\begin{equation}\label{thermal_energy_dist}
F_{\rm{thermal}}=\iint_{-\infty}^\infty {c^2f}dv_1dv_3
\end{equation}
at the upstream locations $x_2/L=-5$ and $-7$, where the deviations in temperature and heat flux are large. It can be found that the number density distributions are nearly the same for different collision models, while the thermal energy distributions exhibit large discrepancy. The latter is analysed as follows. At $x_2/L=-5$ and $-7$, comparing with the Boltzmann solution, an extra bump around $v_2=-3$ for the thermal energy curve of the $\nu$-model and Shakhov model is observed. This energy peak soon diminishes going upstream in the $\nu$-model, while in the Shakhov model there still exists an obvious energy peak even at the very upstream location $x_2/L= -7$. This suggests that, in the Shakhov model, molecules with large negative velocities arising from the  high temperature post-shock gas can travel a very long distance from downstream to upstream, which significantly heats the gas therein. This is why the Shakhov model (and also for the ESBGK model) overpredicts the temperature and heat flux in the upstream. 

Table~\ref{tab:compare_shock_wave_LJ} further quantifies the number density and thermal energy occupied by molecules with $v_2<0$. Although the number of molecules with $v_2<0$ are small (less than 3.76\%), they do carry quite a part of the  energy (up to 34.28\%). It is also shown that, in the upstream region $x_2/L<$ -5, the proportion of thermal energy carried by molecules with $v_2<0$, predicted by the Shakhov model, is much larger than those of the $\nu$-model and Boltzmann equation. 

Based on the above analysis, in order to fix the overprediction of temperature and heat flux on top of the Shakhov model, the collision frequency of molecules with large speed should be increased to prevent high-speed molecules travelling too far to the upstream. Therefore, in our $\nu$-model, we design the velocity-dependent collision frequency based on the equilibrium collision frequency~\eqref{equi_fre_chapter2}, and thus the high-speed molecules have higher collision frequency as shown in figure~\ref{fig:frequency}, which effectively suppresses the heating of upstream gas due to the high speed $v_2<0$ molecules from the shock downstream.

\begin{figure}[t]
	\centering
	\subfloat[$\rm{Ma}=5$, ${Kn}_{\rm{VHS}}=0.1$]{\includegraphics[width=0.35\textwidth]{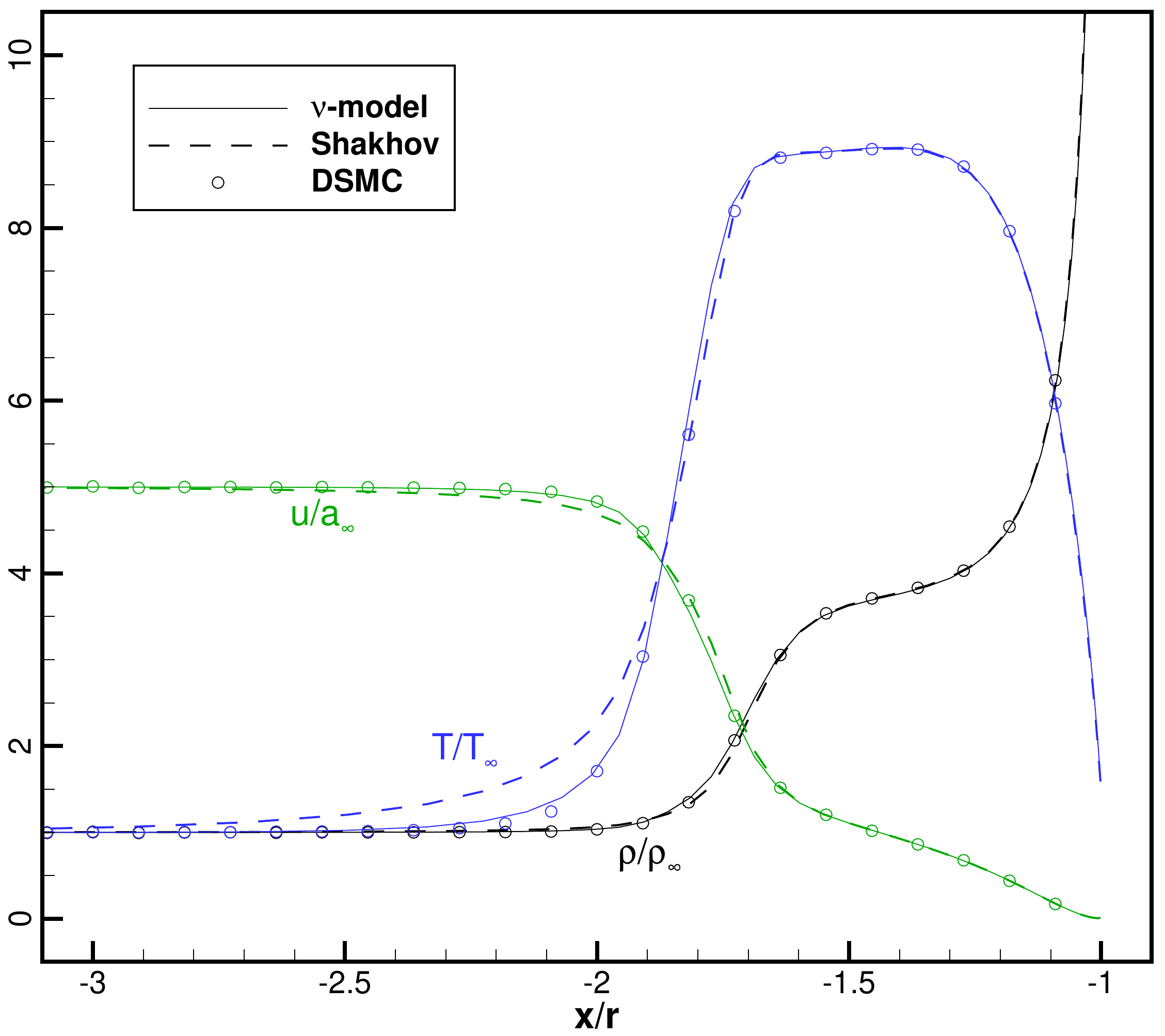}}\hskip 0.8cm
	\subfloat[$\rm{Ma}=5$, ${Kn}_{\rm{VHS}}=1$]{\includegraphics[width=0.35\textwidth]{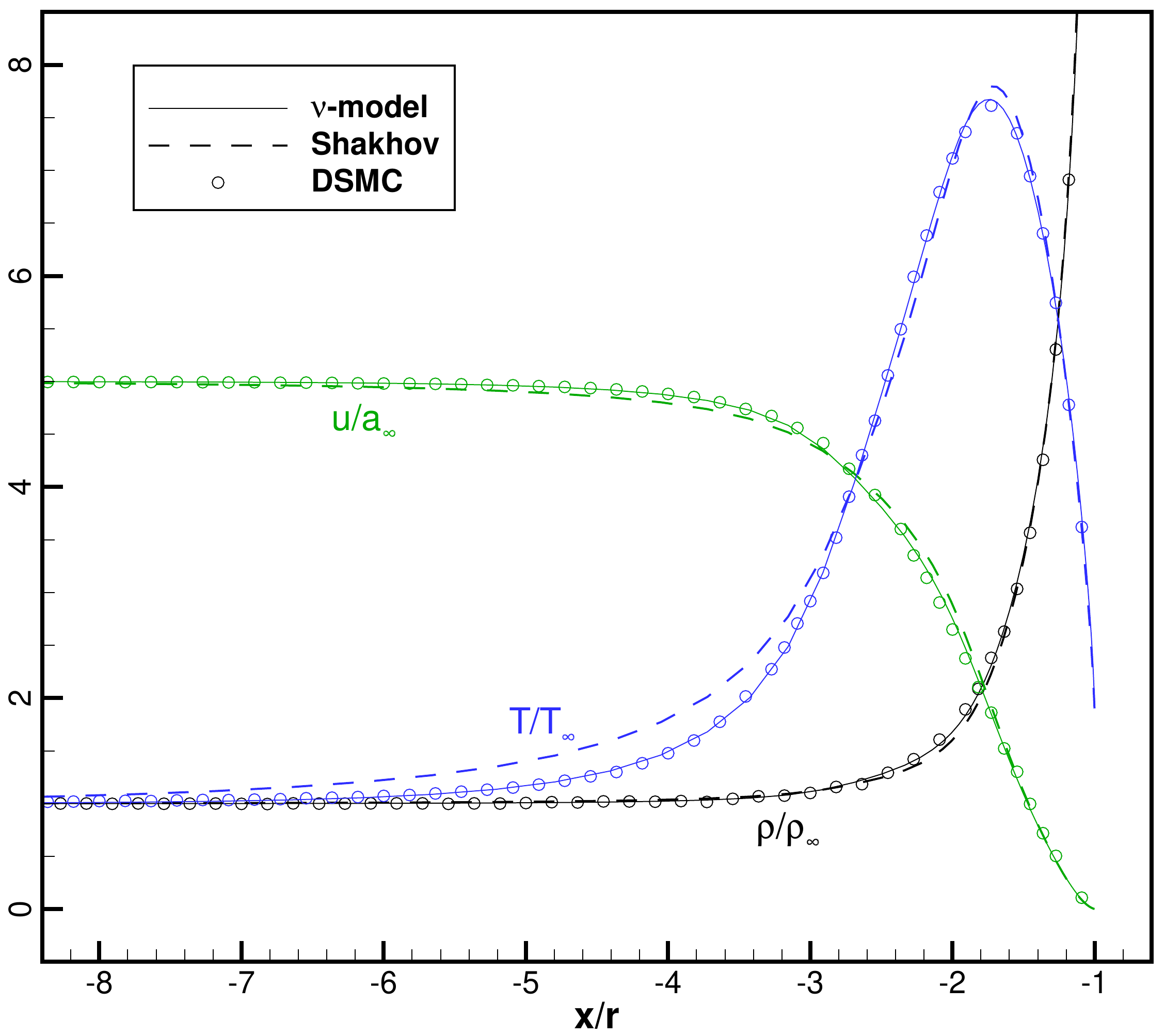}}\\
	\subfloat[$\rm{Ma}=20$, ${Kn}_{\rm{VHS}}=0.1$]{\includegraphics[width=0.35\textwidth]{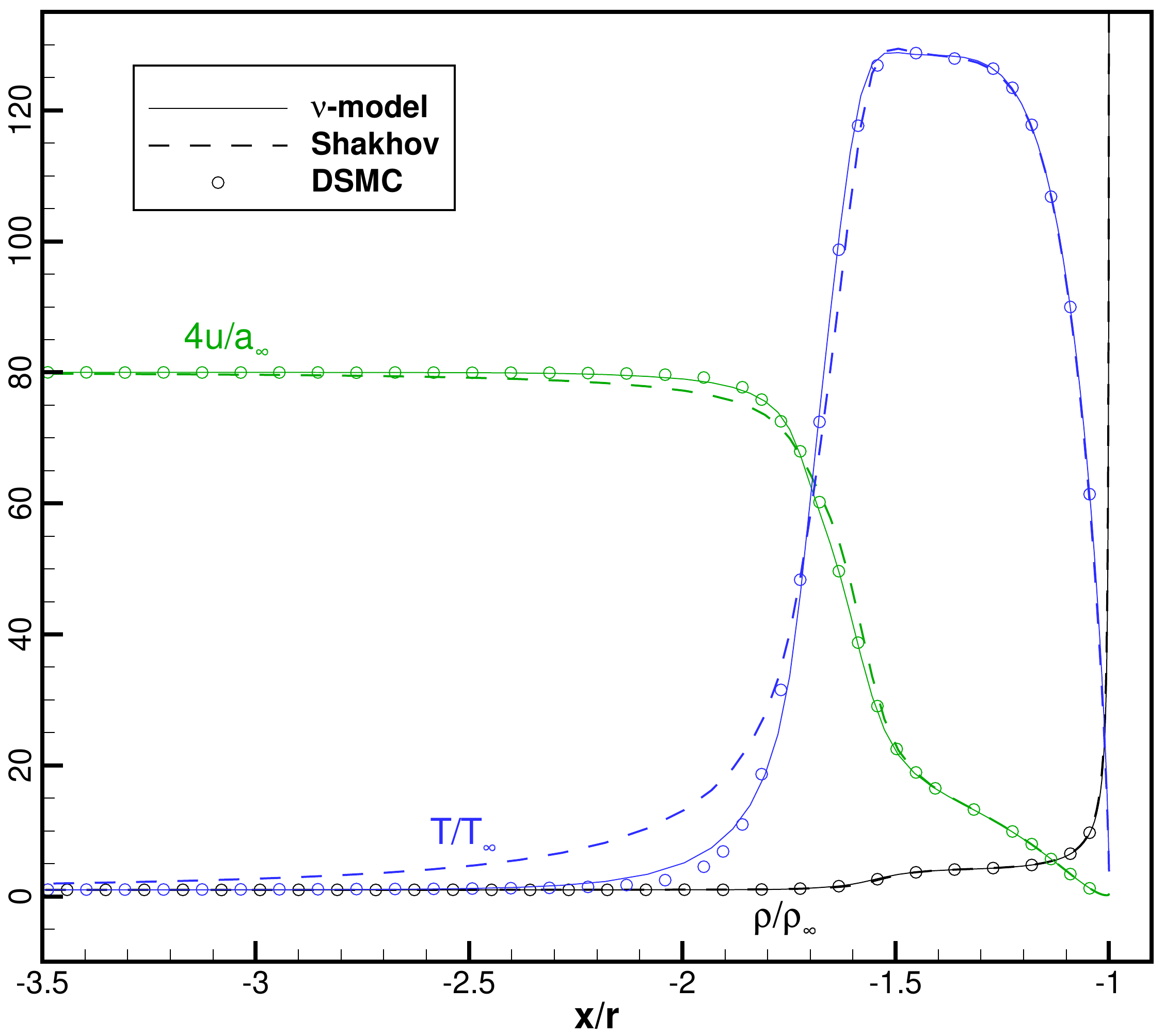}}\hskip 0.8cm
	\subfloat[$\rm{Ma}=20$, ${Kn}_{\rm{VHS}}=1$]{\includegraphics[width=0.35\textwidth]{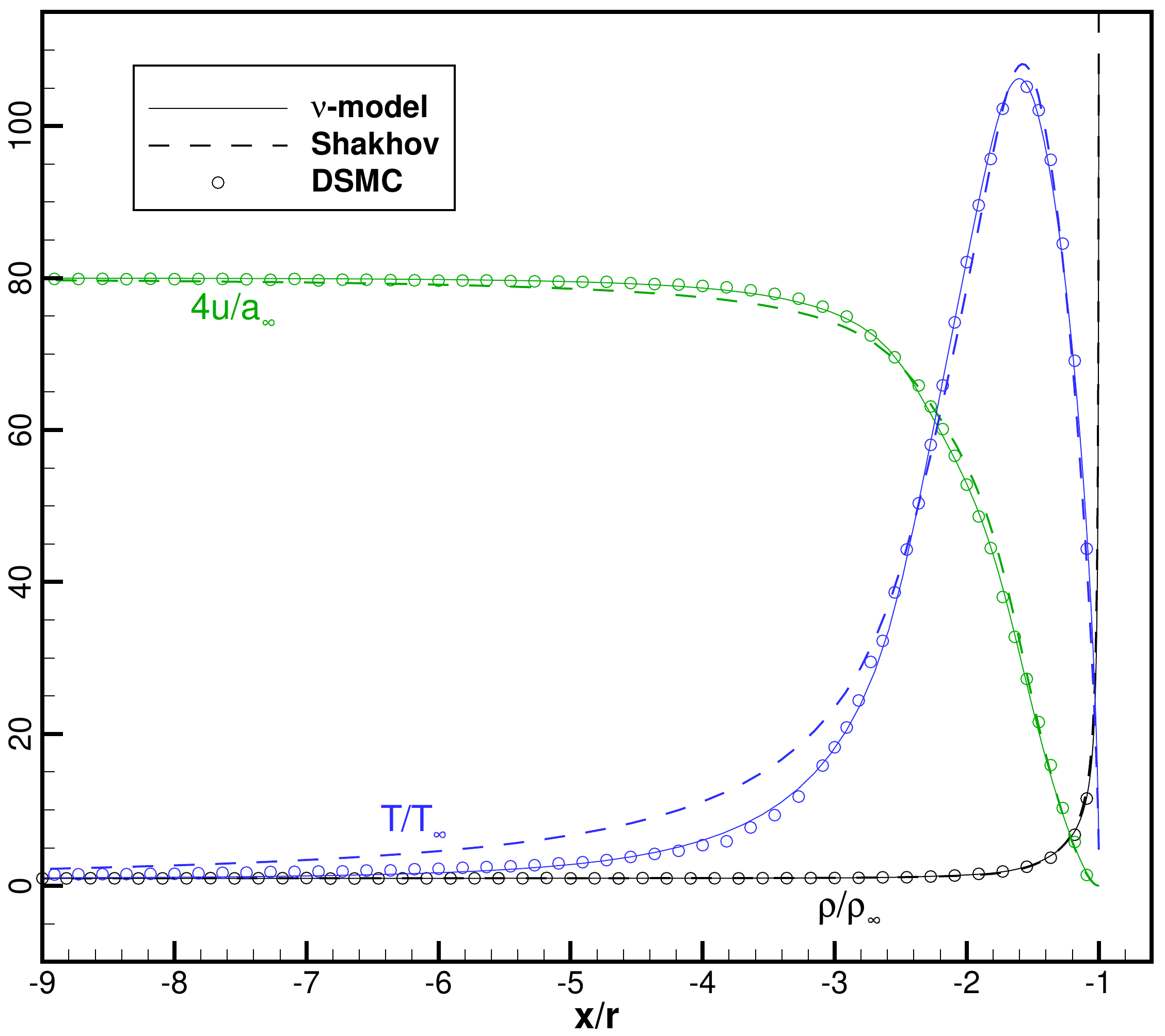}}
	\caption{\label{fig:cyd_ctline} Density, velocity and temperature variables along the central horizontal  line in the front of disc, in
	hypersonic flows of hard-sphere gas around a disc. DSMC results.
	}
\end{figure}

\subsection{Hypersonic flow around a disc}

The hypersonic flow around a disc is simulated to further assess the performance of our $\nu$-model. The inverse power-law potentials with $\omega=0.81$ and $\omega=0.5$ are considered, and due to limited space only the results of $\omega=0.5$ (the hard-sphere gas) are shown here. Results of $\omega=0.81$ from the $\nu$-model show similar accuracy as that of $\omega=0.5$. Four free stream conditions, $\rm{Ma}=5,20$ and ${Kn}_{\rm{VHS}}=0.1,1$ are considered, where the Knudsen number ${Kn}_{\rm{VHS}}$ is defined by the disc radius $r$ and the mean free path $L_{\rm{VHS}}$ is based on the VHS model of DSMC, i.e.
\begin{equation}  
{L_{{\rm{VHS}}}} = \frac{{2(7 - 2\omega )(5 - 2\omega )}}{{15}}\frac{\mu(T_\infty) }{{\rho \sqrt {2\pi RT_\infty} }}.
\end{equation}
The full diffuse reflection condition is imposed on the surface of the disc and the wall temperature is fixed at the freestream temperature: $T_{\rm{w}}=T_{\infty}$. For the discretization of physical space, the structured mesh in polar coordinates is used. The mesh size in the normal direction is refined approaching the disc surface, with the minimum mesh height set as $0.004r$ for $\rm{Ma}=5$ and $0.0006r$ for $\rm{Ma}=20$ to ensure the grid independence of surface stress and heat flux. Due to the multiscale and implicit nature of our numerical scheme, the computational cost is kept small. For the discretization of velocity space, $90 \times 90 \times 50$ uniform points in the velocity range $[-15a_{\infty},15a_{\infty}]$ and $160 \times 160 \times 128$ uniform points in the velocity range $[-55a_{\infty},55a_{\infty}]$ are adopted for $\rm{Ma}=5$ and $\rm{Ma}=20$, respectively, where $a_{\infty}$ is the freestream acoustic velocity.

\begin{figure}
\centering
\subfloat[$\rm{Ma}=5$, ${Kn}_{\rm{VHS}}=0.1$]{\includegraphics[width=0.38\textwidth]{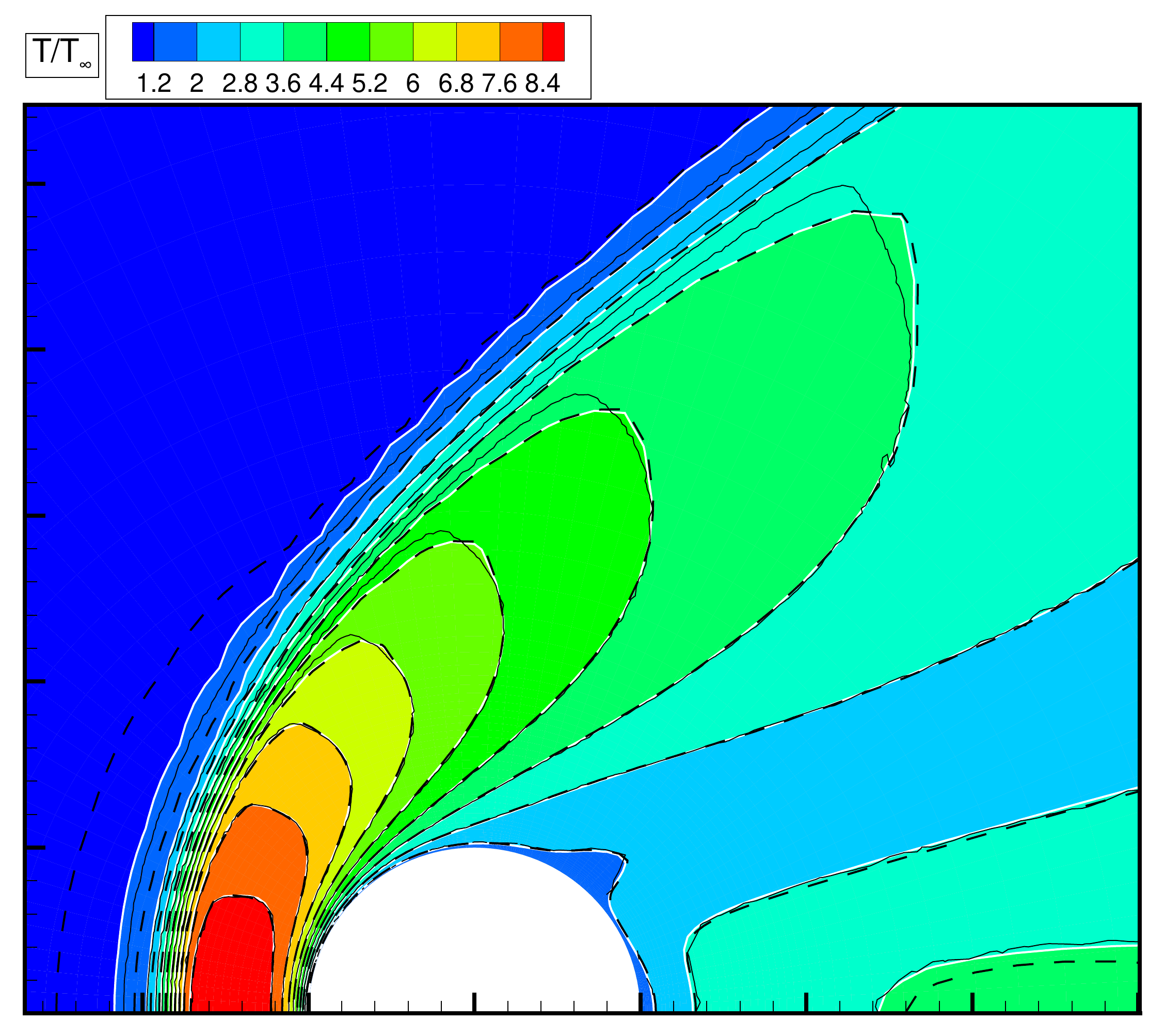}}\hskip 0.8cm
\subfloat[$\rm{Ma}=5$, ${Kn}_{\rm{VHS}}=1$]{\includegraphics[width=0.38\textwidth]{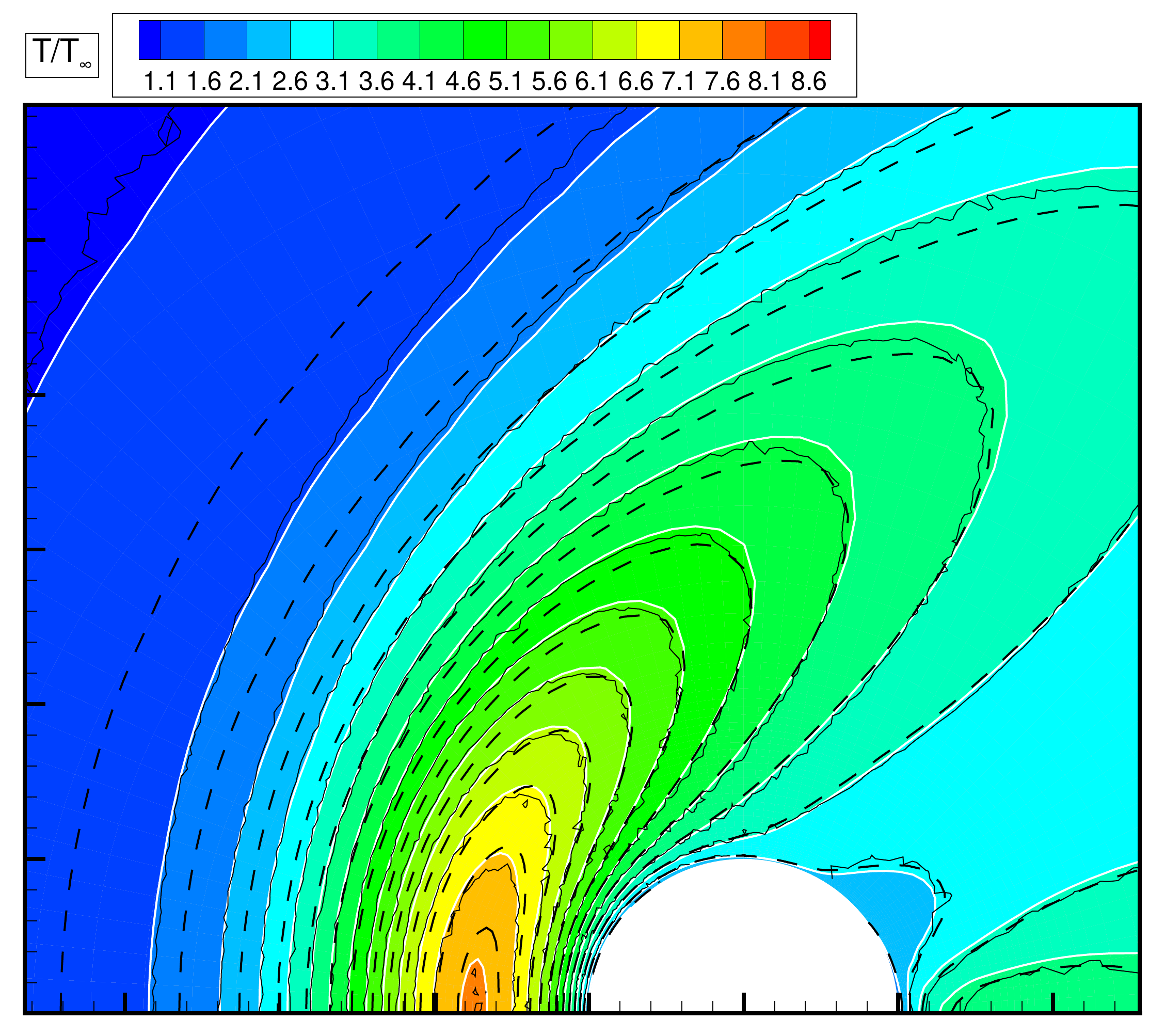}}\\
\subfloat[$\rm{Ma}=20$, ${Kn}_{\rm{VHS}}=0.1$]{\includegraphics[width=0.38\textwidth]{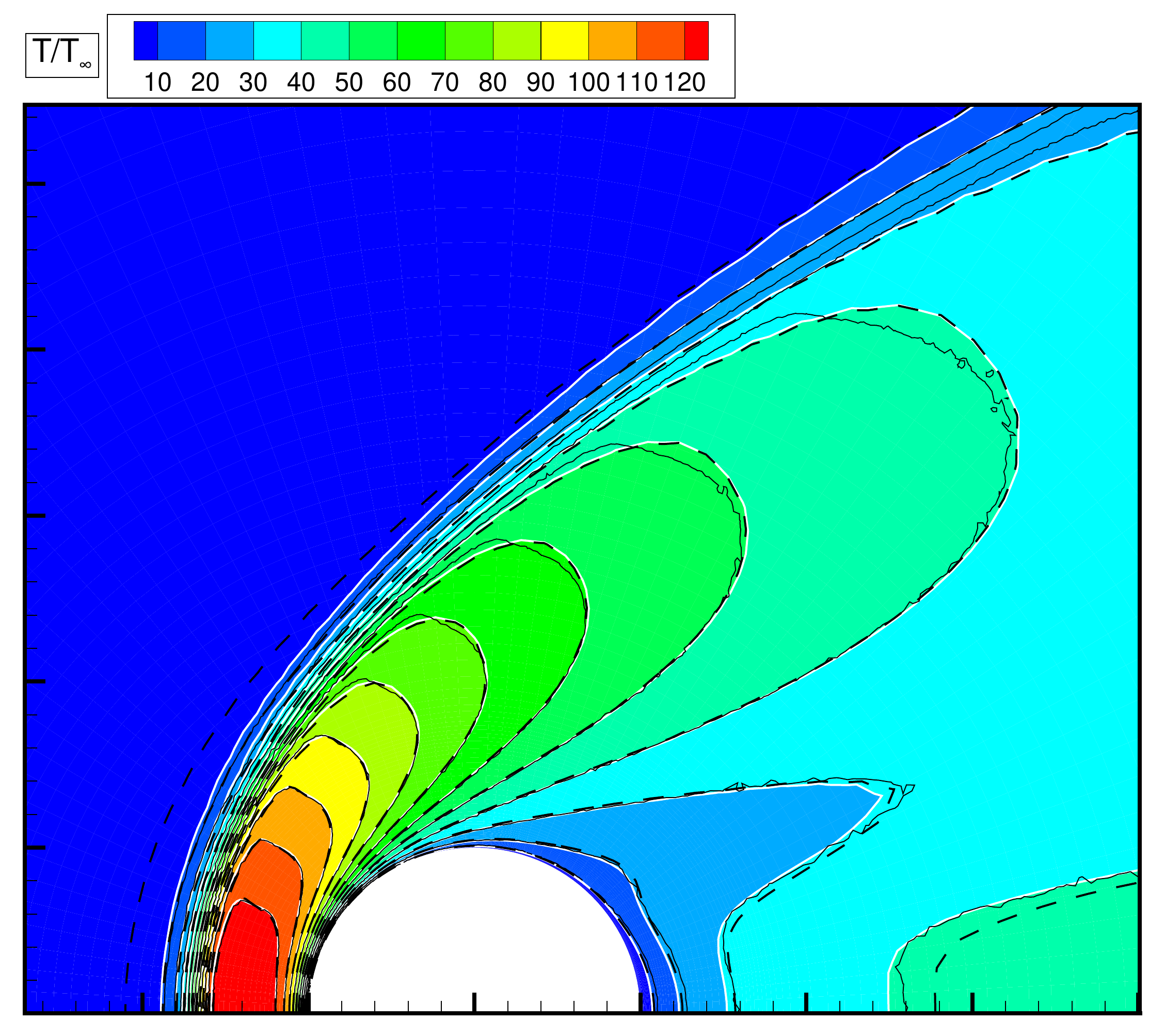}}\hskip 0.8cm
\subfloat[$\rm{Ma}=20$, ${Kn}_{\rm{VHS}}=1$]{\includegraphics[width=0.38\textwidth]{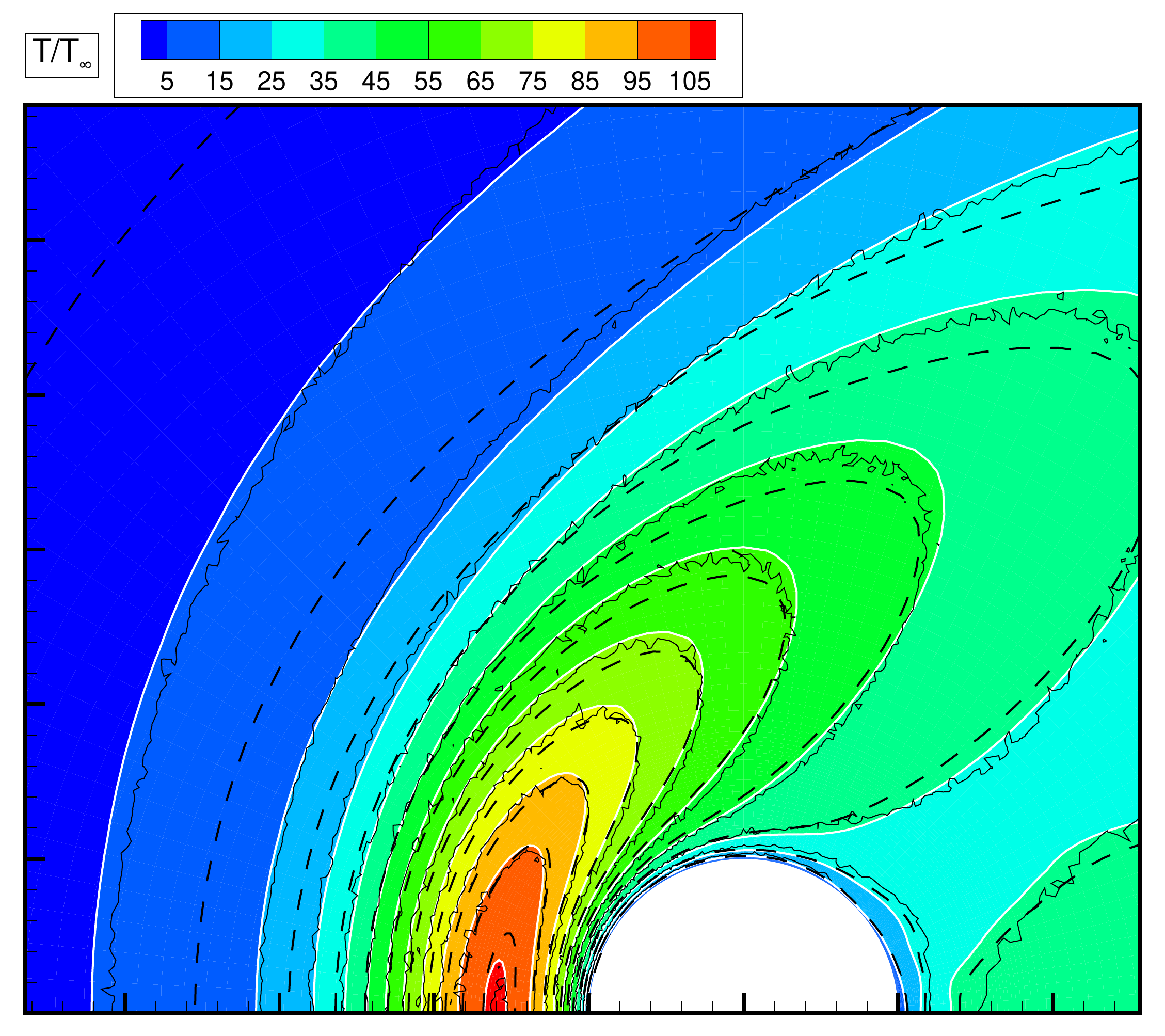}}
\caption{\label{fig:cyd_contourT} Temperature contours in  hypersonic flows of hard-sphere gas around a disc. Colour bands: $\nu$-model. Black dashed lines: Shakhov model. Black solid lines: DSMC results.}
\end{figure}

Numerical results of the flow variable distributions along the central horizontal line are shown in figure~\ref{fig:cyd_ctline}. It is seen that the $\nu$-model predicts quite satisfactory results in consistence with the DSMC results calculated by the DS2V code~\citep{bird2005ds2v}. For the Shakhov model, the accuracy in velocity profiles deteriorates slightly, and the upstream temperature is significantly overpredicted.  Figure~\ref{fig:cyd_contourT} shows that the temperature distributions around the disc obtained from the $\nu$-model agree well with the DSMC results, while the  Shakhov model exhibits large deviation, especially in the upstream of bow shock.


\begin{figure}
	\vskip 0.5cm
\centering
{\includegraphics[width=0.4\textwidth]{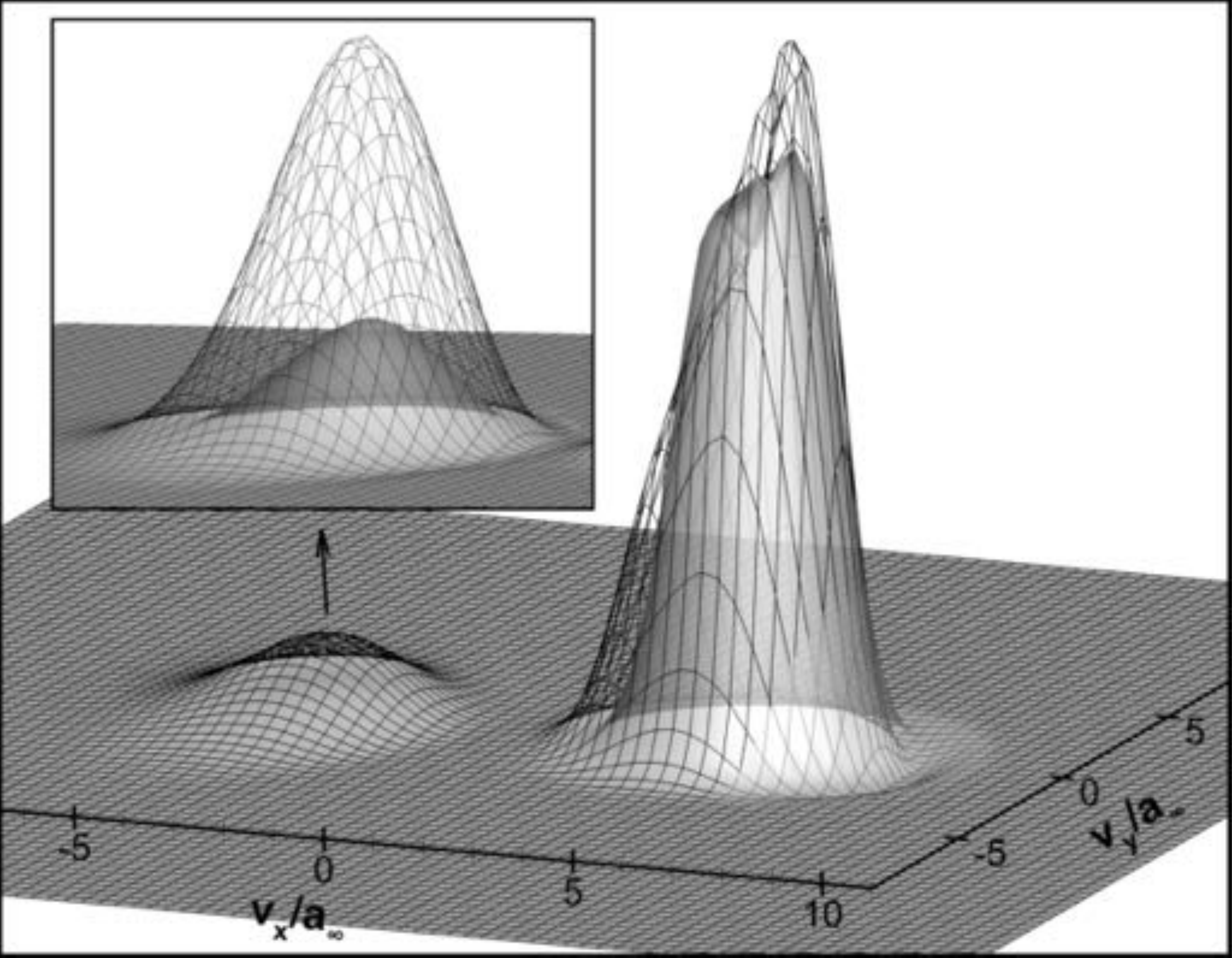}}
\hskip 0.3cm
{\includegraphics[width=0.4\textwidth]{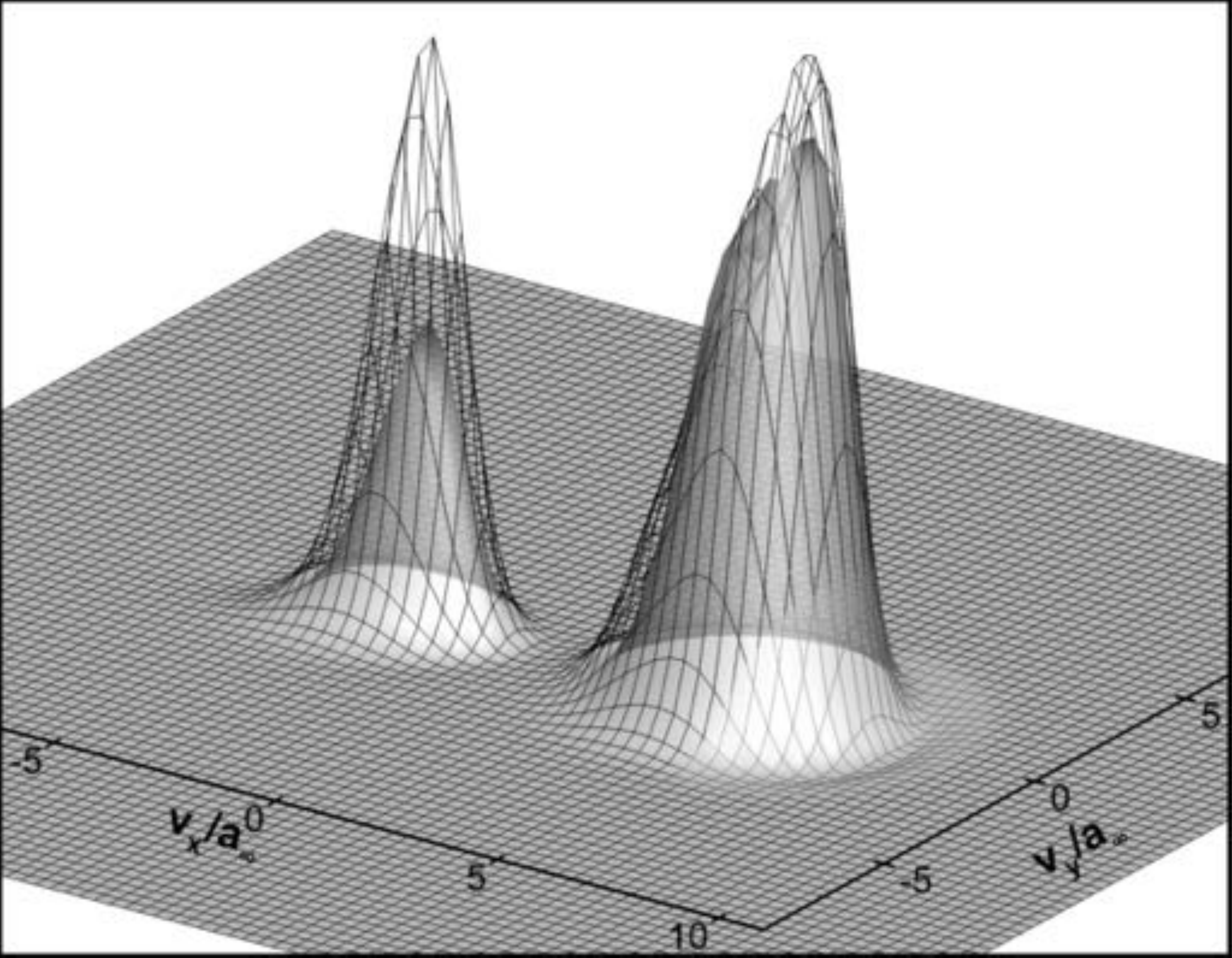}}
\caption{\label{fig:cyd_vdf}
	Thermal energy distributions $\int {c^2fd{v_z}}$ in hypersonic flows of hard-sphere gas of $\rm{Ma}=5$ around a disc, at locations before the bow shock. (Left) ${Kn}_{\rm{VHS}}=0.1$ at  $(x,y)=(-2.2r,-0.055r)$. Molecules with $v_x<0$  occupy 15.38\% and 5.65\% of the total thermal energy in the Shakhov model and $\nu$-model, respectively. (Right) ${Kn}_{\rm{VHS}}=1$ at  $(x,y)=(-4.4r,-0.11r)$. Molecules with $v_x<0$ occupy 23.63\% and 14.70\% of the total thermal energy in the Shakhov model and $\nu$-model, respectively. Gray surface: $\nu$-model. Wire frame: Shakhov model. 
}
\end{figure}

To further investigate the mechanism of such an improvement of the $\nu$-model for temperature prediction, the thermal energy distributions in the upstream of the bow shock for $\rm{Ma}=5$ are shown in figure~\ref{fig:cyd_vdf}. For this set of figures we sum up the following notable points:
\begin{enumerate}
\item Molecules with $v_x<0$  form an obvious energy peak, especially in the case of ${Kn}_{\rm{VHS}}=1$.
As supplement to the data shown in figure~\ref{fig:cyd_vdf} when $\rm{Ma}=5$, at $\rm{Ma}=20$ in the temperature-early-rising region, the Shakhov model predicts the proportions of thermal energy occupied by molecules with $v_x<0$  to be 42.03\% when ${Kn}_{\rm{VHS}}=0.1$ and 61.62\% when ${Kn}_{\rm{VHS}}=1$, while in the $\nu$-model these data are 9.38\% and 37.18\%, respectively. This suggests that the high speed (large peculiar velocity) $v_x<0$ molecules arising from the post-shock gas have a big impact on the thermal energy of the upstream pre-shock gas and cause a significant heating.
\item The thermal energy peak due to the high speed $v_x<0$ molecules predicted by the $\nu$-model is much lower than that predicted by the Shakhov model. This is because that in the $\nu$-model we adopt the velocity-dependent collision frequency~\eqref{nu_freq_power}, where the molecule with larger peculiar velocity has higher collision frequency; and intensive collisions prevent them from transporting upstream too far, and thus the overprediction of upstream temperature observed in the Shakhov model is suppressed in the $\nu$-model. This also suggests that, when the viscosity index $\omega$ approaches $0.5$ and when the Mach number gets larger, temperature-overprediction by the Shakhov model will become more severe due to the steeper collision frequency curve (figure~\ref{fig:frequency}) and higher peculiar velocity of $v_x<0$ molecules.
\end{enumerate}


\begin{figure}
	\centering
	\subfloat[$\rm{Ma}=5$, ${Kn}_{\rm{VHS}}=0.1$]{\includegraphics[width=0.35\textwidth]{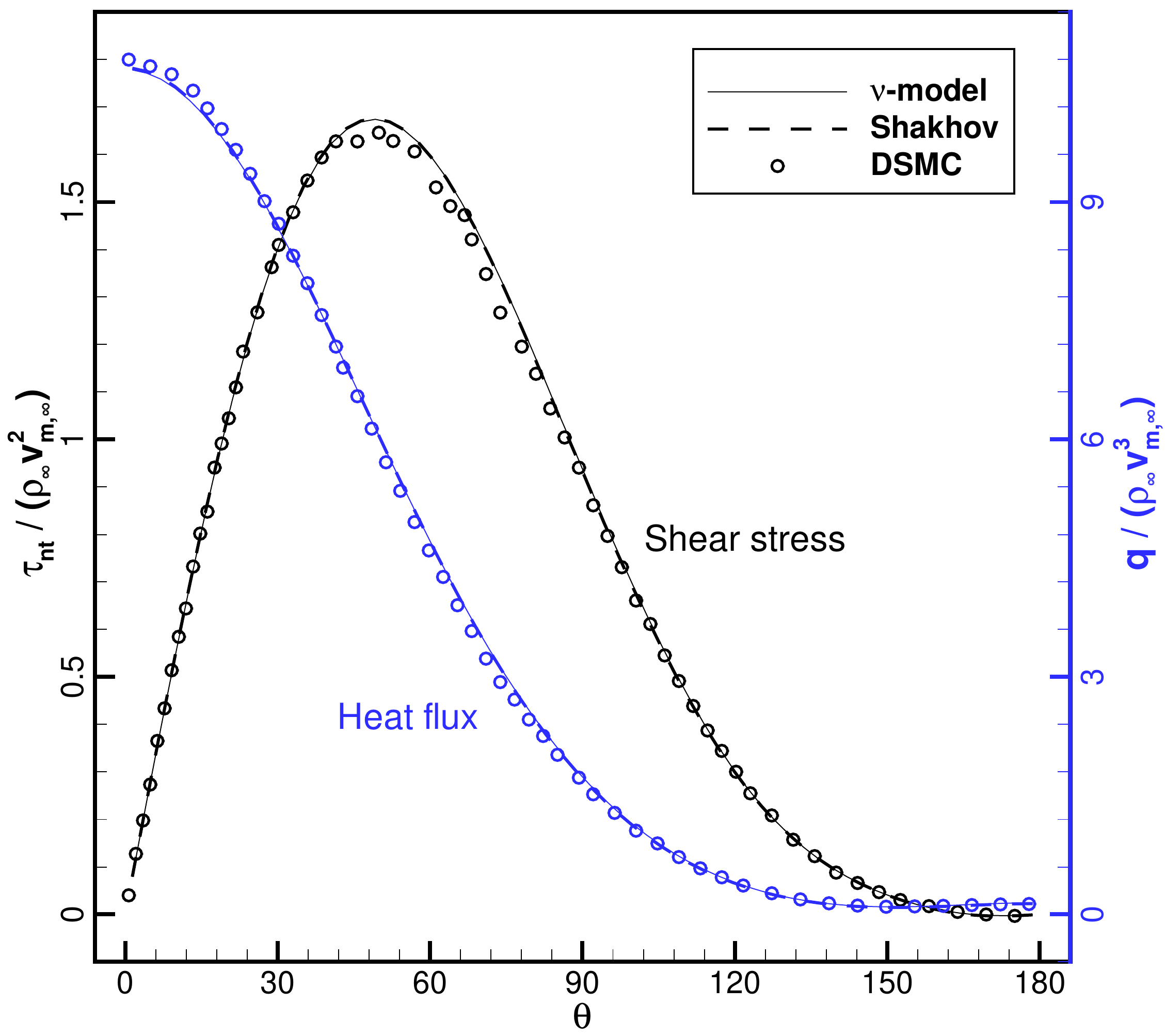}}\hskip 0.3cm
	\subfloat[$\rm{Ma}=5$, ${Kn}_{\rm{VHS}}=1$]{\includegraphics[width=0.35\textwidth]{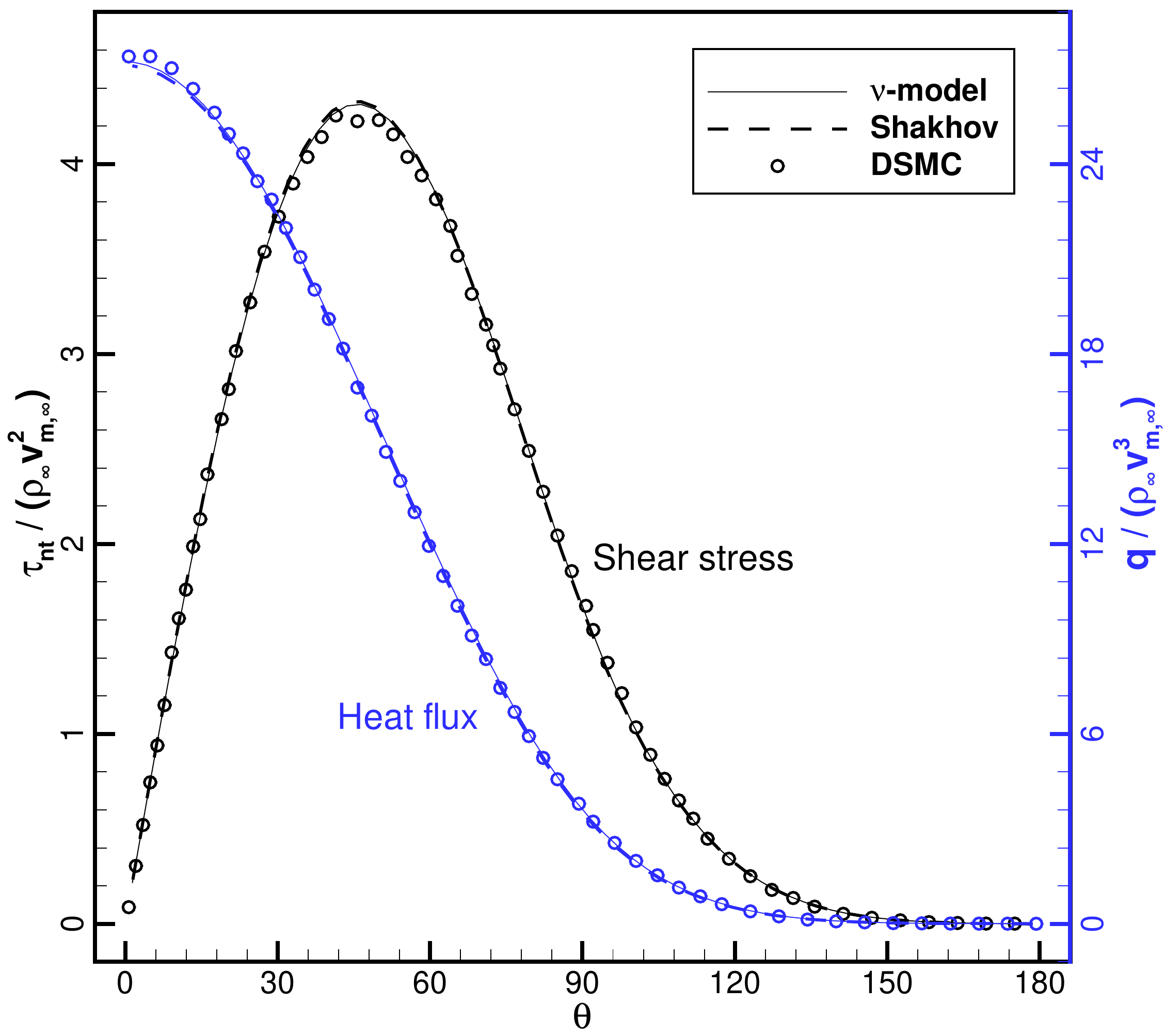}}\\
	\subfloat[$\rm{Ma}=20$, ${Kn}_{\rm{VHS}}=0.1$]{\includegraphics[width=0.35\textwidth]{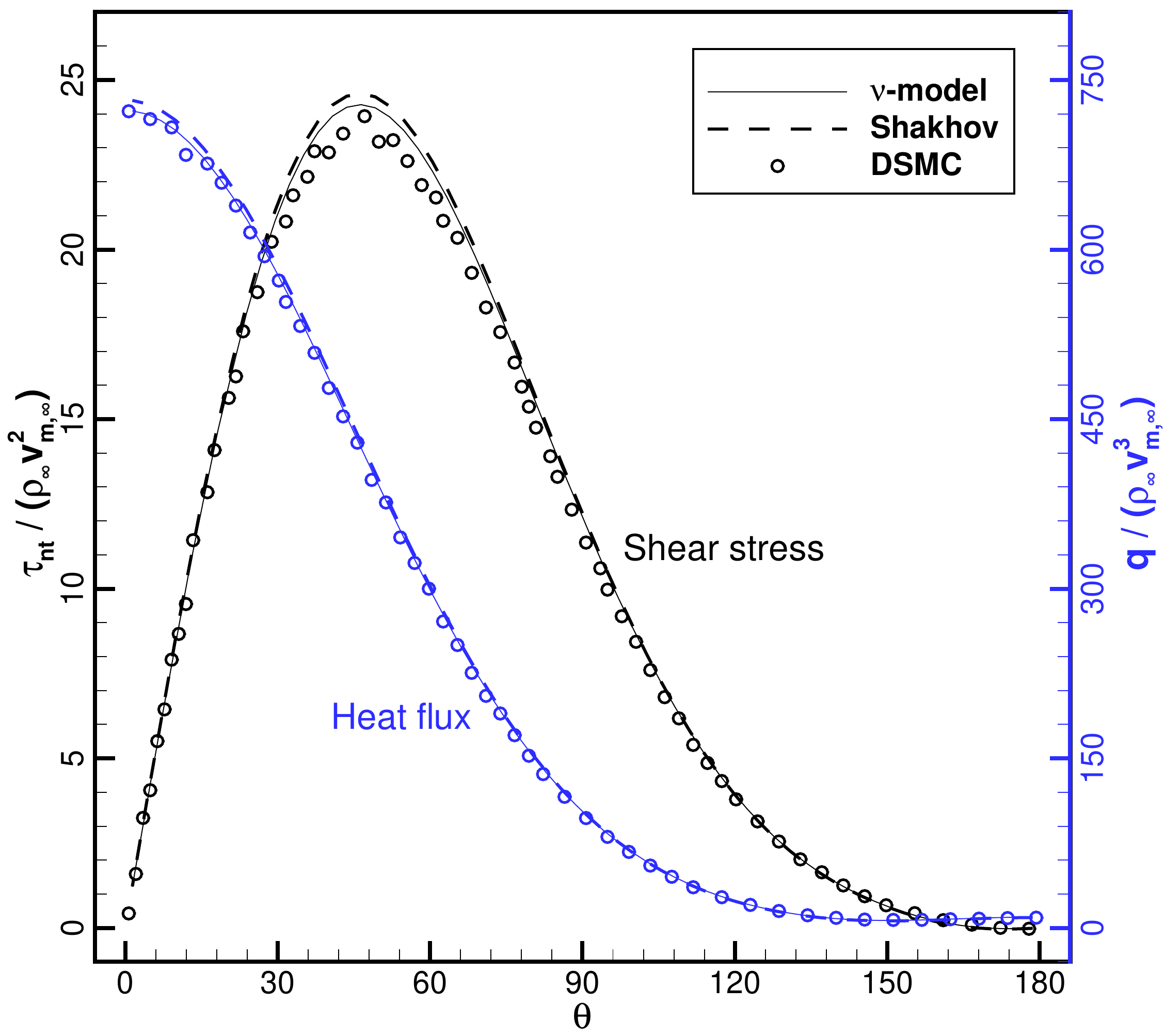}}\hskip 0.3cm
	\subfloat[$\rm{Ma}=20$, ${Kn}_{\rm{VHS}}=1$]{\includegraphics[width=0.35\textwidth]{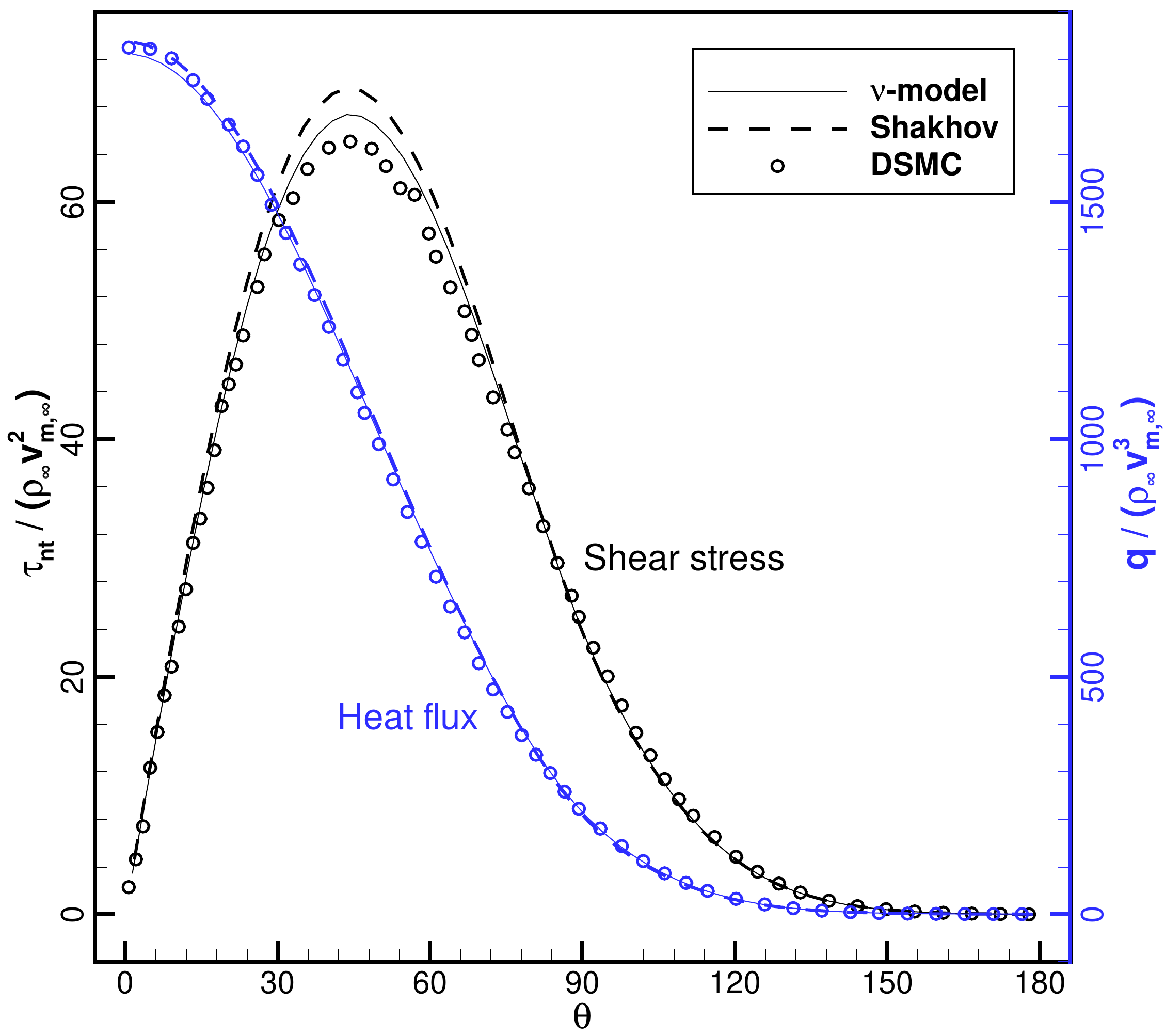}}
	\caption{Distributions of the shear stress and heat flux  along the disc surface, in hypersonic flows of hard-sphere gas around a disc.
	}
	\label{fig:cyd_sfqt}
\end{figure}

Distributions of the shear stress and heat flux on the disc surface are shown in figure~\ref{fig:cyd_sfqt}. When ${\rm{Ma}}=5$, the $\nu$-model and the Shakhov model predict almost the same results and they both agree well with DSMC. This is because for the high-temperature post-shock gas, there is less molecules with large peculiar velocity and the collision frequency in the Shakhov model  is comparable with that in the $\nu$-model. When ${\rm{Ma}}=20$, a certain degree of discrepancy exists between the results of the Shakhov and $\nu$-models, and the $\nu$-model shows better agreement with DSMC. 

\section{Numerical results in micro-flows} \label{num_micro_flows}

In this section we assess the accuracy of the $\nu$-model in canonical rarefied micro-flows, with the velocity-dependent collision frequency determined from the strong normal shock waves.

\subsection{Planar Couette flow}

\begin{figure}[t]
	\vskip 0.5cm
	\centering
	\subfloat[Maxwell gas]{	\includegraphics[height=6cm]{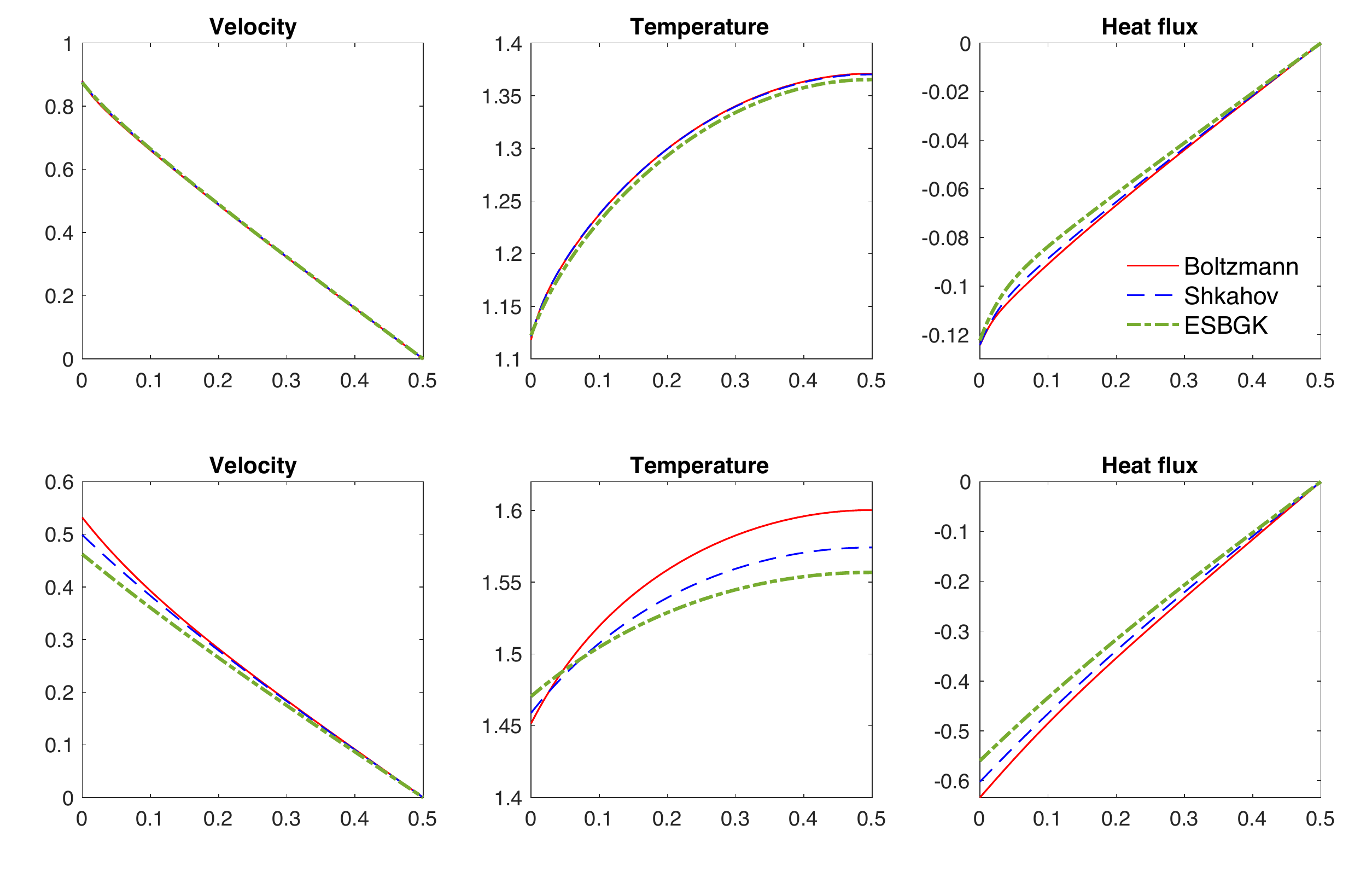}
	}\\
	\subfloat[Hare sphere gas]{
		\includegraphics[height=6cm]{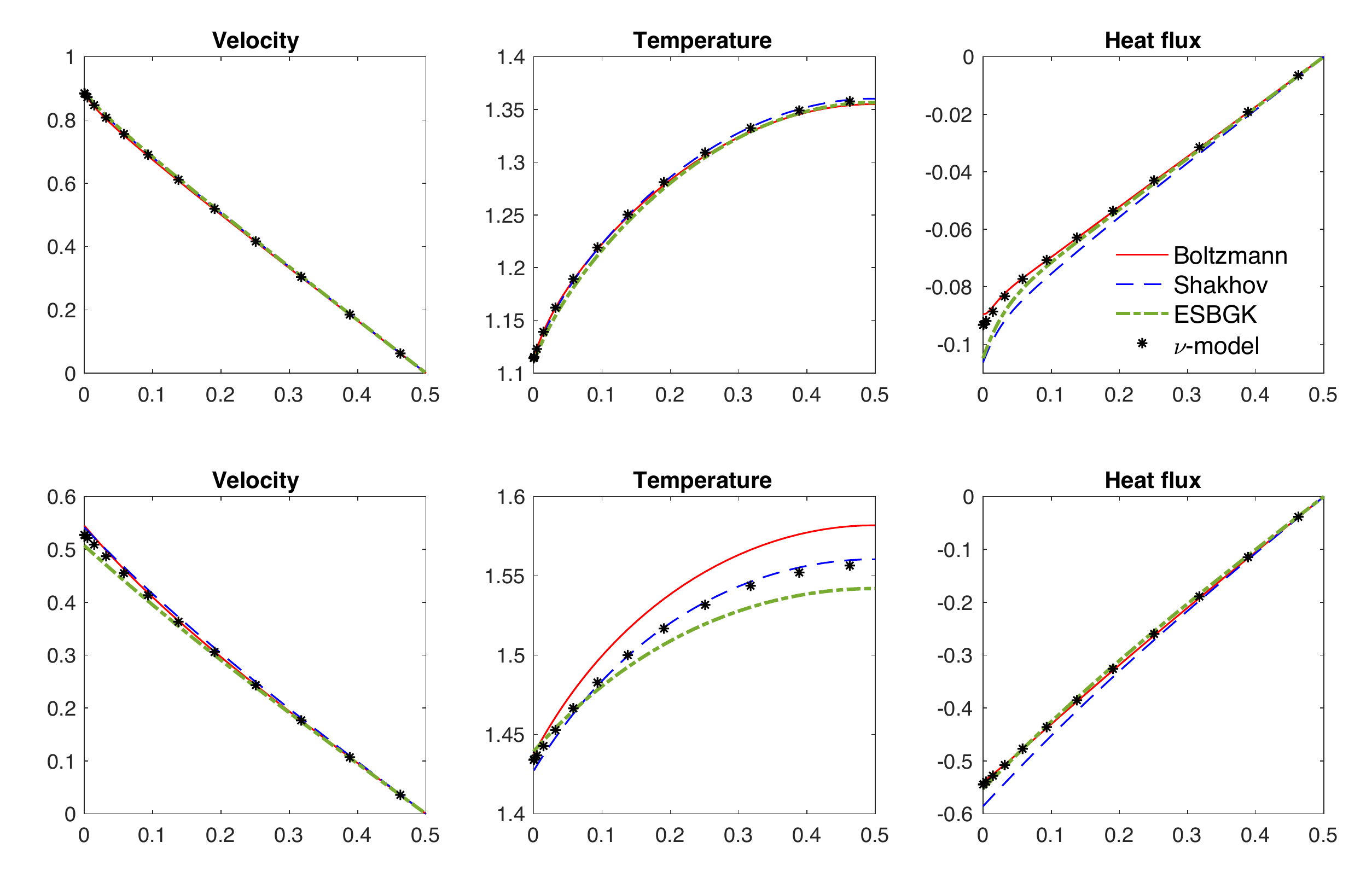}
	}
	\caption{ 
		Couette flow. First and third rows: $Kn=0.1$. Second and fourth rows:  $Kn=1$. The abscissas $x_2$ are for the spatial coordinate, which is in the direction perpendicular to the two plates and normalized by the wall distance. The two plates are located in $x_2=0$ and $x_2=1$. The heat flux is parallel to the wall velocity.  Due to symmetry, only the half spatial region is shown. 
	}
	\label{compare_couette_combined}
\end{figure}

Unlike the normal shock wave that is dominated by the effects of compressibility, the Couette flow is shear-dominated. It is a typical rarefied gas flows, since the heat flux parallel to the plates, is not zero, in sharp contrast to the Navier-Stokes-Fourier equations.  Here we consider the Couette flow between two parallel plates with temperature $T_0$, where the wall speed is equal to the most probable speed of gas molecules at $T_0$. For simplicity we only consider the Maxwellian and hard-sphere gases, since for other gases the viscosity satisfies $0.5\le\omega\le1$, and the results fall between these of Maxwellian and hard-sphere gases. The characteristic flow length $L$ in~\eqref{Knudsen_number} is chosen to be the distance between two plates.

For the Maxwellian gas, when ${Kn}=0.1$, figure~\ref{compare_couette_combined}(a) shows that the Shakhov model produces close results to those of the Boltzmann equation, while the ESBGK model has some slight errors in temperature and heat flux. When $Kn=1$, the difference between the Shakhov/ESBGK model and the Boltzmann equation increases, but we see that the Shakhov model is better than the ESBGK model, in velocity, temperature, and heat flux. 
However, when the hard-sphere gas is considered, figure~\ref{compare_couette_combined}(b) shows that the Shakhov model is better than the ESBGK model in terms of temperature, but is worsen in heat flux.

When the $\nu$-model is used,  we find that its heat flux agrees well with the solution of the Boltzmann equation. However, there is no improvement in the temperature profile as compared to the Shakhov model; nevertheless, the relative error in temperature to that of the Boltzmann equation is within 3\%.  It is also worth noting that the $\nu$-BGK and $\nu$-ESBGK models~\citep{Mieussens2004,Zheng2005} predict even worse results than the standard ESBGK, and they are not suggested for Couette flow~\citep{Zheng2005}. 

\subsection{Thermal transpiration}

Another typical phenomena in rarefied gas dynamics is the thermal transpiration, where the gas moves towards a hotter region even in the absence of a pressure gradient~\citep{Reynolds1879,Maxwell1879vii}. Harnessing this unique property leads to the design of Knudsen compressor that pumps the gas without any moving mechanical part~\citep{vargo1999knudsen,gupta2008thermal}. This problem is a good test case since even when the value of viscosity is same, different intermolecular potentials yield different thermal slip velocity~\citep{Wang2020PoF} and mass flow rate~\citep{Sharipov2009,wuPoF2015}; and this can be captured neither by the relaxation model~\eqref{overall_kinetic_model} with velocity-independent collision frequency, nor by the Fokker-Planck model.

Here we assess the performance of our $\nu$-model in the thermal transpiration between two parallel plates and focus on the steady-state solutions. The governing equation reads
\begin{equation}  
\begin{aligned}
v_2\frac{\partial
	{f}}{\partial{x_2}}=Q+\text{Source},
\end{aligned}
\end{equation}
where the source term is $-a_0v_1(c^2-5/2)F_{eq}$, with $a_0$ being a small constant related to the temperature gradient along the solid wall. The induced flow velocity due to rarefaction effects is proportional to $a_0$, and the final result will be further normalized by $a_0$. 


\begin{figure}
	\centering
	\subfloat[$Kn=0.01$]{	\includegraphics[height=4cm]{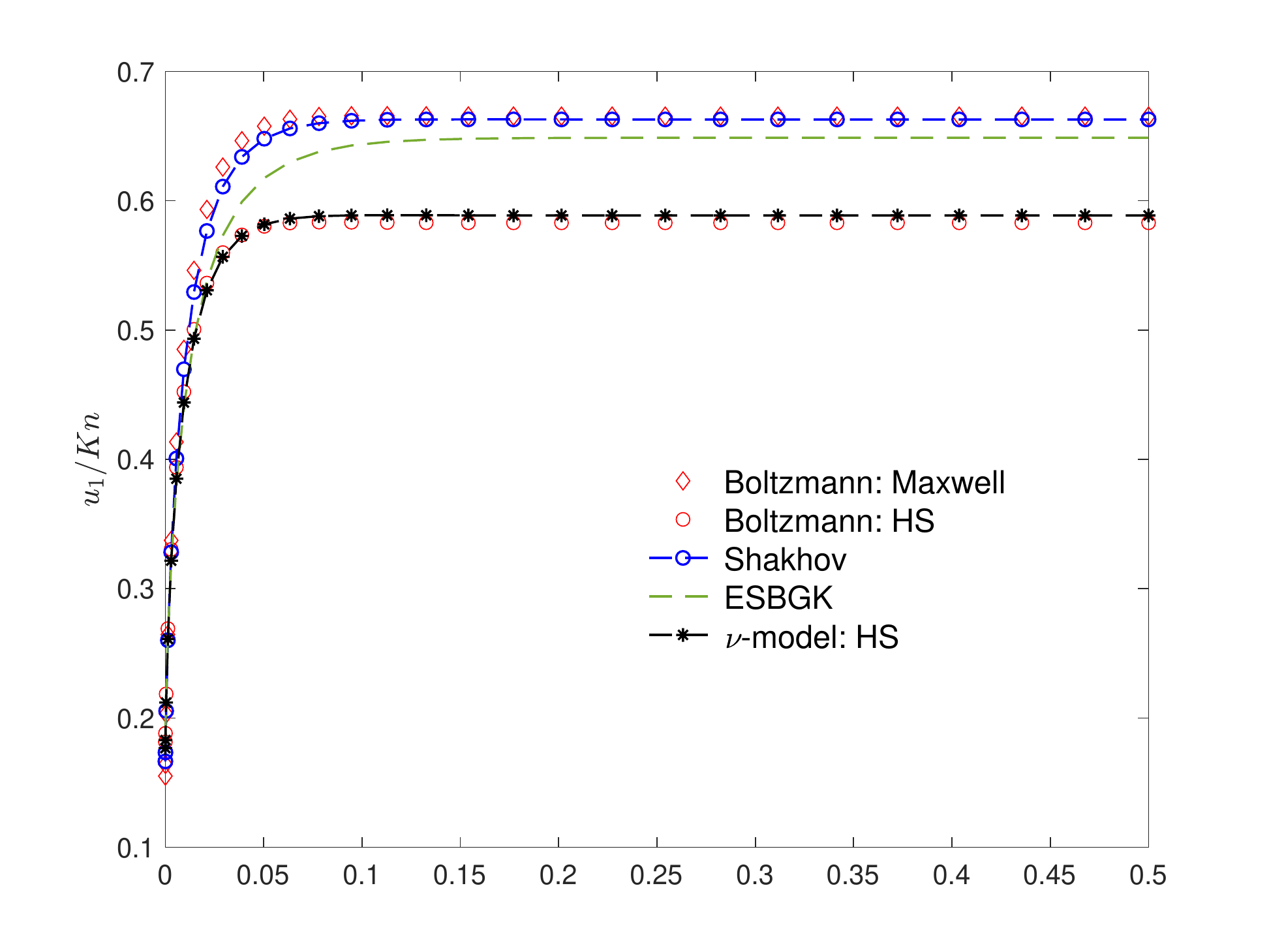}
	}
	\subfloat[$Kn=0.1$]{\includegraphics[height=4cm]{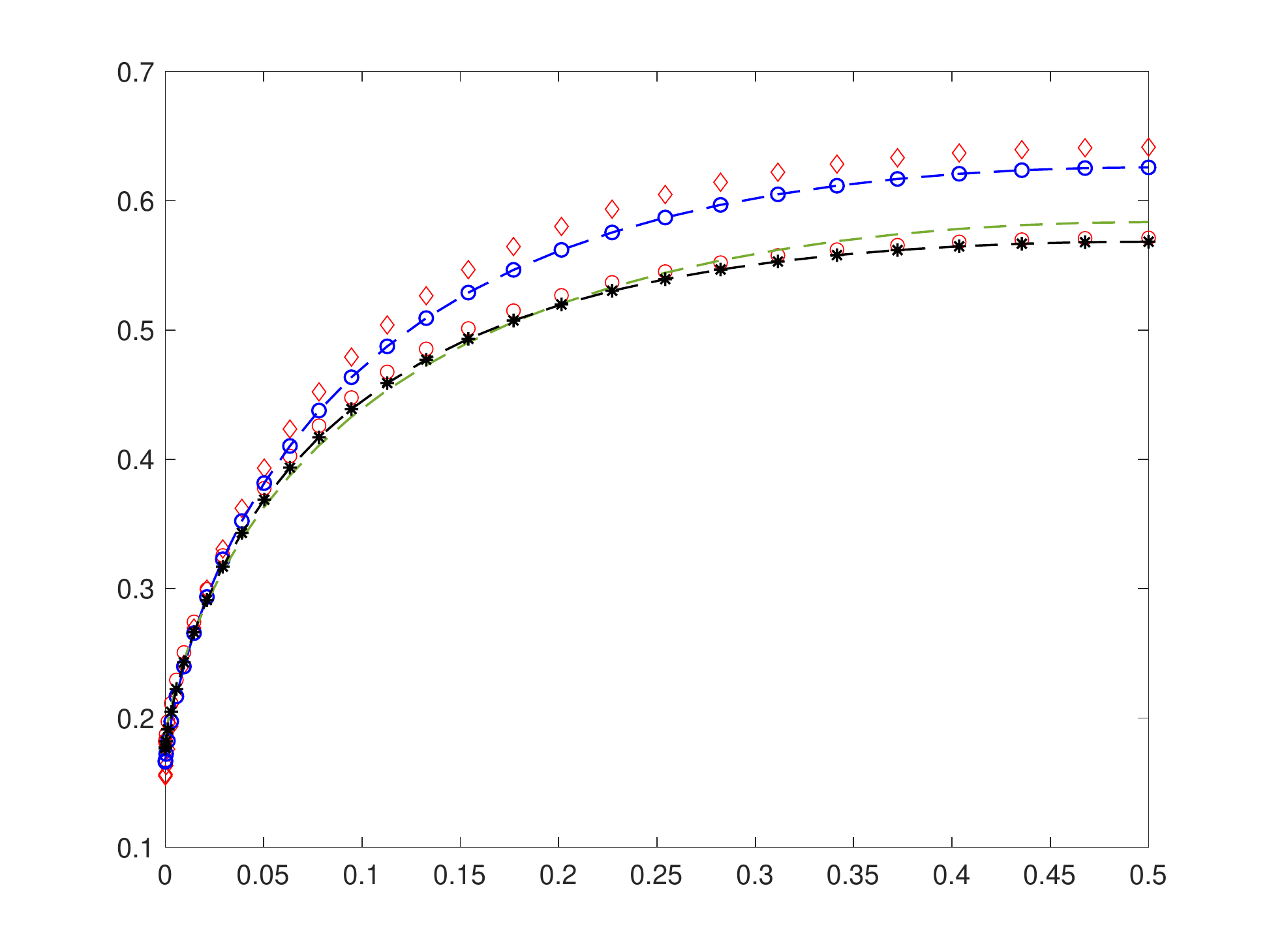}
	}\\
	\subfloat[$Kn=1$]{
		\includegraphics[height=4cm]{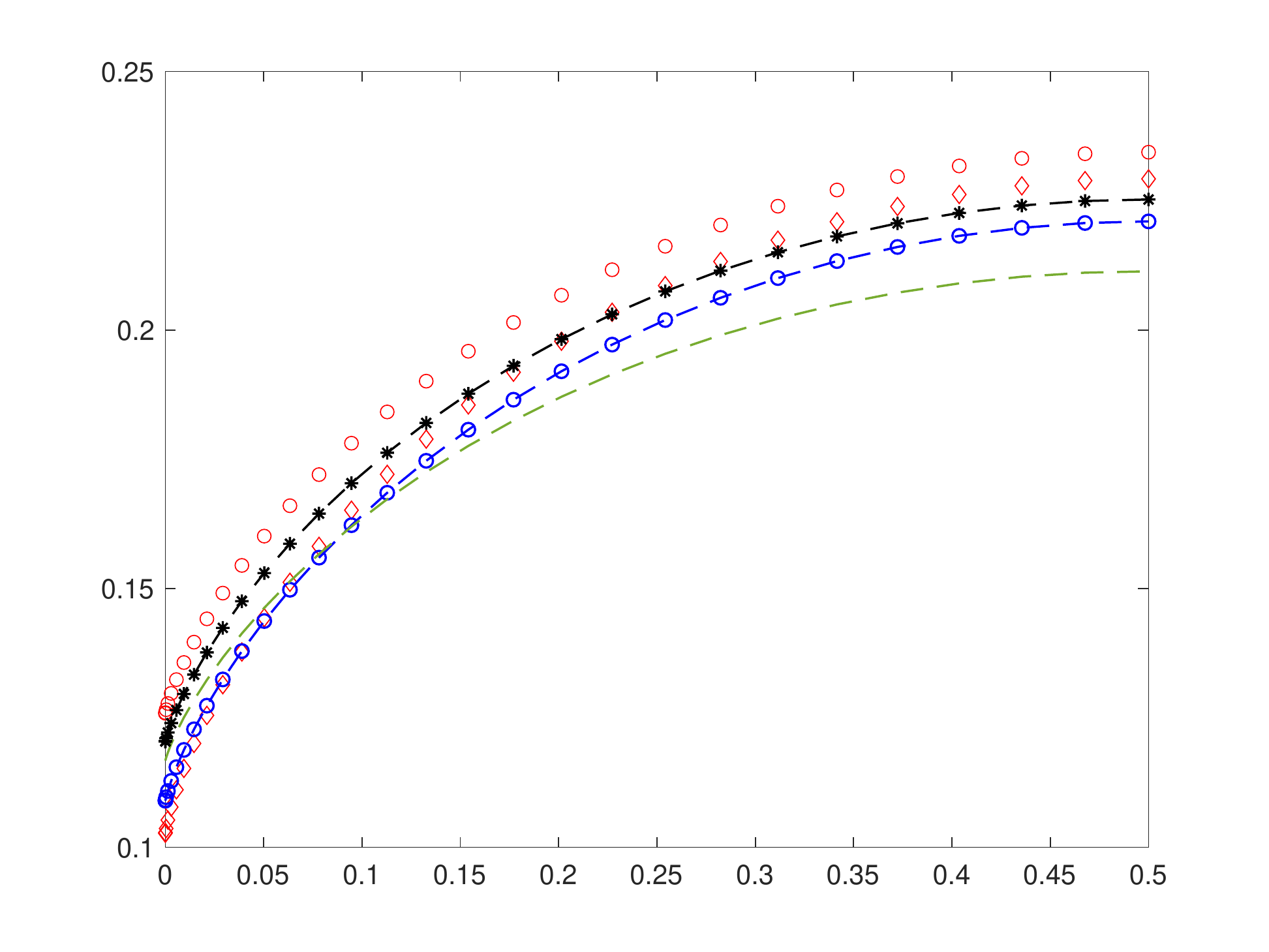}
	}
	\caption{ 
		Velocity profiles in the thermal transpiration between two parallel plates. Abscissas are for the spatial coordinate $x_2$ which is perpendicular to the two plates and normalized by the wall distance.  The two plates are located in $x_2=0$ and 1. Due to symmetry, only half spatial region is shown. Note that the Shakhov and ESBGK models do not distinguish the influence of intermolecular potential, and the $\nu$-model is reduced to the Shakhov model for Maxwellian gas.
	}
	\label{compare_thermal_transpiration}
\end{figure}

Figure~\ref{compare_thermal_transpiration} shows the induced velocity for Maxwell and hard-sphere gases. Numerical solutions of the Boltzmann equation with different values of the viscosity index $\omega$ are different. However, the viscosity index does not affect the solution in the Shakhov and ESBGK models. This is because the gas temperature does not change in the direction perpendicular to the solid wall, so that the coefficient in the collision operator~\eqref{normalized_model} has nothing to do with the viscosity index $\omega$. That is,
\begin{equation}
\begin{aligned}[b]
\frac{\sqrt{\pi}}{2Kn}\times\frac{\mu(T_0)}{n_0k_BT_0}\times\nu\left(\frac{c}{T}\right)= \frac{\sqrt{\pi}}{2Kn}.
\end{aligned}
\end{equation}
Thus, the Shakhov and ESBGK models with molecular-velocity-independent collision frequency don't have the degree of  freedom to describe the change of the intermolecular potential, while for the $\nu$-model the intermolecular potential has an impact on the velocity-dependent collision frequency~\eqref{nu_freq_power} and it predicts different results. 

When $Kn=0.01$, it is seen from figure~\ref{compare_thermal_transpiration}(a) that the Shakhov model well predicts the velocity profile of the Maxwell gas, while the ESBGK model predict a slight low velocity. However, both kinetic models cannot predict the velocity profile of hard-sphere gas. This problem is fixed in the $\nu$-model. When the Knudsen number is increased to 0.1, the Shakhov and ESBGK models predict a close velocity profile to that of the Maxwell and hard-sphere gas, respectively. When the $\nu$-model is used, good agreement with the Boltzmann equation solution is observed. When the Knudsen number further increases, the $\nu$-model always predicts better velocity profiles than the Shakhov and ESBGK models. 

From this test case we can clearly see that there are more degrees of freedom in the $\nu$-model to recover more details of the intermolecular collision, and thus yield more accurate results than the standard Shakhov and ESBGK models with velocity-independent collision frequency.


\subsection{Thermal transpiration in cavity}

We further investigate the thermal transpiration of a hard-sphere gas in a two-dimensional cavity with a length-to-width ratio of 5. The temperature at the right side is set to be twice that of the left side, while the temperature of the top and bottom walls varies linearly along the channel. The Knudsen number $Kn$ is defined at the average temperature of the left and right walls, the average molecular number density $n$, and the cavity height ${L}$. Due to symmetry, only the half spatial region $0\le{}y\le{L}/2$ is considered. 

\begin{figure}[ht]
	\centering
	\includegraphics[width=0.7\textwidth]{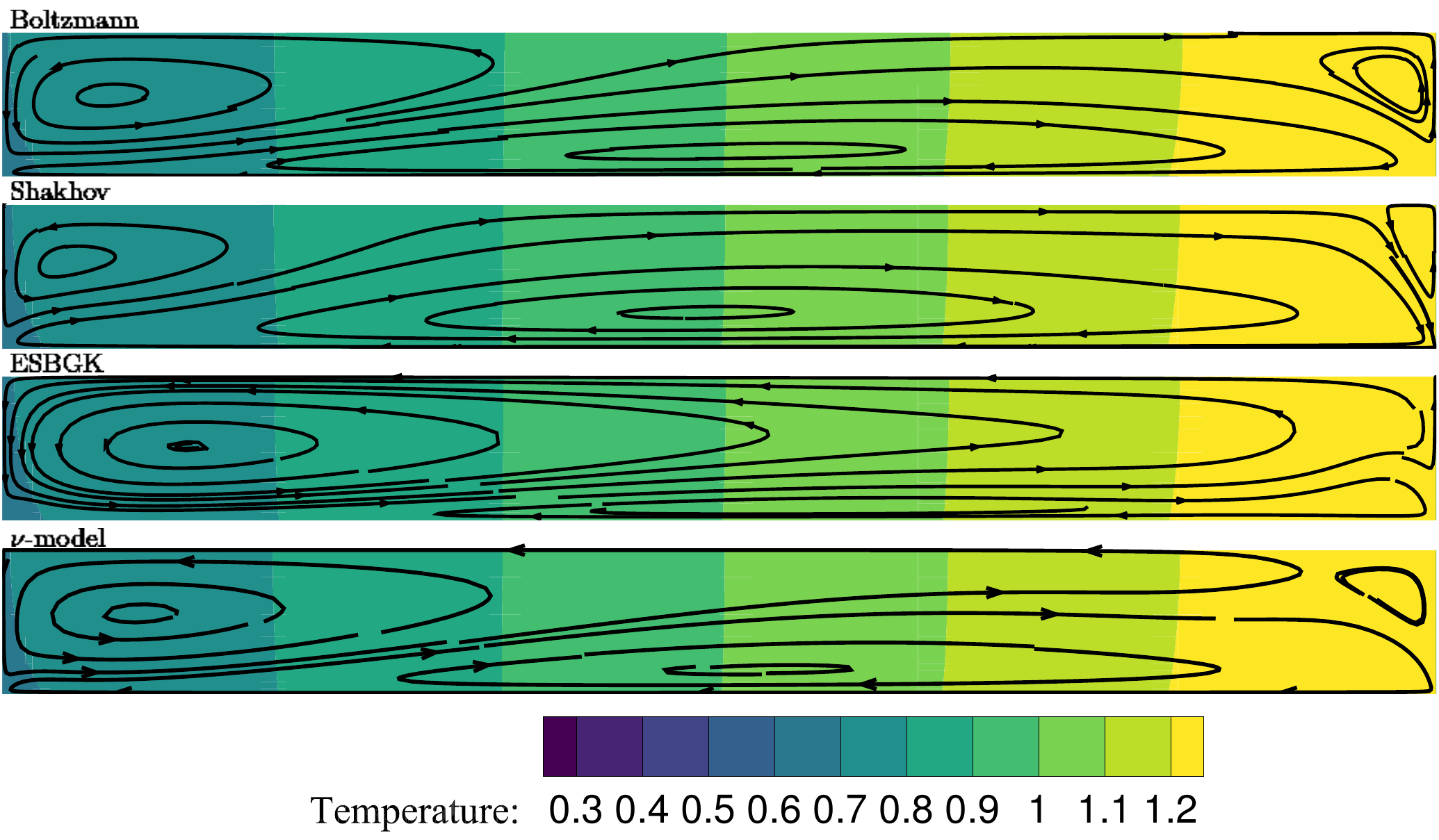}
	\caption{
		Thermal transpiration of hard-sphere gas in a rectangular cavity of aspect ratio 5: temperature fields and streamlines (in half of the channel) calculated by Boltzmann equation and different kinetic models. The Knudsen number is 0.5. 
	}
	\label{tc_nu_model_field}
\end{figure}

\begin{figure}[t]
	\centering
	\includegraphics[width=0.8\textwidth]{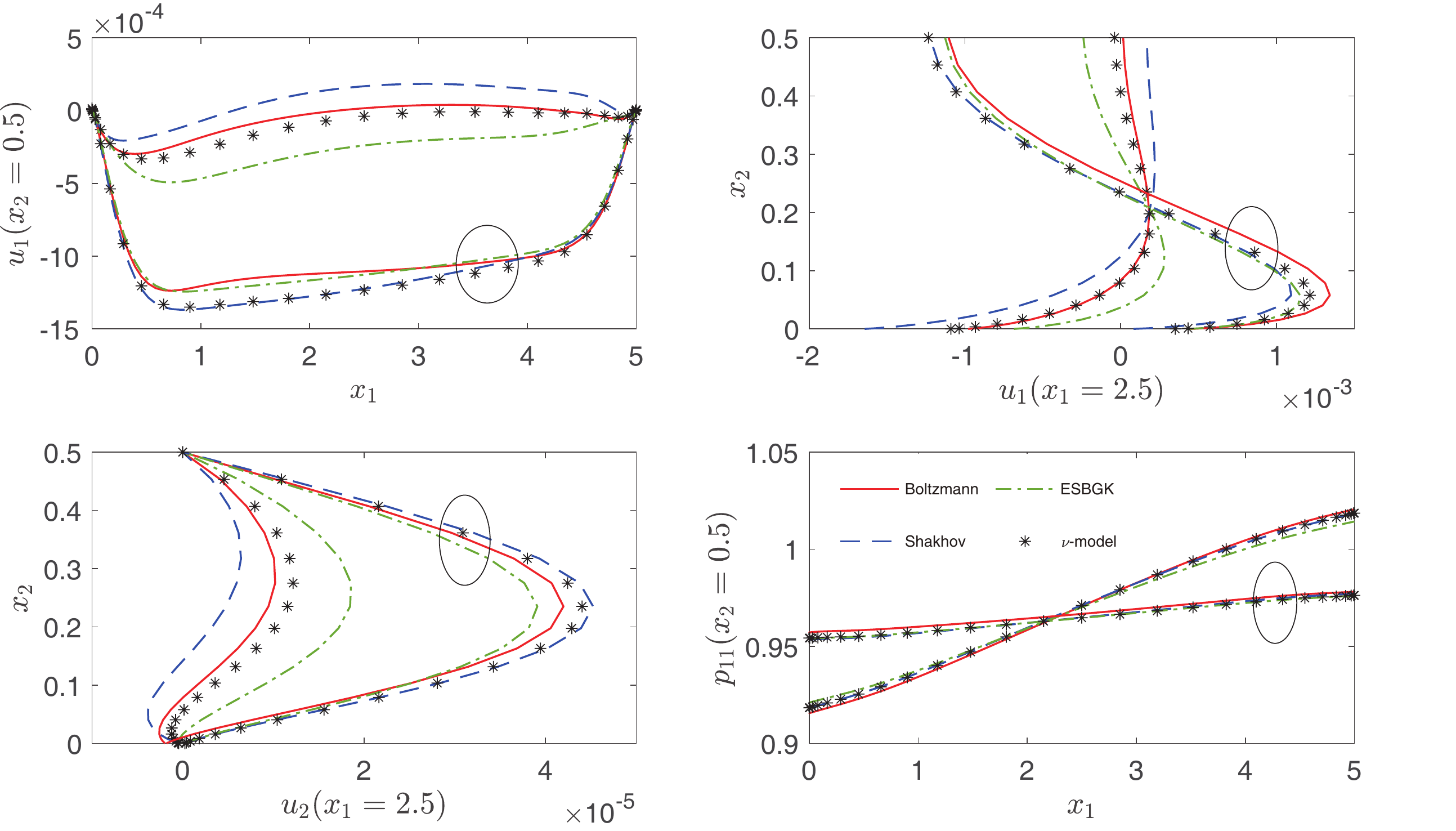}
	\caption{
		Thermal transpiration of hard-sphere gas in a rectangular cavity: velocity and normal stress profiles calculated by Boltzmann equation and different kinetic models. The cavity aspect ratio is 5, Knudsen numbers are 0.1 (results passing through circles) and 0.5, respectively.
	}
	\label{tc_nu_model_profile}
\end{figure}

The temperature fields and the streamlines obtained from the Boltzmann equation, Shakhov model, ESBGK model and $\nu$-model are compared in figure~\ref{tc_nu_model_field}, when $Kn=0.5$. All kinetic models predict good temperature field  with the Boltzmann solution, but not for velocity. For the Boltzmann solution the flow is characterised by three vortexes: the left vortex, bottom vortex and right vortex adjoining the left wall, bottom wall and right wall, respectively. For the Shakhov and ESBGK models their streamlines deviate largely from the Boltzmann solution in different trend. The Shakhov model predicts larger  bottom vortex but smaller right and left vortexes, while the ESBGK model predicts much larger left vortex with significantly shrunken bottom vortex and the right vortex completely disappears. By contrast, the $\nu$-model predicts nearly the same flow pattern with the Boltzmann solution. The velocity and normal stress profiles are further shown in figure~\ref{tc_nu_model_profile}, when $Kn=0.1$ and $0.5$. The profiles coincide with the above observations about the flow fields data that all kinetic models predict similar normal stress profiles agreeing well with the Boltzmann solution, but quite different velocity profiles where only those from the $\nu$-model show good agreement with the Boltzmann solution at different $Kn$ numbers. 

\section{Conclusions}\label{sec:conclusion}


The $\nu$-model  has been developed to better approximate the Boltzmann collision operator while keeping the computational cost at the same level with traditional gas kinetic models. The new model takes the relaxation-time approximation, where the target VDF to which the VDF relaxes is as simple as that in the Shakhov model, and the  collision frequency is a function of the molecular velocity. A multiscale numerical method is used to solve the proposed model equation deterministically. 

Based on the numerical simulation of normal shock waves, semi-empirical formula for the collision frequency are proposed for different intermolecular potentials, which showed certain universality for other rarefied gas flows. Specifically, in hypersonic flows, the overprediction of temperature and heat flux in the upstream of shock wave caused by the heating of high-speed reflected molecules is suppressed or even eliminated; in thermal transpiration, the $\nu$-model captures more derails of the intermolecular collision and predicts better results, while the Shakhov and ESBGK models with velocity-independent collision frequency cannot distinguish the influence of intermolecular potential.


In summary, the $\nu$-model is able to recover more details of the intermolecular collision and predict satisfactory results in a wide range of flow cases with various intermolecular potentials. In view of its good accuracy and easy implementation, we expect that it can be extended to better model rarefied flows of polyatomic gas and gas mixtures.

\section*{Acknowledgements}
This work is supported by the National Natural Science Foundation of China under the grant number 12172162 and the Guangdong-Hong Kong-Macao Joint Laboratory for Data-Driven Fluid Mechanics and Engineering Applications in China under grant 2020B1212030001.

\section*{Declaration of interests} 
The authors report no conflict of interest.


\bibliographystyle{jfm}
\bibliography{thesisBib}

\begin{thebibliography}{55}
\expandafter\ifx\csname natexlab\endcsname\relax\def\natexlab#1{#1}\fi
\def\au#1{#1} \def\ed#1{#1} \def\yr#1{#1}\def\at#1{#1}\def\jt#1{\textit{#1}}
  \def\bt#1{#1}\def\bvol#1{\textbf{#1}} \def\vol#1{#1} \def\pg#1{#1}
  \def\publ#1{#1}\def\arxiv#1{#1}\def\org#1{#1}\def\st#1{\textit{#1}}

\bibitem[Andries {\em et~al.\/}(2000)Andries, Tallec, Perlat \&
  Perthame]{Andries2000}
{\sc \au{Andries, P.}, \au{Tallec, P.~Le}, \au{Perlat, J.} \& \au{Perthame,
  B.}} \yr{2000}  \at{{The Gaussian-BGK model of Boltzmann equation with small
  Prandtl number}}.  \jt{Eur. J. Mech. B Fluids}  \bvol{19},  \pg{813--830}.

\bibitem[Bhatnagar {\em et~al.\/}(1954)Bhatnagar, Gross \&
  Krook]{Bhatnagar1954}
{\sc \au{Bhatnagar, P.~L.}, \au{Gross, E.~P.} \& \au{Krook, M.}} \yr{1954}
  \at{A model for collision processes in gases. {I}. {S}mall amplitude
  processes in charged and neutral one-component systems}.  \jt{Phys. Rev.}
  \bvol{94},  \pg{511--525}.

\bibitem[Bird(1963)]{Bird1963}
{\sc \au{Bird, G.~A.}} \yr{1963}  \at{Approach to translational equilibrium in
  a rigid sphere gas}.  \jt{Physics of Fluids}  \bvol{6}~(10),
  \pg{1518--1519}.

\bibitem[Bird(1994)]{Bird1994}
{\sc \au{Bird, G.~A.}} \yr{1994} {\em Molecular Gas Dynamics and the Direct
  Simulation of Gas Flows\/}.  \publ{Oxford University Press Inc, New York:
  Oxford Science Publications}.

\bibitem[Bird(2005)]{bird2005ds2v}
{\sc \au{Bird, G.~A.}} \yr{2005} {The DS2V/3V program suite for DSMC
  calculations}.  \bt{In {\em AIP conference proceedings\/}}, ,  \vol{vol.
  762},  \pg{pp. 541--546}. American Institute of Physics.

\bibitem[Cercignani(1966)]{Cercignani1966}
{\sc \au{Cercignani, C.}} \yr{1966}  \at{{The method of elementary solutions
  for kinetic models with velocity-dependent collision frequency}}.  \jt{Ann.
  Phys.}  \bvol{40},  \pg{469}.

\bibitem[Cercignani(1975)]{cercignani1975theory}
{\sc \au{Cercignani, Carlo}} \yr{1975} {\em Theory and application of the
  {B}oltzmann equation\/}.  \publ{Scottish Academic Press}.

\bibitem[Cercignani(2000)]{Cercignani2000Book}
{\sc \au{Cercignani, C.}} \yr{2000} {\em Rarefied Gas Dynamics From Basic
  Concepts to Actual Calculations\/}.  \publ{Cambridge University Press}.

\bibitem[Chapman \& Cowling(1970)]{CE}
{\sc \au{Chapman, S.} \& \au{Cowling, T.G.}} \yr{1970} {\em {The Mathematical
  Theory of Non-uniform Gases}\/}.  \publ{Cambridge University Press}.

\bibitem[Chen {\em et~al.\/}(2015)Chen, Xu \& Cai]{Chensz2013}
{\sc \au{Chen, S.~Z.}, \au{Xu, K.} \& \au{Cai, Q.~D.}} \yr{2015}  \at{A
  comparison and unification of ellipsoidal statistical and {Shakhov BGK}
  models}.  \jt{Advances in Applied Mathematics and Mechanics}  \bvol{7},
  \pg{245--266}.

\bibitem[Fei {\em et~al.\/}(2020)Fei, Liu, Liu \& Zhang]{Fei2020AIAA}
{\sc \au{Fei, F.}, \au{Liu, H.~L.}, \au{Liu, Z.~H.} \& \au{Zhang, J.}}
  \yr{2020}  \at{A benchmark study of kinetic models for shock waves}.
  \jt{AIAA Journal}  \bvol{58},  \pg{2596--2608}.

\bibitem[Gorji \& Jenny(2013)]{Gorji2013}
{\sc \au{Gorji, M.~H.} \& \au{Jenny, P.}} \yr{2013}  \at{A {Fokker-Planck}
  based kinetic model for diatomic rarefied gas flows}.  \jt{Phys. Fluids}
  \bvol{25},  \pg{062002}.

\bibitem[Gorji {\em et~al.\/}(2011)Gorji, Torrilhon \& Jenny]{Gorji2011}
{\sc \au{Gorji, M.~H.}, \au{Torrilhon, M.} \& \au{Jenny, P.}} \yr{2011}
  \at{{Fokker-Planck model for computational studies of monatomic rarefied gas
  flows}}.  \jt{J. Fluid Mech.}  \bvol{680},  \pg{574--601}.

\bibitem[Gross \& Jackson(1959)]{GrossJackson1959}
{\sc \au{Gross, E.~P.} \& \au{Jackson, E.~A.}} \yr{1959}  \at{Kinetic models
  and the linearized {B}oltzmann equation}.  \jt{Phys. Fluids}  \bvol{2}~(4),
  \pg{432--441}.

\bibitem[Gupta \& Gianchandani(2008)]{gupta2008thermal}
{\sc \au{Gupta, N.~K.} \& \au{Gianchandani, Y.~B.}} \yr{2008}  \at{{Thermal
  transpiration in zeolites: A mechanism for motionless gas pumps}}.  \jt{Appl.
  Phys. Lett.}  \bvol{93}~(19),  \pg{193511}.

\bibitem[Holway(1966)]{Holway1966}
{\sc \au{Holway, L.~H.}} \yr{1966}  \at{New statistical models for kinetic
  theory: methods of construction}.  \jt{Phys. Fluids}  \bvol{9},
  \pg{1658--1673}.

\bibitem[Ivanov \& Gimelshein(1998)]{Ivanov1998Review}
{\sc \au{Ivanov, M.~S.} \& \au{Gimelshein, S.~F.}} \yr{1998}  \at{Computational
  hypersonic rarefied flows}.  \jt{Ann. Rev. Fluid Mech.}  \bvol{30},
  \pg{469--505}.

\bibitem[Jenny {\em et~al.\/}(2010)Jenny, Torrilhon \& Heinz]{Jenny2010JCP}
{\sc \au{Jenny, P.}, \au{Torrilhon, M.} \& \au{Heinz, S.}} \yr{2010}  \at{A
  solution algorithm for the fluid dynamic equations based on a stochastic
  model for molecular motion}.  \jt{J. Comput. Phys.}  \bvol{229},
  \pg{1077--1098}.

\bibitem[Karniadakis {\em et~al.\/}(2005)Karniadakis, Beskok \&
  Aluru]{Karniadakis2005}
{\sc \au{Karniadakis, G.}, \au{Beskok, A.} \& \au{Aluru, N.}} \yr{2005} {\em
  Microflows and Nanoflows: Fundamentals and Simulation\/}.  \publ{233 Spring
  St, New York: Springer Science+Business Media, Inc.}

\bibitem[Krook(1959)]{krook1959continuum}
{\sc \au{Krook, Max}} \yr{1959}  \at{Continuum equations in the dynamics of
  rarefied gases}.  \jt{Journal of Fluid Mechanics}  \bvol{6}~(4),
  \pg{523--541}.

\bibitem[Larina \& Rykov(2007)]{larina2007models}
{\sc \au{Larina, Irine~Nikolaevna} \& \au{Rykov, Vladimir~Alekseevich}}
  \yr{2007}  \at{Models of a linearized {B}oltzmann collision integral}.
  \jt{Computational Mathematics and Mathematical Physics}  \bvol{47}~(6),
  \pg{983--997}.

\bibitem[Liu {\em et~al.\/}(2019)Liu, Yuan, Javid \& Zhong]{Liu2019PRE}
{\sc \au{Liu, S.}, \au{Yuan, R.~F.}, \au{Javid, U.} \& \au{Zhong, C.~W.}}
  \yr{2019}  \at{{Conservative discrete-velocity method for the ellipsoidal
  Fokker-Planck equation in gas-kinetic theory}}.  \jt{Phys. Rev. E}
  \bvol{100},  \pg{033310}.

\bibitem[Liu \& Zhong(2014)]{liu2014investigation}
{\sc \au{Liu, Sha} \& \au{Zhong, Chengwen}} \yr{2014}  \at{Investigation of the
  kinetic model equations}.  \jt{Physical Review E}  \bvol{89}~(3),
  \pg{033306}.

\bibitem[Loyalka \& Ferziger(1967)]{Loyalka1967}
{\sc \au{Loyalka, S.~K.} \& \au{Ferziger, J.~H.}} \yr{1967}  \at{Model
  dependence of the slip coefficient}.  \jt{Phys. Fluids}  \bvol{10},
  \pg{1833--1839}.

\bibitem[Loyalka \& Ferziger(1968)]{Loyalka1968}
{\sc \au{Loyalka, S.~K.} \& \au{Ferziger, J.~H.}} \yr{1968}  \at{Model
  dependence of the temperature slip coefficient}.  \jt{Phys. Fluids}
  \bvol{11},  \pg{1168--1671}.

\bibitem[Maxwell(1879)]{Maxwell1879vii}
{\sc \au{Maxwell, J.~C.}} \yr{1879}  \at{{VII}. {On} stresses in rarified gases
  arising from inequalities of temperature}.  \jt{Proc. Royal Soc. Lond.}
  \bvol{170},  \pg{231--256}.

\bibitem[Mieussens(2000{\natexlab{{\em a\/}}})]{Luc2000JCP}
{\sc \au{Mieussens, L.}} \yr{2000{\natexlab{{\em a\/}}}}  \at{{ Discrete
  velocity models and numerical schemes for the Boltzmann-BGK equation in plane
  and axisymmetric geometries}}.  \jt{J. Comput. Phys.}  \bvol{162},
  \pg{429--466}.

\bibitem[Mieussens(2000{\natexlab{{\em b\/}}})]{mieussens2000discrete}
{\sc \au{Mieussens, Luc}} \yr{2000{\natexlab{{\em b\/}}}}  \at{Discrete
  velocity model and implicit scheme for the {BGK} equation of rarefied gas
  dynamics}.  \jt{Mathematical Models and Methods in Applied Sciences}
  \bvol{10},  \pg{1121--1149}.

\bibitem[Mieussens \& Struchtrup(2004)]{Mieussens2004}
{\sc \au{Mieussens, L.} \& \au{Struchtrup, H.}} \yr{2004}  \at{{Numerical
  comparison of Bhatnagar-Gross-Krook models with proper Prandtl number}}.
  \jt{Phys. Fluids}  \bvol{16}~(8),  \pg{2297--2813}.

\bibitem[Reynolds(1879)]{Reynolds1879}
{\sc \au{Reynolds, O.}} \yr{1879}  \at{On certain dimensional properties of
  matter in the gaseous state}.  \jt{Philos. Trans. R. Soc. Part 1}
  \bvol{170},  \pg{727--845}.

\bibitem[Rogers(1995)]{Rogers1995Comparison}
{\sc \au{Rogers, Stuart~E}} \yr{1995}  \at{Comparison of implicit schemes for
  the incompressible {Navier-Stokes} equations}.  \jt{AIAA Journal}
  \bvol{33}~(11),  \pg{2066--2072}.

\bibitem[Shakhov(1968{\natexlab{{\em a\/}}})]{Shakhov1968}
{\sc \au{Shakhov, E.~M.}} \yr{1968{\natexlab{{\em a\/}}}}  \at{Approximate
  kinetic equations in rarefied gas theory}.  \jt{Fluid Dynamics}  \bvol{3},
  \pg{112--115}.

\bibitem[Shakhov(1968{\natexlab{{\em b\/}}})]{Shakhov_S}
{\sc \au{Shakhov, E.~M.}} \yr{1968{\natexlab{{\em b\/}}}}  \at{{Generalization
  of the Krook kinetic relaxation equation}}.  \jt{Fluid Dyn.}  \bvol{3}~(5),
  \pg{95--96}.

\bibitem[Sharipov \& Benites(2017)]{Sharipov_quantum_DCS2017}
{\sc \au{Sharipov, F.} \& \au{Benites, V.~J.}} \yr{2017}  \at{Transport
  coefficients of helium-neon mixtures at low density computed from ab initio
  potentials}.  \jt{J. Chem. Phys}  \bvol{147},  \pg{224302}.

\bibitem[Sharipov \& Bertoldo(2009)]{Sharipov2009}
{\sc \au{Sharipov, F.} \& \au{Bertoldo, G.}} \yr{2009}  \at{{Poiseuille flow
  and thermal creep based on the Boltzmann equation with the Lennard-Jones
  potential over a wide range of the Knudsen number}}.  \jt{Phys. Fluids}
  \bvol{21},  \pg{067101}.

\bibitem[Sharipov \& Seleznev(1998)]{sharipov1998data}
{\sc \au{Sharipov, F.} \& \au{Seleznev, V.}} \yr{1998}  \at{Data on internal
  rarefied gas flows}.  \jt{J. Phys. Chem. Ref. Data}  \bvol{27},
  \pg{657--706}.

\bibitem[Sone(2002)]{Sone2002Book}
{\sc \au{Sone, Y.}} \yr{2002} {\em Kinetic theory and fluid dynamics\/}.
  \publ{Birkhauser Boston}.

\bibitem[Struchtrup(1997)]{Struchtrup_velocity}
{\sc \au{Struchtrup, H.}} \yr{1997}  \at{{The BGK-model with velocity-dependent
  collision frequency}}.  \jt{Cont. Mech. Theromodyn.}  \bvol{9},  \pg{23--32}.

\bibitem[Struchtrup(2005)]{henning}
{\sc \au{Struchtrup, H.}} \yr{2005} {\em Macroscopic Transport Equations for
  Rarefied Gas Fows: Approximation Methods in Kinetic Theory\/}.
  \publ{Heidelberg, Germany: Springer}.

\bibitem[Takata \& Funagane(2011)]{Takata2011}
{\sc \au{Takata, S.} \& \au{Funagane, H.}} \yr{2011}  \at{Poiseuille and
  thermal transpiration flows of a highly rarefied gas: over-concentration in
  the velocity distribution function}.  \jt{J. Fluid Mech.}  \bvol{669},
  \pg{242--259}.

\bibitem[Valentini \& Schwartzentruber({2009})]{Valentini2009}
{\sc \au{Valentini, P.} \& \au{Schwartzentruber, T.~E.}} \yr{{2009}}
  \at{{Large-scale molecular dynamics simulations of normal shock waves in
  dilute argon}}.  \jt{Phys. Fluids}  \bvol{{21}}~({6}).

\bibitem[Vargo {\em et~al.\/}(1999)Vargo, Muntz, Shiflett \&
  Tang]{vargo1999knudsen}
{\sc \au{Vargo, S.~E.}, \au{Muntz, E.~P.}, \au{Shiflett, G.~R.} \& \au{Tang,
  W.~C.}} \yr{1999}  \at{Knudsen compressor as a micro-and macroscale vacuum
  pump without moving parts or fluids}.  \jt{J. Vac. Sci. Technol. A: Vacuum,
  Surfaces, and Films}  \bvol{17},  \pg{2308--2313}.

\bibitem[Wang {\em et~al.\/}(2018)Wang, Ho, Wu, Guo \& Zhang]{WANG201833}
{\sc \au{Wang, P.}, \au{Ho, M.~T.}, \au{Wu, L.}, \au{Guo, Z.~L.} \& \au{Zhang,
  Y.~H.}} \yr{2018}  \at{A comparative study of discrete velocity methods for
  low-speed rarefied gas flows}.  \jt{Computers \& Fluids}  \bvol{161},  \pg{33
  -- 46}.

\bibitem[Wang {\em et~al.\/}(2020)Wang, Su \& Wu]{Wang2020PoF}
{\sc \au{Wang, P.}, \au{Su, W.} \& \au{Wu, L.}} \yr{2020}  \at{Thermal
  transpiration in molecular gas}.  \jt{Phys. Fluids}  \bvol{32},  \pg{082005}.

\bibitem[Wu {\em et~al.\/}(2017)Wu, Ho, Germanou, Gu, Liu, Xu \&
  Zhang]{Wu2017JFM2}
{\sc \au{Wu, L.}, \au{Ho, M.~H.}, \au{Germanou, L.}, \au{Gu, X.~J.}, \au{Liu,
  C.}, \au{Xu, K.} \& \au{Zhang, Y.~H.}} \yr{2017}  \at{On the apparent
  permeability of porous media in rarefied gas flows}.  \jt{J. Fluid Mech.}
  \bvol{822},  \pg{398--417}.

\bibitem[Wu {\em et~al.\/}(2016)Wu, Liu, Reese \& Zhang]{Wu2016JFM}
{\sc \au{Wu, L.}, \au{Liu, H.~H.}, \au{Reese, J.~M.} \& \au{Zhang, Y.~H.}}
  \yr{2016}  \at{Non-equilibrium dynamics of dense gas under tight
  confinement}.  \jt{J. Fluid Mech.}  \bvol{794},  \pg{252--266}.

\bibitem[Wu {\em et~al.\/}(2015{\natexlab{{\em a\/}}})Wu, Liu, Zhang \&
  Reese]{wuPoF2015}
{\sc \au{Wu, L.}, \au{Liu, H.~H.}, \au{Zhang, Y.~H.} \& \au{Reese, J.~M.}}
  \yr{2015{\natexlab{{\em a\/}}}}  \at{{Influence of intermolecular potentials
  on rarefied gas flows: Fast spectral solutions of the Boltzmann equation}}.
  \jt{Phys. Fluids}  \bvol{27},  \pg{082002}.

\bibitem[Wu {\em et~al.\/}(2014)Wu, Reese \& Zhang]{lei_Jfm}
{\sc \au{Wu, L.}, \au{Reese, J.~M.} \& \au{Zhang, Y.~H.}} \yr{2014}
  \at{{Solving the Boltzmann equation by the fast spectral method: application
  to microflows}}.  \jt{J. Fluid Mech.}  \bvol{746},  \pg{53--84}.

\bibitem[Wu {\em et~al.\/}(2013)Wu, White, Scanlon, Reese \& Zhang]{Lei2013}
{\sc \au{Wu, L.}, \au{White, C.}, \au{Scanlon, T.~J.}, \au{Reese, J.~M.} \&
  \au{Zhang, Y.~H.}} \yr{2013}  \at{{Deterministic numerical solutions of the
  {B}oltzmann equation using the fast spectral method}}.  \jt{J. Comput. Phys.}
   \bvol{250},  \pg{27--52}.

\bibitem[Wu {\em et~al.\/}(2015{\natexlab{{\em b\/}}})Wu, White, Scanlon, Reese
  \& Zhang]{LeiJFM2015}
{\sc \au{Wu, L.}, \au{White, C.}, \au{Scanlon, T.~J.}, \au{Reese, J.~M.} \&
  \au{Zhang, Y.~H.}} \yr{2015{\natexlab{{\em b\/}}}}  \at{A kinetic model of
  the {Boltzmann equation for} non-vibrating polyatomic gases}.  \jt{J. Fluid
  Mech.}  \bvol{763},  \pg{24--50}.

\bibitem[Yang {\em et~al.\/}(2019)Yang, Shu, Yang, Wu \&
  Zhang]{yang2019numerical}
{\sc \au{Yang, LM}, \au{Shu, C}, \au{Yang, WM}, \au{Wu, J} \& \au{Zhang, MQ}}
  \yr{2019}  \at{Numerical investigation on performance of three solution
  reconstructions at cell interface in {DVM} simulation of flows in all
  {Knudsen} number regimes}.  \jt{International Journal for Numerical Methods
  in Fluids}  \bvol{90}~(11),  \pg{545--563}.

\bibitem[Yuan(2002)]{Yuan2002Comparison}
{\sc \au{Yuan, L.}} \yr{2002}  \at{Comparison of implicit multigrid schemes for
  three-dimensional incompressible flows}.  \jt{J. Comput. Phys.}
  \bvol{177}~(1),  \pg{134--155}.

\bibitem[Yuan \& Zhong(2019)]{yuan2019conservative}
{\sc \au{Yuan, R.~F.} \& \au{Zhong, C.~W.}} \yr{2019}  \at{A conservative
  implicit scheme for steady state solutions of diatomic gas flow in all flow
  regimes}.  \jt{Computer Physics Communications}  \pg{p. 106972}.

\bibitem[Zheng \& Struchtrup({2005})]{Zheng2005}
{\sc \au{Zheng, Y.~S.} \& \au{Struchtrup, H.}} \yr{{2005}}  \at{{Ellipsoidal
  statistical Bhatnagar-Gross-Krook model with velocity-dependent collision
  frequency}}.  \jt{Phys. Fluids}  \bvol{{17}}~({12}),  \pg{127103}.

\bibitem[Zhu {\em et~al.\/}(2016)Zhu, Zhong \& Xu]{zhuyajun2016}
{\sc \au{Zhu, Y.~J.}, \au{Zhong, C.~W.} \& \au{Xu, K.}} \yr{2016}  \at{Implicit
  unified gas-kinetic scheme for steady state solutions in all flow regimes}.
  \jt{J. Comput. Phys.}  \bvol{315},  \pg{16--38}.

\end{thebibliography}

\end{document}